\documentclass[aps,prb,reprint,superscriptaddress,longbibliography]{revtex4-1}

\usepackage{graphicx}
\usepackage{latexsym}
\usepackage{amsmath}
\usepackage{amssymb}
\usepackage{amsfonts}
\usepackage{color}
\usepackage{bm}
\usepackage{verbatim}
\usepackage[]{color}
\definecolor{red}{rgb}{1,0,0}
\usepackage{framed}
\definecolor{shadecolor}{RGB}{222,222,221}

\begin{document}

\title{N\'eel proximity effect in superconductor/antiferromagnet heterostructures}

 \date{\today}
 
\author{I.V. Bobkova}
\affiliation{Moscow Institute of Physics and Technology, Dolgoprudny, 141700 Russia}

\author{G.A. Bobkov}
\affiliation{Moscow Institute of Physics and Technology, Dolgoprudny, 141700 Russia}

\author{V.M. Gordeeva}
\affiliation{Moscow Institute of Physics and Technology, Dolgoprudny, 141700 Russia}

\author{A.M. Bobkov}
\affiliation{Moscow Institute of Physics and Technology, Dolgoprudny, 141700 Russia}


 \begin{abstract}
 
It is well-known that the cornerstone of the proximity effect in superconductor/ferromagnet heterostructures is a generation of triplet Cooper pairs from singlet Cooper pairs inherent in a conventional superconductor. This proximity effect brought a lot of new exciting physics and gave a powerful impulse to development of superconducting spintronics. Nowadays a new direction of spintronics is actively developing, which is based on antiferromagnets and their heterostructures. It is called antiferromagnetic spintronics. By analogy with an important role played by triplet Cooper pairs in conventional superconducting spintronics based on ferromagnets the question arises: does the triplet proximity effect exist in superconductor/antiferromagnet heterostructures and, if so, what are the properties of the induced triplet correlations and the prospects for use in superconducting spintronics? Recent theoretical findings predict that despite the absence of a net magnetization, the N\'eel magnetic order of the antiferromagnet does give rise to specific spin-triplet correlations at superconductor/antiferromagnet interfaces. They were called N\'eel triplet correlations. The goal of this review is to discuss
the current understanding of the fundamental physics of
these N\'eel triplet correlations and their physical manifestations.

 \end{abstract}

 \maketitle
 
\tableofcontents{}

\section{Introduction}

Superconducting proximity effect in mesoscopic heterostructures composed of conventional superconductors and normal, that is nonsuperconducting and nonmagnetic, metals (S/N heterostructures) is a penetration of Cooper pairs from a superconductor into an adjacent nonsuperconducting material with partial suppression of superconducting order parameter in the superconductor near the interface.   Conventional superconductors are formed by spin-singlet Cooper pairs~\cite{Cooper1956,Bardeen1957} and, therefore, they induce spin-singlet superconducting correlations in the adjacent normal metal. If a normal metal is replaced with a ferromagnet, the spin-singlet pairs, which penetrate into the ferromagnet, are partially converted  into their spin-triplet counterparts due to the presence of a macroscopic spin-splitting field in it \cite{Bergeret2005,Buzdin2005,Bergeret2018}. Simultaneously the triplet pairs are also induced in the superconductor due to an inverse proximity effect. The same effect occurs if a thin superconducting film is subjected to a parallel magnetic field or if the superconductor is contacted with a ferromagnetic insulator \cite{Eschrig2015}. The triplet pairs are produced at the expense of the singlet ones. This weakens the conventional superconducting state and lowers the critical temperature~\cite{Chandrasekhar1962,Clogston1962,Sarma1963}. The generation of triplet Cooper pairs in superconductor/ferromagnet (S/F) heterostructures brought a lot of new exciting physics \cite{Buzdin2005,Bergeret2005} and gave a powerful impulse to development of superconducting spintronics \cite{Linder_review,Eschrig_review,Bergeret2018}.

Now what is about the proximity effect in superconductor/antiferromagnet (S/AF) heterostructures? Naively, since the net magnetization in an antiferromagnet averaged over the size of a typical Cooper pair vanishes, one should expect that S/AF heterostructures behave like S/N heterostructures from the point of view of the proximity effect. This means penetration of only singlet pairs into the antiferromagnetic metal and that the superconductor  interfaced to the antiferromagnetic metal or insulator via a compensated interface is expected to experience no net spin-splitting or reduction in critical temperature~\cite{Werthammer1966}. Any macroscopic spin-splitting in S/AF heterostructures is only expected via an uncompensated (non-zero) interface magnetization. Indeed, it was predicted that the uncompensated interface induces a spin-splitting field in thin-film S/AF bilayers \cite{Kamra2018}, which can result in some physical effects similar to thin-film S/F heterostructures \cite{Bergeret2018}, for example in the giant thermoelectric effect \cite{Bobkov2021}, the anomalous phase shift \cite{Rabinovich2019} and anisotropy of the critical current in S/AF/S Josephson
junctions with spin-orbit coupling \cite{Falch2022}. 

However, it was realized long ago that in fact antiferromagnetism influences superconductivity not only via the uncompensated interface magnetization. In particular, it was reported that in antiferromagnetic superconductors the staggered exchange field suppresses superconductivity due to
changes in the density of states and due to atomic oscillations of electronic wave functions \cite{Buzdin1986}. The atomic oscillations of the electronic wave functions in antiferromagnetic materials also lead to the fact that nonmagnetic impurities in antiferromagnetic superconductors behave like effectively magnetic and additionally suppress superconductivity \cite{Buzdin1986}. Further the theory taking into account the suppression of superconductivity by magnetic impurities was also developed for S/AF heterostructures \cite{Fyhn2022,Fyhn2022_1}. Several experiments have found that AFs lower the critical temperature of an S layer~\cite{Bell2003,Hubener2002,Wu2013,Seeger2021}, despite there is no net spin-splitting. In some cases, the effect has been comparable to or even larger than that induced by a ferromagnet layer~\cite{Wu2013}. Of course, a number of physical reasons can contribute to this observation. First, an AF doubles the spatial period of the lattice due to its antiparallel spins on the two sublattices. This can open a bandgap in the adjacent conductor, which may reduce the normal-state density of states in S and thus suppress superconductivity~\cite{Buzdin1986,Krivoruchko1996}. Second, partially the suppresion can result from the uncompensated magnetization of the S/AF  interface, which seems to be common in experiments~\cite{Belashchenko2010,Kamra2018,He2010,Manna2014} and induces a spin-splitting and spin-flipping disorder in the superconductor, just like a ferromagnet \cite{Kamra2018}. Furthermore, the nonmagnetic disorder in the S/AF system also suppresses superconductivity, as it was mentioned above. Although all these physical effects are likely present in real systems, they are not associated with a physics of proximity-induced triplet superconducting correlations, which are a cornerstone of the physics and applications of S/F hybrids. Therefore, an important question arises: is it true that triplet correlations in superconductor/antiferromagnet hybrid systems can arise only due to uncompensated surface magnetization or the N\'eel magnetism itself is
capable of generating new types of superconducting proximity effect?

Some unconventional for S/N heterostructures physical effects in S/AF hybrids with compensated interfaces were reported in the literature. For example, unconventional Andreev reflection and bound states at such S/AF interfaces have been  predicted~\cite{Bobkova2005,Andersen2005}. The atomic-thickness $0-\pi$ transitions in S/AF/S Josephson junctions were investigated theoretically~\cite{Andersen2006,Enoksen2013,Bulaevskii2017}, an analysis of a hybrid comprising a ferromagnet and a compensated AF interfaced with an S suggested the interface to be spin-active~\cite{Johnsen2021}. All these results indicated that a key piece of  understanding of the physics of S/AF hybrids was missing.

Further it was found that in spite of the absence of a net magnetization, the N\'eel magnetic order of the AF induces spin-triplet correlations at S/AF interfaces, which penetrate into the superconductor and into the antiferromagnet (if it is metallic) \cite{Bobkov2022}. Their amplitude flips sign from one lattice site to the next, just like the N\'eel magnetic order in the AF. These correlations were called N\'eel triplet Cooper pairs. The goal of this review is to discuss the current understanding of the fundamental physics of these N\'eel triplet correlations and their physical manifestations.  

\section{N\'eel triplet correlations at AF/S interfaces}

\subsection{Bogolubov-de Gennes visualization of the N\'eel triplets}

\begin{figure}[tbh]
	\begin{center}
		\includegraphics[width=85mm]{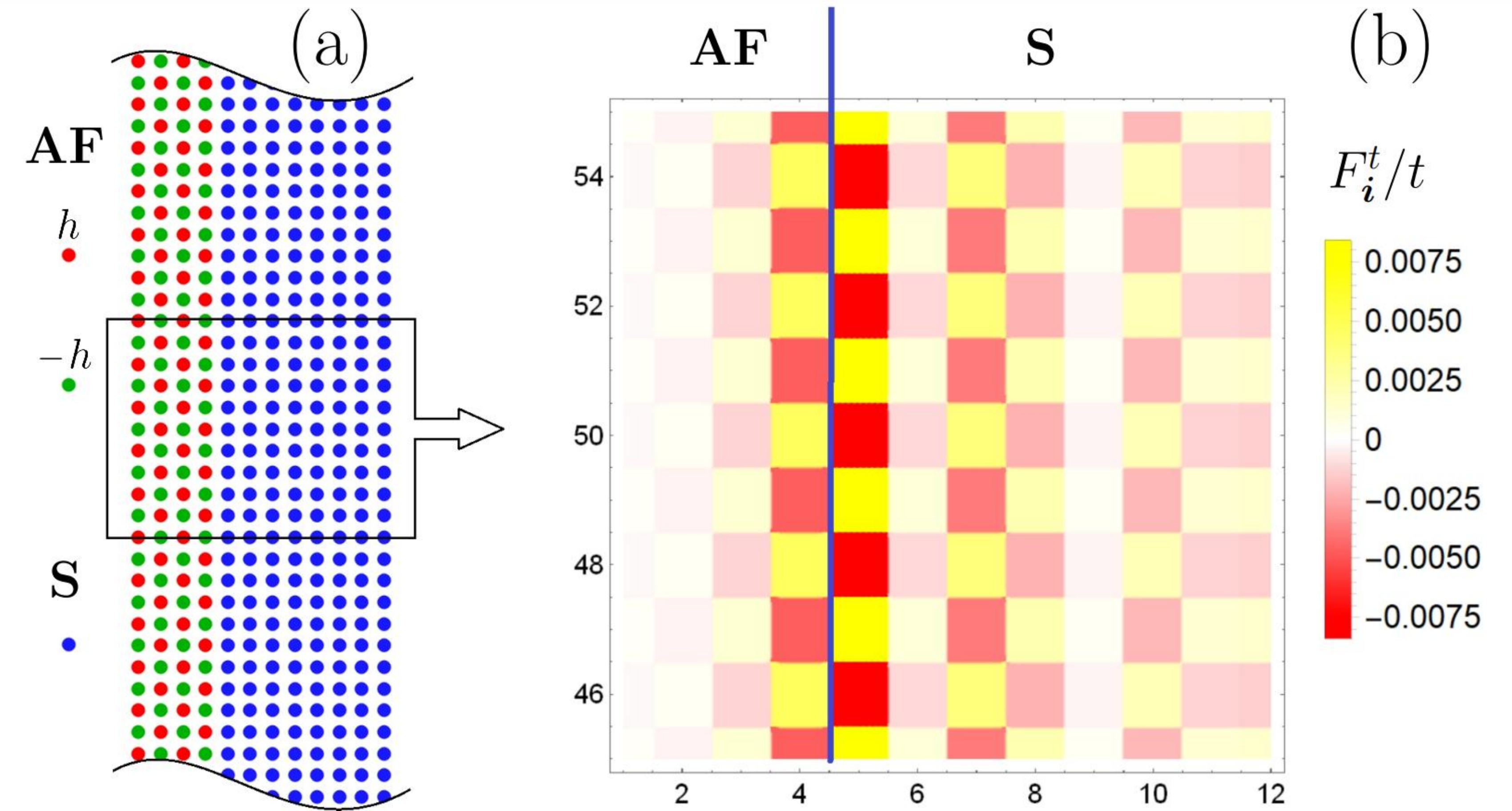}
		\caption{(a) Sketch of the antiferromagnetic insulator interfaced via a compensated interface to the thin superconductor.  (b) Spatial variation of the triplet correlations amplitude $F_{\bm i}^t$. Each colored square codes the value of $F_{\bm i}^t$ at a given site. An alternating sign of the correlations in S commensurate with the N\'eel order in the AF can be seen along the interfacial direction. The picture is adopted from Ref.~\onlinecite{Bobkov2022}.} \label{fig:trip}
	\end{center}
\end{figure}

Let us consider an antiferromagnetic insulator interfaced via a compensated interface to a thin conventional $s$-wave superconductor [Fig.~\ref{fig:trip}(a)]. The system can be described by the following tight-binding Hamiltonian:
\begin{align}
H= - t \sum \limits_{\langle \bm{i}\bm{j}, \rangle \sigma} \hat c_{\bm{i} \sigma}^\dagger \hat c_{\bm{j} \sigma} + \sum \limits_{ \bm{i}} (\Delta_{\bm{i}} \hat c_{\bm{i}\uparrow}^\dagger \hat c_{\bm{i}\downarrow}^\dagger + H.c.) - \nonumber \\
\mu \sum \limits_{\bm{i} \sigma} \hat n_{\bm{i}\sigma} + \sum \limits_{\bm{i},\alpha \beta} \hat c_{\bm{i}\alpha}^\dagger (\bm{h}_{\bm{i}} \bm{\sigma})_{\alpha \beta} \hat c_{\bm{i}\beta} ,
\label{ham}
\end{align}
where $\bm i = (i_x,i_y)^T$ is the radius vector of the site and greek letters correspond to the spin indices. $\langle \bm i \bm j \rangle$ means summation over the nearest neighbors. $\Delta_{\bm i}$ and $\bm h_{\bm i}$ denote the on-site superconducting order parameter and the magnetic exchange field at site $\bm i$, respectively. $\hat c_{\bm i \sigma}^\dagger (\hat c_{\bm i \sigma})$ creates (annihilates) an electron of spin $\sigma = \uparrow,\downarrow$ on the site $\bm i$, $t$ denotes the nearest-neighbor hopping integral, $\mu$ is the filling factor. $\hat n_{\bm i \sigma} = \hat c_{\bm i \sigma}^\dagger \hat c_{\bm i \sigma}$ is the particle number operator at site $\bm i$. We also define the vector of Pauli matrices in spin space $\bm \sigma = (\sigma_x, \sigma_y, \sigma_z)^T$. We assume that the antiferromagnet is of G-type. Then the exchange field can be taken in the form $\bm h_{\bm i} =  (-1)^{i_x+i_y} \bm h$ inside the antiferomagnet.  $x$- and $y$-axes are taken normal to the AF/S interface and parallel to it, respectively. 

The Hamiltonian (\ref{ham}) can be diagonalized by the Bogolubov transformation \cite{Bobkov2022}. Further one can investigate the structure of superconducting  correlations  at the AF/S interface using the solutions of the resulting Bogolubov-de Gennes equation. The anomalous Green's function in Matsubara representation can be calculated as $F_{\bm i, \alpha \beta} = - \langle \hat c_{\bm i \alpha}(\tau) \hat c_{\bm i \beta}(0) \rangle$, where $\tau$ is the imaginary time. The component of this anomalous Green's function for a given Matsubara frequency $\omega_m = \pi T(2m+1)$ is calculated as follows: 
\begin{align}
F_{\bm i,\alpha\beta}(\omega_m)= \sum\limits_n (\frac{  u_{n,\alpha}^{\bm i} v_{n,\beta}^{\bm i*}}{i \omega_m -\varepsilon_n}+\frac{ u_{n,\beta}^{\bm i} v_{n,\alpha}^{\bm i*}}{i \omega_m +\varepsilon_n}),
\end{align}
where $u_{n,\alpha}^{\bm i}$ and $v_{n,\alpha}^{\bm i}$ are electron and hole parts of the two-component eigenfunction of the Bogolubov-de Gennes equation, corresponding to the $n$-th eigenstate, $\varepsilon_n$ is the eigenenergy of this state, and $\alpha$, $\beta$ are
spin indices. Only off-diagonal in spin space components, corresponding to opposite-spin pairs, are nonzero for the case under consideration. The singlet (triplet) correlations are described by $F_{\bm i}^{s,t}(\omega_m) = F_{\bm i,\uparrow \downarrow}(\omega_m) \mp F_{\bm i,\downarrow \uparrow}(\omega_m)$. The  on-site triplet correlations are odd in Matsubara frequency, as it should be according to the general fermionic symmetry. The spin-triplet correlations amplitude $F_{\bm i}^t$ at each lattice site with the radius-vector $\bm i$ is evaluated by summing the anomalous Green's function over the positive Matsubara frequencies $F_{\bm i}^t = \sum \limits_{\omega_m>0} F_{\bm i}^t(\omega_m) $.

Figure \ref{fig:trip}(b) plots the spatially resolved spin-triplet pairing amplitude in the AF/S bilayer. A clear imprinting of the AF N\'eel order is seen on the triplet pairing amplitude in the direction parallel to the interface: an alternating sign of the correlations in the S layer is seen. This perfect picture is disturbed in the direction perpendicular to the interface due to the absence of the translational invariance in this direction. We will get back to physical reasons of the disturbance later, in Sec.~\ref{finite_momentum} of this review.

The physics related to the proximity-induced N\'eel triplet correlations can be also described in the framework of the two-sublattice quasiclassical theory in terms of the Eilenberger equations, which was developed in Ref.~\onlinecite{Bobkov2022}.  

\subsection{Qualitative physical picture of the N\'eel triplets' origin}
\label{Neel_origin}

What is the physical origin of the N\'eel Cooper pairs? The essential physics is captured already within a one-dimensional (1D) model \cite{Bobkov2022}. In this model, a 1D AF is interfaced to a 1D superconductor running along the AF. Therefore, the electron wavevector bears only one component which is along the interface.

\begin{figure}[tb]
	\centering
	\includegraphics[width=85mm]{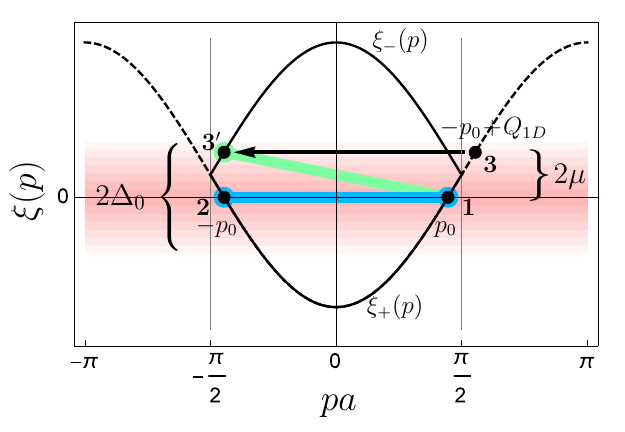}
	\caption{Electron dispersion $\xi_{\pm}(p) = \mp 2t \cos pa - \mu$  of the normal-state S in the reduced Brillouin zone (BZ) $pa \in [-\pi/2,\pi/2]$ considering a 1D system with two sites in the unit cell (solid curves). The electron energy defined by $\xi_{\pm}(p)$ is counted from the Fermi surface. The reciprocal lattice vector due to the periodicity enforced by the AF is $Q_{1D} = \pi/a$. The spectrum branches are doubled in the BZ due to the reduction of the BZ volume. The spectrum of the original 1D superconductor with one site in the unit cell in the BZ $pa \in [-\pi,\pi]$  is shown by dashed curves. The blue line indicates ordinary pairing between $p_0$ (1) and $- p_0$ (2) electrons corresponding to the zero total pair momentum. The green line indicates N\'eel pairing between $p_0$(1) and $- p_0 + Q_{1D}$ (3) corresponding to the total pair momentum $ Q_{1D}$. The picture is adopted from Ref.~\onlinecite{Bobkov2022}.}
	\label{fig:pairing} 
\end{figure}

In the absence of the antiferromagnet the normal-state electronic dispersion of S can be depicted as in Fig.~\ref{fig:pairing} with a Brillouin zone (BZ) $pa \in [-\pi,\pi]$, where $a$ is the lattice constant. Within this single-sublattice dispersion  the N\'eel magnetic order in AF causes scattering between electronic states that differ by the wavenumber $Q_{1D} = \pi/a$ (Umklapp scattering \cite{Cheng2014,Takei2014,Baltz2016}) at the S/AF interface.  Thus, the AF converts conventional spin-singlet pairing between  electrons with momenta $+p_0$ and $-p_0$ at the Fermi surface (blue line in Fig.~\ref{fig:pairing}) into the N\'eel spin-triplet pairing between, for example, $+p_0$ and $-p_0 + Q_{1D}$ (green line in Fig.~\ref{fig:pairing}). In real space such a pairing changes sign from a site to its nearest neighbor similar to the N\'eel order with the wavenumber $Q_{1D}$. The antiferromagnetic gap opening has been disregarded in the present simplified figure. 

Strictly speaking, to describe the whole S/AF system, we should use a unit cell with two sites in it, corresponding to two sublattices of the antiferromagnet with opposite magnetizations. Within this two-sublattice picture, we now have two bands in the electronic dispersion. What appeared as pairing between $+p_0$ and $-p_0 + Q_{1D}$ states in the single-sublattice picture is actually pairing between the $+p_0$ state from one band with the $-p_0$ state from the other band, as depicted in Fig.~\ref{fig:pairing}. Therefore, we conclude that the N\'eel pairing is the interband pairing.

\subsection{Dependence of the N\'eel triplet pairing on the chemical potential}

\label{Neel_chemical}

Due to its interband origin the N\'eel triplet pairing is very sensitive to the value of the chemical potential in the material, where it is induced. It is in sharp contrast with the usual triplet pairing induced by the proximity effect at S/F interfaces, which is insensitive to the value of the chemical potential because of its intraband nature. To see the dependence of the N\'eel pairing on the chemical potential, let us have another look at Fig.~\ref{fig:pairing}. 
Taking into account that $p_0$ is defined from the condition $\xi_+(p) = -2 t \cos p_0 a - \mu = 0$ one immediately obtains that $\xi_1 - \xi_{3'} = 2\mu$. That is, the energy difference between (1) and (3') electrons grows with $\mu$ thus reducing the efficiency of pairing. This qualitative picture was supported in Ref.~\onlinecite{Bobkov2022} by strict calculations performed in the framework of the two-sublattice quasiclassical theory. 

\begin{figure}[tb]
	\begin{center}
		\includegraphics[width=85mm]{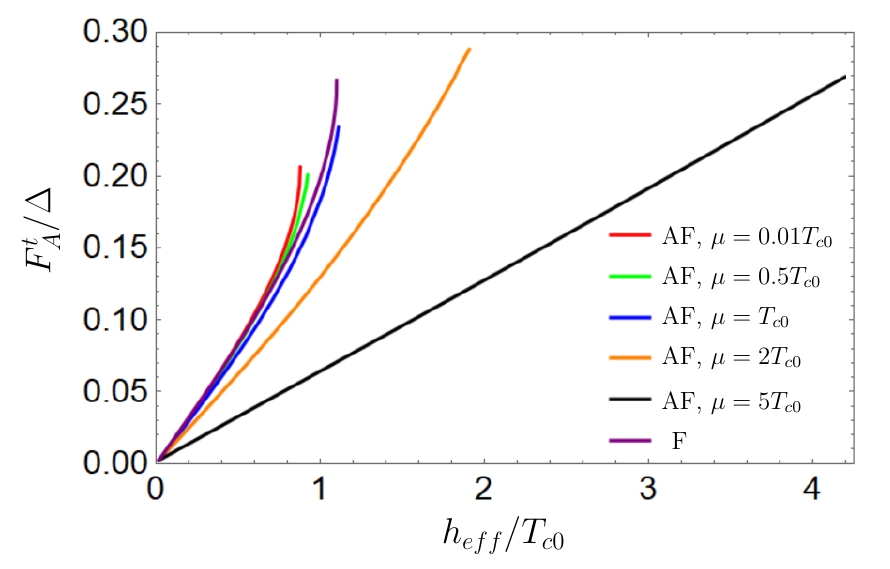}
		\caption{Anomalous Green's function of the N\'eel triplet correlations as a function of $h_{eff}$ for different values of  $\mu$. Each line ends at the critical value of $h_{eff}$ corresponding to the full suppression of superconductivity. $T_{c0}$ is the critical temperature of the superconductor without proximity to a magnet. The picture is adopted from Ref.~\onlinecite{Bobkov2022}.} \label{triplets_mu}
	\end{center}
\end{figure}

Fig.~\ref{triplets_mu} represents the results of calculation of the N\'eel triplet amplitude $F^t_A$ in a thin superconductor proximitized by an AF insulator. Such a thin-film superconductor with the thickness $d_S \ll \xi_S$, where $\xi_S$ is the superconducting coherence length, can be considered as a homogeneous superconductor under the influence of an effective exchange field of the N\'eel type $\bm h_{eff}$ \cite{Kamra2018,Bobkov2022}, which accounts for the influence of the AF insulator on the superconducting film. In Fig.~\ref{triplets_mu} the N\'eel triplet amplitude as a function of the effective exchange field $h_{eff} \equiv |\bm h_{eff}|$ is plotted for different values of the chemical potential $\mu$ in the superconductor. It is seen that for a given value of $h_{eff}$ the N\'eel triplet amplitude indeed weakens upon increase in $\mu$. The amplitude of conventional triplet correlations induced in the same thin-film superconductor by proximity to a ferromagnetic insulator producing the effective exchange field of the same amplitude $h_{eff}$ (homogeneous, not N\'eel) is also shown for comparison. 

\begin{figure}[tb]
	\begin{center}
		\includegraphics[width=88mm]{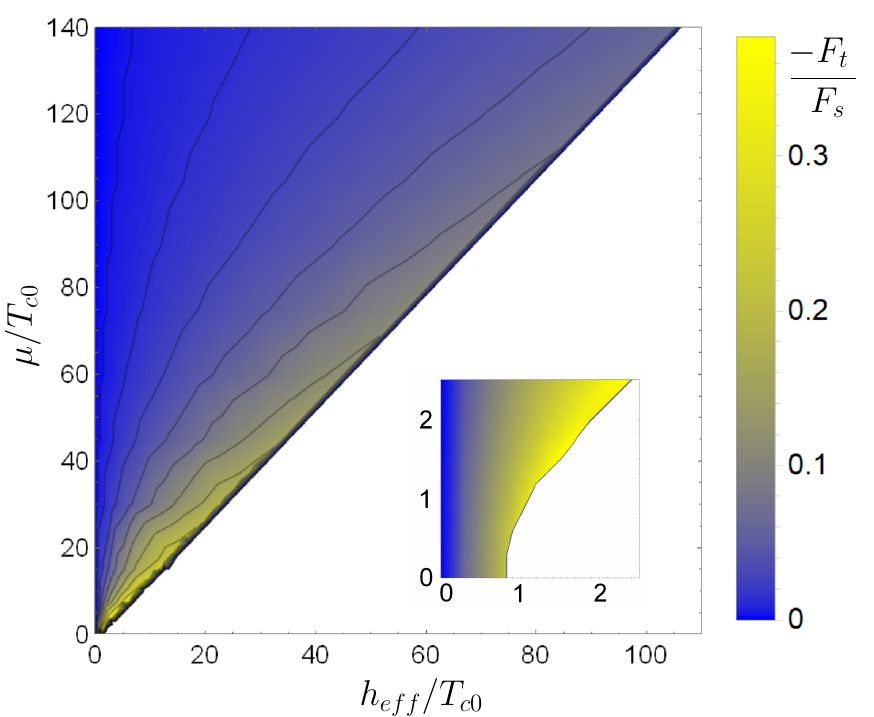}
		\caption{Amplitude of the triplet correlations relative to the singlet amplitude as function of $(h_{eff},\mu)$. Inset: region $(h_{eff},\mu) \sim T_{c0}$ on a larger scale. $T \to T_c$. The picture is adopted from Ref.~\onlinecite{Bobkov2023_impurities}. }
        \label{fig:triplets}
	\end{center}
\end{figure}

The weakening of the N\'eel triplet correlations with increasing chemical potential does not mean that they are completely suppressed at high values of the chemical potential $\mu \gg T_{c0}$, when the filling factor in the superconductor is far from half filling of the conduction band. Here $T_{c0}$ is the critical temperature of the superconductor in the absence of the proximity to the AF. The amplitude of the N\'eel triplet correlations increases with increasing effective exchange field $h_{eff}$, as it is seen from Fig.~\ref{triplets_mu}. In Ref.~\onlinecite{Bobkov2023_impurities} it was shown that at $(h_{eff}, \mu) \gg T_{c0}$ the amplitude of the N\'eel triplet correlations is only determined by the ratio $h_{eff}/\mu$ and reaches its maximal value at the line $h_{eff}/\mu = const \approx 0.8$. This tendency is demonstrated in Fig.~\ref{fig:triplets}. For larger values of this parameter the superconductivity in the system is fully suppressed.

\subsection{Suppression of the critical temperature of thin-film S/AF bilayers by the N\'eel triplets}

Now we demonstrate that the N\'eel triplet correlations suppress the superconductivity at S/AF interfaces and, in particular, they suppress the superconducting critical temperature of thin-film superconductors proximitized by antiferromagnetic insulators. The effect is analogous to the well-known suppression of superconductivity by proximity-induced triplets at S/F interfaces \cite{Buzdin2005}. Again we discuss a thin-film superconductor with $d_S \ll \xi_S$ in proximity to an antiferromagnetic insulator. 

\begin{figure}[tb]
	\begin{center}
		\includegraphics[width=85mm]{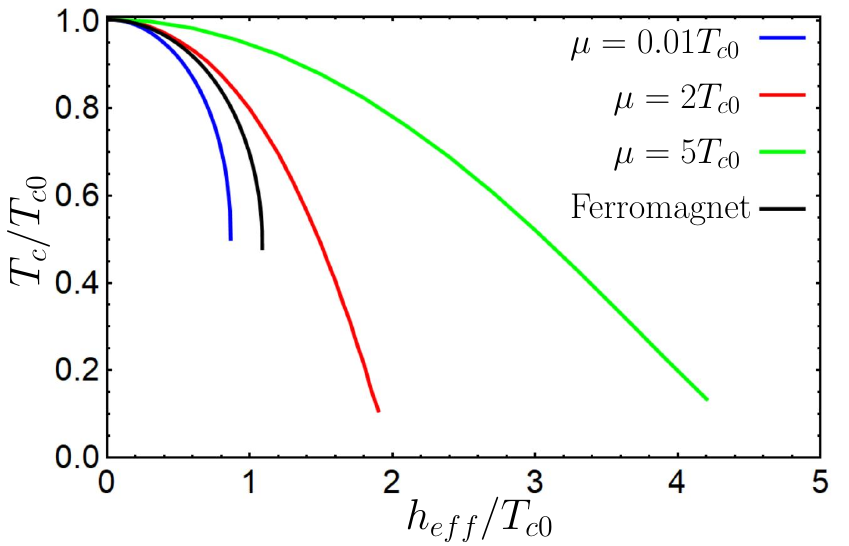}
		\caption{Critical temperature of the AF/S bilayer as a function of $h_{eff}$ for different values of $\mu$. Black line represents $T_c(h_{eff})$ for an S/F interface with a ferromagnetic
insulator producing the same value of the effective exchange field (but homogeneous, not staggered)
in the superconductor. The picture is partially overlapped with an analogous picture from Ref.~\onlinecite{Bobkov2022}.} \label{Fig:Tc_mu}
	\end{center}
\end{figure}

The dependence of the critical temperature of the AF/S bilayer on the effective exchange field is presented in Fig.~\ref{Fig:Tc_mu} for different values of the chemical potential.  The critical temperature is indeed suppressed by the staggered exchange field $h_{eff}$. The efficiency of suppression by the staggered field is of the same order, and even higher, as the suppression by the ferromagnet with the same absolute value of the exchange field. The stronger suppression of the superconductivity by the staggered exchange as compared to the uniform ferromagnetic exchange field is explained by the presence of the antiferromagnetic gap at the Fermi surface, which prevents electronic states inside this gap from superconducting pairing. The superconductivity suppression for a given $h_{eff}$ becomes weaker for larger values of the chemical potential, what is explained by weakening of the N\'eel triplet correlations upon increase of $\mu$.

\subsection{Proximity effect produced by canted antiferromagnets: mixture of N\'eel and conventional triplets}

\label{canted}

\begin{figure}[tb]
	\begin{center}
		\includegraphics[width=85mm]{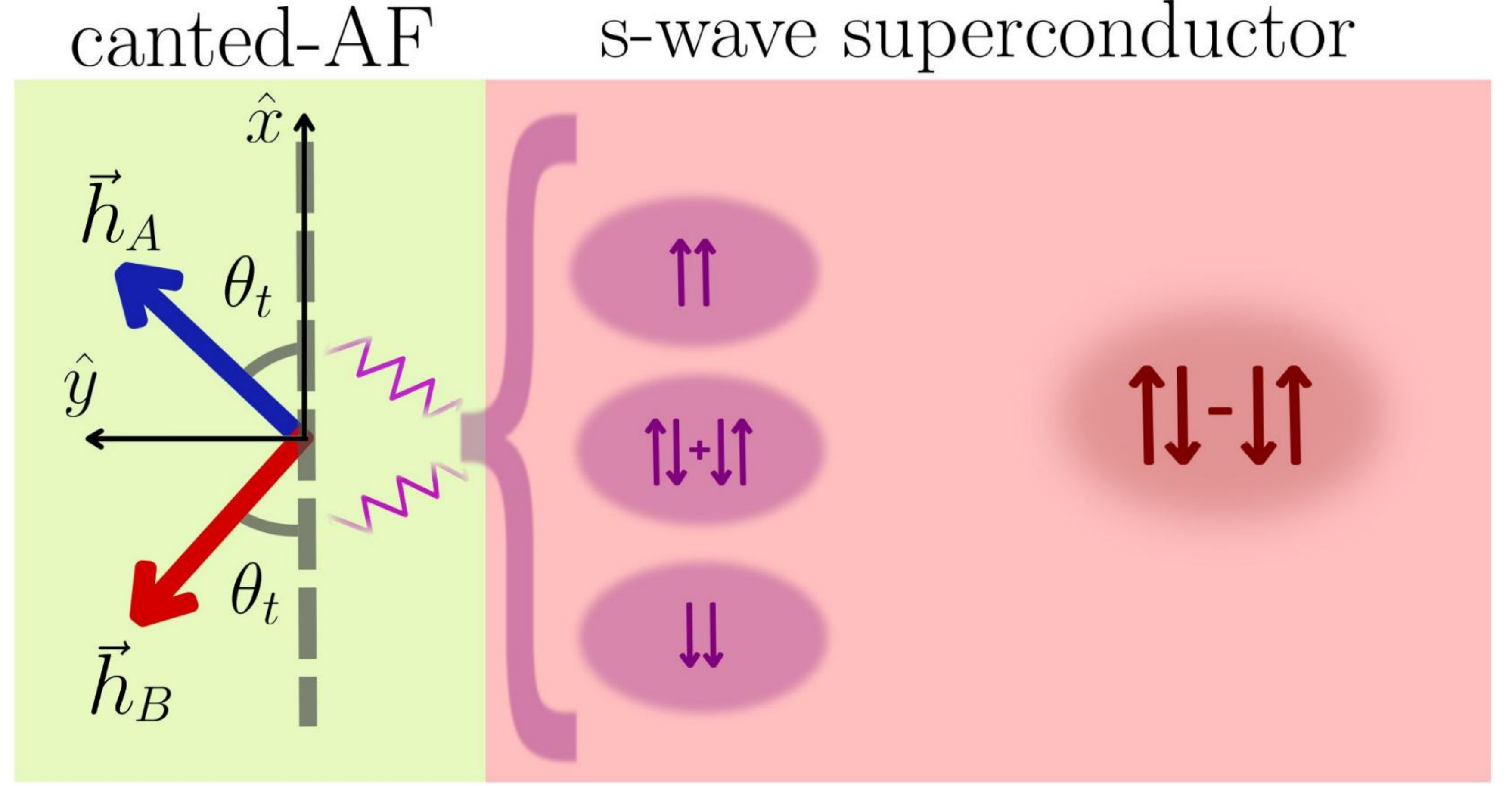}
		\caption{Sketch of the system and key physics of the proximity effect at S/canted AF interfaces. Equal-spin and zero-spin triplet correlations are generated in a conventional $s$-wave spin-singlet superconductor when it is interfaced with a canted antiferromagnet (canted-AF). The equal-spin triplet correlations result from the intrinsic noncollinearity between the two AF sublattice magnetizations. The canting angle $\theta_t$ allows one to vary the magnet from being a collinear AF ($\theta_t = 0$) to a ferromagnet ($\theta_t = \pi/2$). The picture is redrawn after Ref.~\onlinecite{Chourasia2023}.}
		\label{fig:1}
	\end{center}
\end{figure}

Now we discuss, following Ref.~\onlinecite{Chourasia2023} what happens with the proximity effect in heterostructures composed of superconductors and canted antiferromagnets. In that paper a bilayer structure consisting of an insulating AF with canted sublattice magnetizations exchange coupled to an adjacent S was considered. The sketch of the system is shown in Fig.~\ref{fig:1}. The canting angle is $\theta_t$. For $\theta_t=0$ the considered AF becomes a collinear antiferromagnet with the axis of magnetic moments along the $x$-direction. As we increase the value of $\theta_t$, the canted-AF acquires a net magnetization along the $y$-direction. So the canted-AF can be decomposed into an antiferromagnetic component (along the $x$-axis) and a ferromagnetic component (along the $y$-axis). 

In Ref.~\onlinecite{Chourasia2023} Chourasia {\it et.~al} investigated the spin-triplet correlations and the critical temperature  of the S as  functions of the canting angle in the AF, which allows us to continuously tune the AF from its collinear antiparallel state to it effectively becoming an F. In general, the vector amplitude of triplet superconducting correlations induced in the S by proximity to the AF can be written as $\vec{F^t_j}=F^{t,x}_j \hat{x} + F^{t,y}_j \hat{y} + F^{t,z}_j \hat{z}$. It always has a component aligning with the local exchange field  whether the magnetization of a magnet (ferromagnet or antiferromagnet) is homogeneous or inhomogeneous. If the magnetization is inhomogeneous, other components, not coinciding with the direction of the local exchange field, appear \cite{Bergeret2005}. In the case under consideration the magnetization is obviously inhomogeneous due to canting. For this reason all three components of the triplet vector are nonzero.

\begin{figure*}[tb]
	\begin{center}
		\includegraphics[width=\linewidth]{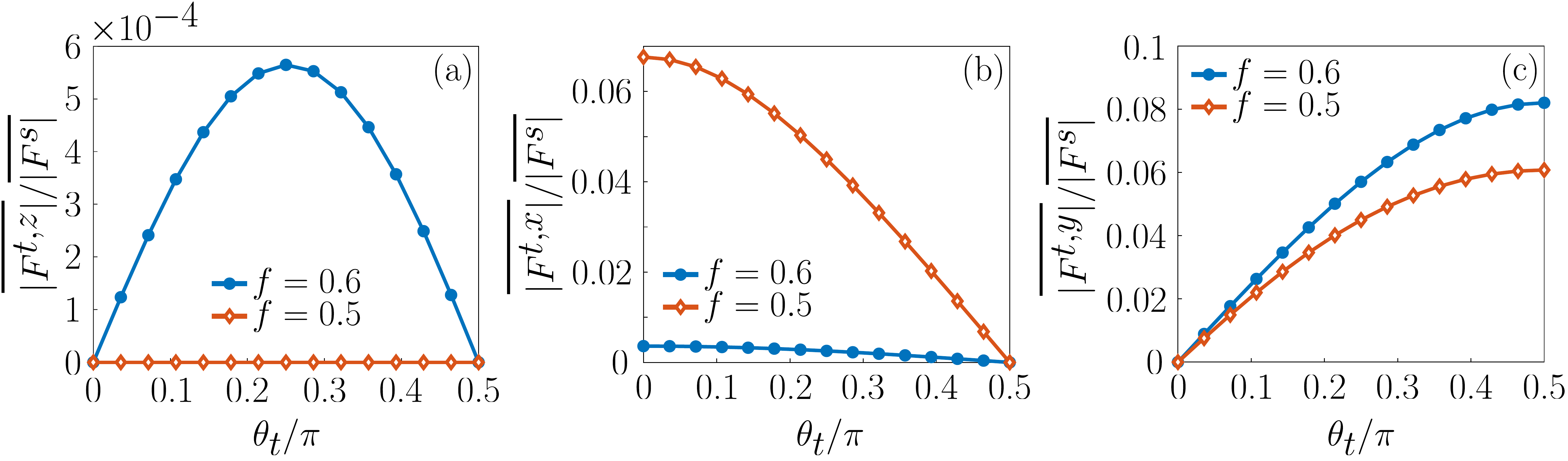}
		\caption{Variation of triplet correlations with canting angle $\theta_t$ for filling factors $f=0.5$ $(\mu=0)$ and $f=0.6$ $(\mu/T_{c0} \approx 65)$. (a) The average  magnitude of the normalized spin-triplet correlation $F^{t,z}$. (b) The average magnitude of the normalized spin-triplet correlation $F^{t,x}$. (c) The average magnitude of the normalized spin-triplet correlation $F^{t,y}$. The averages are taken over all superconducting sites and are denoted via an overhead bar. The picture is adopted from Ref.~\onlinecite{Chourasia2023}.}
		\label{fig:3}
	\end{center}
\end{figure*}

The dependence of these spin-triplets on the canting angle is presented in Fig.~\ref{fig:3}. At $\theta_t = 0$ only $F^{t,x}$ component is nonzero. It corresponds to the pure N\'eel triplet order discussed above. The N\'eel structure of this component was demonstrated explicitly \cite{Chourasia2023}. This component decreases as $\theta_t$ goes from 0 to $\pi/2$ [Fig.~\ref{fig:3}(b)] and vanishes at $\theta_t = \pi/2$ because the N\'eel triplets are absent in the purely ferromagnet state. It is also seen that the amplitude of this component is smaller at larger chemical potential in agreement with the general dependence of the N\'eel triplets on the chemical potential, discussed above. [Here the chemical potential is determined via the filling factor $f$, which is the fraction of filled electronic states in the system. $f=0.5$ (half-filled band) corresponds to $\mu=0$.]

The component $F^{t,y}$ increases with the canting angle and it appears to be caused primarily by the net magnetization. It was directly checked that it is a conventional triplet component without the N\'eel structure. It is maximal in the purely ferromagnetic state. 
$F^{t,z}$ is also found to be of the N\'eel type. It appears due to the noncollinearity of two sublattices. Its N\'eel character can be understood in terms of this noncollinearity. From one lattice site to the next, the angle between the spin-splitting at the site and its direct neighbors is changing sign. It is interesting that it vanishes identically at $\mu=0$ ($f=0.5$) for all canting angles. The physical reason is that for $\mu \approx 0$ the antiferromagnetic gap opens in the superconductor in the vicinity of the normal state Fermi surface. In this case the most important contribution to
the pairing correlations is given by the electronic states
at the edge of the gap. They correspond to $\xi_\pm \approx 0$, what
means that the electrons are practically fully localized at one of the sublattices. Consequently, they only feel the
magnetization of the corresponding sublattice, which is homogeneous. The noncollinearity does not work.  However, for non-zero $\mu$ (away from the half-filling case), $F^{t,z}$ increases from 0 to a finite value as we go from a collinear antiferromagnetic alignment to maximal noncollinearity between the sublattice magnetic moments, and decreases back to zero in the ferromagnetic alignment [Fig.~\ref{fig:3}(a)]. 

\begin{figure*}[tb]
	\begin{center}
		\includegraphics[width=140mm]{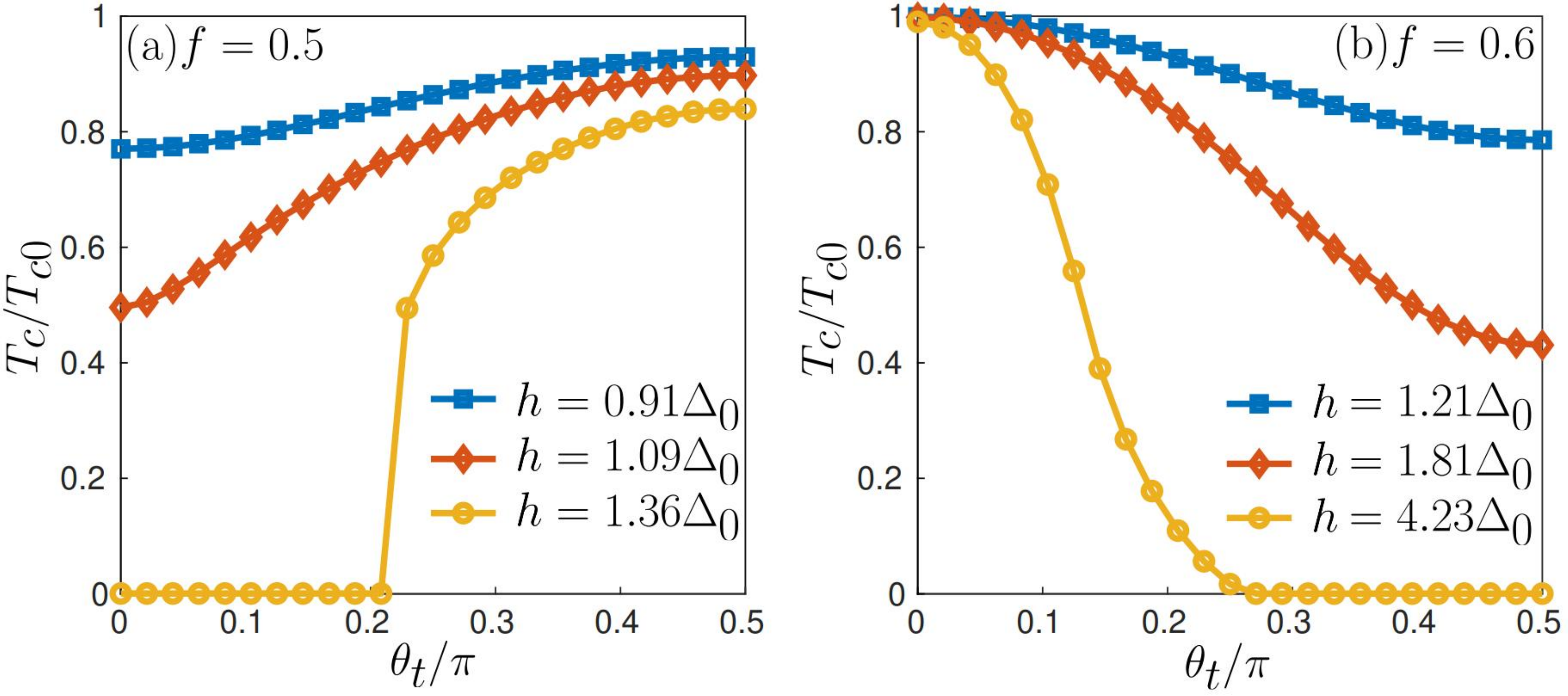}
		\caption{(a) Normalized critical temperature $T_c$ vs.~canting angle $\theta_t$ for filling factor (a) $f=0.5$ and (b) $f = 0.6$ considering different values of the antiferromagnetic exchange field $h$. $\Delta_0$ is the zero-temperature superconducting order parameter of the same superconductor without the AF layer. The picture is adopted from Ref.~\onlinecite{Chourasia2023}.}
		\label{fig:5}
	\end{center}
\end{figure*}

As it was already discussed, critical temperature of superconductor/magnet  heterostructures is suppressed by triplet correlations, both of conventional type and the N\'eel type. Therefore, it is natural to expect that the critical temperature of the AF/S bilayers with canted antiferromagnets is always suppressed with respect to the critical temperature of the isolated superconductor $T_{c0}$. In Ref.~\onlinecite{Chourasia2023} it was found that it is indeed the case. However, the physics of the suppression is very interesting. In the framework of our review here we encounter the first manifestation of the crucial dependence of the N\'eel triplets on the chemical potential and its physical consequence. It was obtained that the dependence of the critical temperature on the canting angle $\theta_t$ is opposite near half-filling and far from half-filling. It is demonstrated in Fig.~\ref{fig:5} for (a) $f = 0.5$ ($\mu=0$) and (b) $f = 0.6$ ($\mu \approx 65 T_{c0}$). It is found that for $\mu=0$, $T_c$ increases with $\theta_t$ while it manifests the opposite dependence for $\mu \neq 0$. 

For the case $\mu = 0$, presented in Fig.~\ref{fig:5}(a), a strong generation of the N\'eel triplets  due to interband pairing leads to maximal $T_c$ suppression at $\theta_t = 0$. Since the $T_c$ suppression is stronger for the pure AF case ($\theta_t = 0$) than for the pure F case (it corresponds to $\theta_t = \pi/2$), the $T_c$ increases with $\theta_t$ . For the case of $f = 0.6$, that is $\mu \gg T_{c0}$ and $h \sim T_{c0}$ ($h \equiv |\bm h|$), the N\'eel triplets generation by the antiferromagnetic order is much weaker. On the other hand, the ordinary spin-triplets generation by a ferromagnet remain of the same order of magnitude as for $f = 0.5$. Thus, $T_c$ is largest for $\theta_t = 0$ and it decreases with $\theta_t$. 

\section{Influence of impurities on the N\'eel triplets and on superconductivity in AF/S heterostructures}

\label{impurities}

In this section we discuss how the superconductivity in AF/S heterostructures is influenced by conventional nonmagnetic impurities and what is the role of the N\'eel triplets in this physics. Here we will have the second example of the crucial influence of the chemical potential on the physics of S/AF heterostructures. 

\subsection{N\'eel triplets and impurities}

\begin{figure}[tb]
	\begin{center}
		\includegraphics[width=85mm]{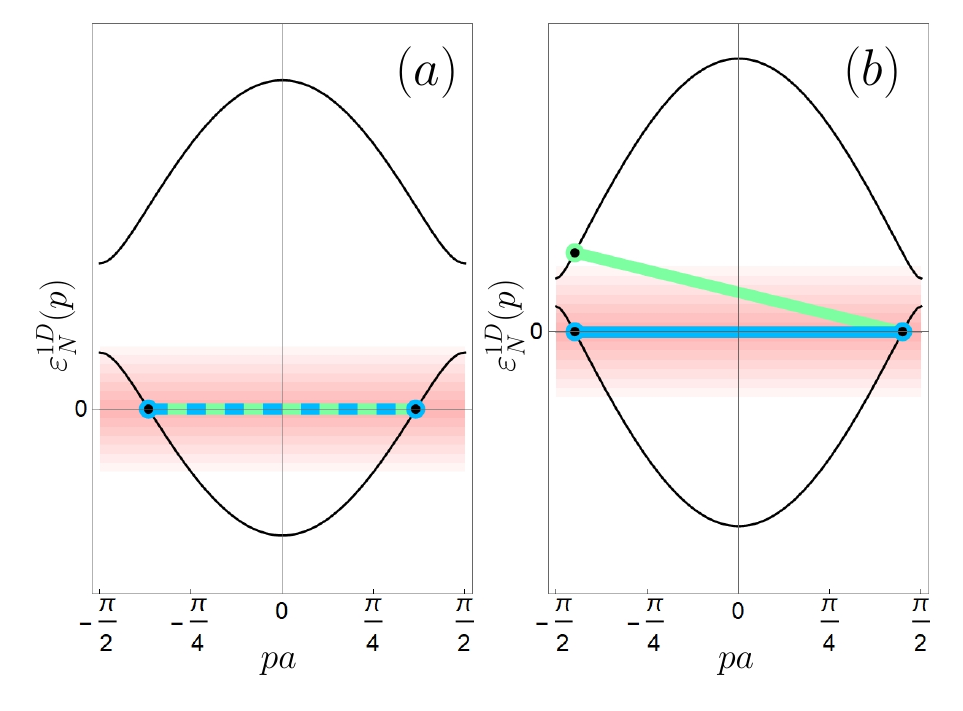}
		\caption{ Electron dispersion of the normal-state in the reduced Brillouin zone (BZ) $pa \in [-\pi/2, \pi/2]$. For simplicity of visualization a 1D system dispersion  $\varepsilon_N^{1D} = \mp \sqrt{\xi^2(p)+h_{eff}^2} - \mu$ is demonstrated instead of a real dispersion $\varepsilon_N^{3D}$. Here we take into account the opening of the antiferromagnetic gap due to $h_{eff}$. (a) Large $\mu$. The electronic states in the vicinity of the Fermi surface $\varepsilon_N^{1D} = 0$ allowed for pairing (pink region) do not involve the second electronic branch. Only $(\bm p, -\bm p)$ intraband singlet and intraband N\'eel triplet pairs are present (dashed blue-green). (b) Small $\mu$ for comparison. Electronic states belonging to the both branches are present in the vicinity of the Fermi surface and are allowed for pairing. Both intraband singlet (blue) and interband N\'eel triplet pairs (green) exist. The picture is adopted from Ref.~\onlinecite{Bobkov2023_impurities}.}
        \label{fig:FS}
	\end{center}
\end{figure}

First of all, we discuss how the N\'eel triplets behave in the presence of impurities. In Ref.~\onlinecite{Bobkov2022} it was shown that near half-filling nonmagnetic disorder destroys the N\'eel spin-triplet correlations. The physical reason of the suppression of the N\'eel triplets by the nonmagnetic disorder is their interband nature. At the same time, at $\mu \gg T_{c0}$ the interband N\'eel triplet pairing is suppressed. However,  the N\'eel triplet correlations can be essential even at large values of $\mu$, as it was demonstrated in Refs.~\onlinecite{Bobkov2023_impurities,Chourasia2023} and was discussed above in this review. In this case all pairing correlations are completely intraband, as it is shown in Fig.~\ref{fig:FS}(a). However, the normal state eigenvectors of the electronic band crossed by the Fermi level represent a mixture of sign-preserving and sign-flipping components between the A and B sites of the same unit cell:
\begin{eqnarray}
\left( \begin{array}{c} \hat \psi_{\bm i \sigma}^A \\ \hat \psi_{\bm i \sigma}^B
\end{array}
\right)(\bm p) = \left[C_{p\sigma} \left( \begin{array}{c} 1 \\ 1
\end{array}
\right) + C_{f\sigma} \left( \begin{array}{c} 1 \\ -1
\end{array}
\right)\right]e^{i \bm p \bm i},
\label{eq:eigenvectors}
\end{eqnarray}
where
\begin{eqnarray}
C_{p(f)\sigma} = \frac{1}{2}\left[ \sqrt{1+\sigma h_{eff}/\mu}\pm \sqrt{1-\sigma h_{eff}/\mu} \right]
\label{eq:cpf}
\end{eqnarray}
are the sign-preserving (flipping) amplitudes of the eigenvector. Due to the presence of the sign-flipping component of the eigenvectors and its spin sensitivity the singlet homogeneous pairing between $\bm p$ and $-\bm p$ states at the Fermi level [see Fig.~\ref{fig:FS}(a)] is inevitably accompanied by the N\'eel sign-flipping triplet component. As it can be seen from Eq.~(\ref{eq:cpf}) the amplitude of the sign-flipping mixture is controlled by $h_{eff}/\mu$ and is suppressed with growing of $\mu$, what is in agreement with the dependence of the N\'eel triplets on the chemical potential, presented in Sec.~\ref{Neel_chemical}.

\begin{figure}[tb]
	\begin{center}
		\includegraphics[width=85mm]{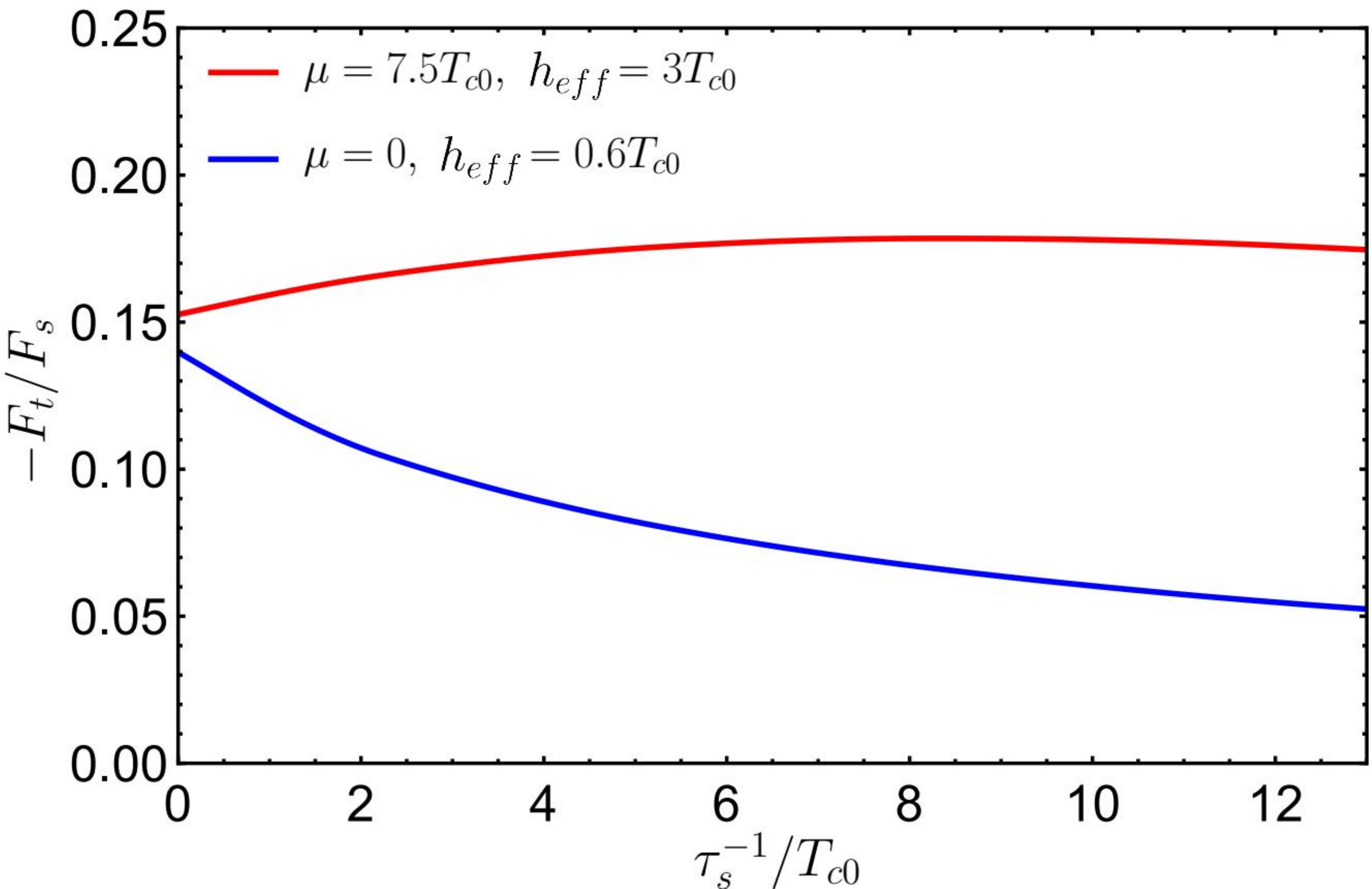}
		\caption{Amplitude of the triplet correlations relative to
the singlet amplitude as a function of the inverse impurity scattering time for a superconductor in the presence of the N\'eel-type effective exchange field $h_{eff}$. Different curves correspond to different values of the chemical potential $\mu$.}
        \label{fig:neel_impurities}
	\end{center}
\end{figure}

It is natural to expect that the $s$-wave intraband triplets are not sensitive to the influence of nonmagnetic impurities. Our calculations, performed in the framework of the non-quasiclassical Green's functions approach, originally presented in Ref.~\onlinecite{Bobkov2023_impurities} confirm these expectations. The dependencies of the N\'eel triplet amplitudes on the nonmagnetic impurity inverse scattering time $\tau_s^{-1}$ are demonstrated in Fig.~\ref{fig:neel_impurities}. It is seen that at $\mu=0$ the amplitude of the N\'eel triplet correlations is suppressed by impurities, as it was found in Ref.~\onlinecite{Bobkov2022}. At large enough $\mu=7.5 T_{c0}$ we clearly see no  suppression. Instead a weak increase of the N\'eel triplet amplitude is observed. We cannot definitely say what is the reason for this weak increase. Probably it is related to the influence of impurities on the normal state DOS.

\subsection{Depairing effect of nonmagnetic impurities in S/AF heterostructures at large chemical potentials}

\label{depairing_imp}

What happens with the critical temperature of thin-film S/AF bilayers in the presence of impurities? From the results discussed in the previous subsection it may seem that it should be enhanced with impurity scattering strength at small chemical potentials because of weakening the triplets, which are generated at the expense of singlets. It is indeed the case, as it is shown in the next subsection. But what happens at large $\mu$? Is the critical temperature only negligibly sensitive to impurities? The answer is that at large chemical potentials the critical temperature is suppressed by nonmagnetic impurities quite strongly. The mechanism of the suppression is not related to the N\'eel triplets.   The amplitude of wave functions of electrons is different for A and B sublattices [see Eq.~(\ref{eq:eigenvectors})]. Physically it is  because of the fact that for an electron with spin up it is energetically favorable to be localized on B sublattice and for an electron with spin down --- on A sublattice. Thus, this sublattice-spin coupling in
the presence of the N\'eel-type exchange field gives an effective magnetic component to the non-magnetic impurities. And it is well-known that magnetic impurities do suppress superconductivity\cite{Abrikosov1961}. This mechanism of superconductivity suppression by nonmagnetic impurities was originally discussed for antiferromagnetic superconductors \cite{Buzdin1986}. Then it was realized that it also works for S/AF heterostructures \cite{Fyhn2022,Fyhn2022_1}, where an appropriate quasiclassical theoretical formalism taking into account the destroying  action of nonmagnetic impurities was developed \cite{Fyhn2022} and the suppression of the critical temperature of S/AF bilayers with metallic antiferromagnets due to the nonmagnetic impurities was found \cite{Fyhn2022_1}. 

\subsection{Dependence of the critical temperature of S/AF heterostructures on impurities: enhancement vs suppression}

The influence of nonmagnetic impurities on the critical temperature of thin-film S/AF bilayers, which can be effectively modelled as superconductors in a homogeneous effective N\'eel exchange field $h_{eff}$, in the full range of parameters of the bilayer was considered in Ref.~\onlinecite{Bobkov2023_impurities}. 

\begin{figure}[tb]
	\begin{center}
		\includegraphics[width=82mm]{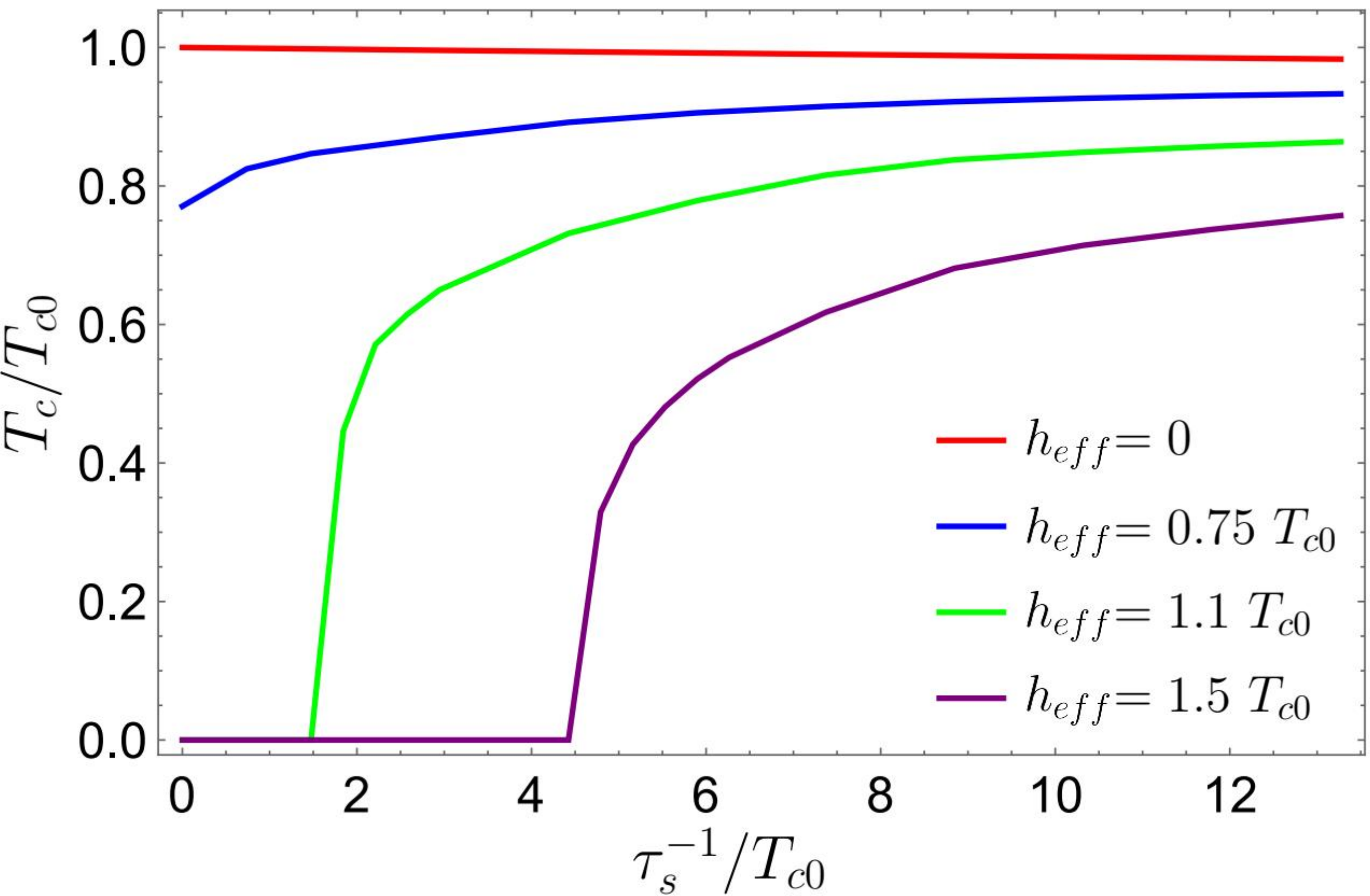}
		\caption{$T_c(\tau_s^{-1})$ at $\mu=0$ for different effective exchange fields $h_{eff}$. $T_c$ is normalized to the value of the critical temperature of the isolated S film $T_{c0}$. The picture is adopted from Ref.~\onlinecite{Bobkov2023_impurities}.}
        \label{fig:fig_A}
	\end{center}
\end{figure}

\begin{figure}[tb]
	\begin{center}
		\includegraphics[width=85mm]{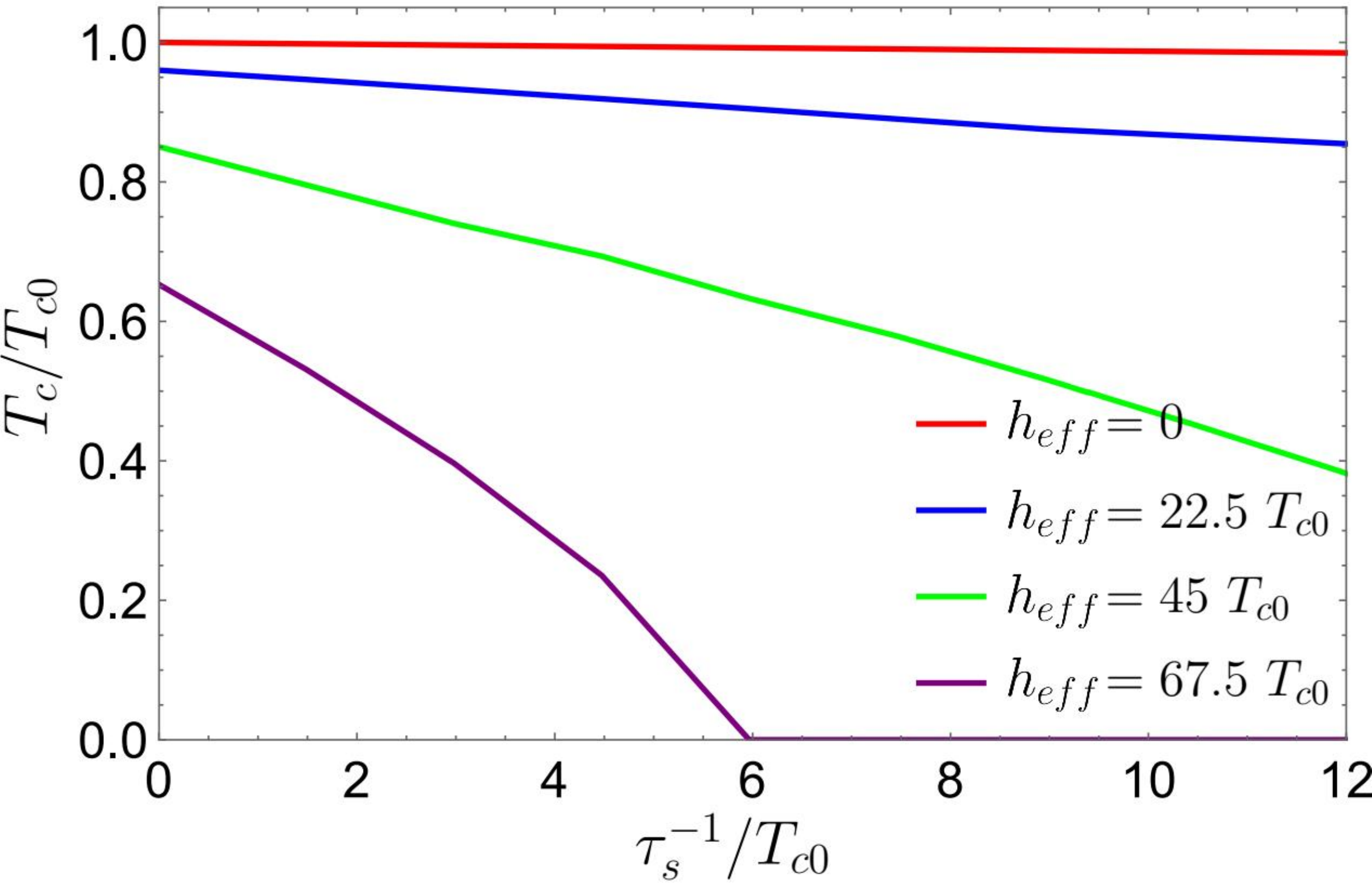}
		\caption{$T_c(\tau_s^{-1})$ at $\mu=150 T_{c0}$ for different effective exchange fields $h_{eff}$. The picture is adopted from Ref.~\onlinecite{Bobkov2023_impurities}.}
        \label{fig:fig_C}
	\end{center}
\end{figure}

At first we discuss the opposite limiting cases. Figs.~\ref{fig:fig_A} and \ref{fig:fig_C} demonstrate the critical temperature of the S/AF bilayer as a function of the inverse impurity scattering time $\tau_s^{-1}$. The results shown in Fig.~\ref{fig:fig_A} are calculated at $\mu=0$ and represent a typical example of the dependence in the regime when the interband N\'eel triplets are strong and play the role of the main depairing mechanism, and the impurities do not really work as effectively magnetic. It can be seen that at $h_{eff} \neq 0$ the critical temperature grows with the disorder strength or even appears at some nonzero $\tau_s^{-1}$. This behavior is explained by the presence of the N\'eel triplets, which suppress the critical temperature of the singlet superconductivity. In the clean limit  $\tau_s^{-1}=0$ their amplitude is the maximal. Due to the interband nature of the N\'eel pairing they are gradually reduced with impurity strength and, consequently, the critical temperature grows. 

Fig.~\ref{fig:fig_C} corresponds to $\mu= 150 T_{c0}$. It represents the opposite limit when the interband N\'eel triplets are suppressed. Intraband N\'eel triplets are still there and they suppress superconductivity of the bilayer with respect to the case of an isolated superconductor, especially at large values of $h_{eff}$, as it is seen at $\tau_s^{-1} = 0$. However, the intraband N\'eel triplets are not sensitive to nonmagnetic impurities. The dependence $T_c(\tau_s^{-1})$ is dominated by the impurity suppression. 

\begin{figure}[tb]
	\begin{center}
		\includegraphics[width=85mm]{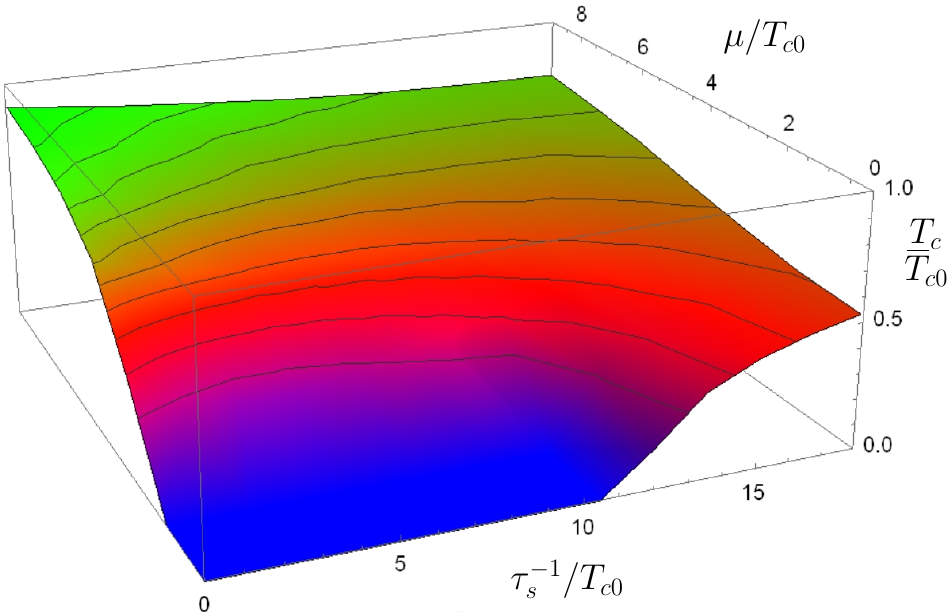}
		\caption{Dependence of the critical temperature on $(\tau_s^{-1},\mu)$ at $h_{eff}=2.25 T_{c0}$. The picture is adopted from Ref.~\onlinecite{Bobkov2023_impurities}.}
        \label{fig:impurities}
	\end{center}
\end{figure}

Now, after considering the limiting cases we discuss the effect of nonmagnetic impurities on the superconductivity of the S/AF hybrids in the entire range of parameters. Fig.~\ref{fig:impurities} represents the behavior of the critical temperature in the plane $(\tau^{-1}_s,\mu)$ for a given $h_{eff}=2.25 T_{c0}$. Front and back edges of the image correspond to opposite limits. Front edge is the limit of small $\mu$, where N\'eel triplets dominate and, consequently, superconductivity is restored with increase in impurity strength. Back edge, corresponding to large $\mu$, represents the suppression of superconductivity by nonmagnetic impurities. For intermediate $\mu$ there is a crossover between them. In particular, for a certain range of $\mu$ a nonmonotonic dependence $T_c(\tau^{-1}_s)$ is observed. The initial suppression of $T_c$ is changed by some growth. This is because the  singlet superconductivity is suppressed by the impurities more rapidly than the N\'eel triplets. This behavior is in sharp contrast with the behavior of a F/S bilayer, where the critical temperature is not sensitive to the {\it nonmagnetic} impurity concentration, and the {\it magnetic} impurities suppress the superconductivity \cite{Abrikosov1961} and, moreover, the physics of S/F bilayers is mainly not sensitive to the deviation from half-filling of the electronic spectrum.

It is worth noting that all the discussed above  results are valid for the case of relatively weak disorder, which can be considered in the Born approximation. The issue about the influence  of the strong disorder on superconductivity in AF/S heterostructures is yet to be explored. For the case of conventional $s$-wave superconductors it is known that a strong disorder
can lead to a metal–insulator transition in the normal state, to the appearance of a pseudogap in spectrum and larger spatial fluctuations of superconductive pairing, what results in increased $\Delta/T_c$ ratio\cite{Goldman1998,Sadovskii1997,Gantmakher2010,Sacepe2008,Sacepe2010}. Furthermore, Anderson localization and phase fluctuations, are more pronounced in low dimensional
structures, leading to suppression of superconductivity. At the same time, the disorder can also result in a remarkable enhancement of superconductivity \cite{Arrigoni2003,Gastiasoro2018,Martin2005,Zhao2019,Petrovic2016,
Peng2018,Neverov2022}. The stronger disorder increases spatial
inhomogeneity, which enhances the local pairing correlations
and superconducting gap, comparing with the clean system. Disorder-related effects are assumed responsible for a large
increase of the critical temperature in the recently discovered
superconducting NbSe2 monolayers. Theoretical analysis attributes
the enhancement to the disorder-induced multi-fractal
structure of the electronic wave functions. Which of the listed possibilities are relevant to S/AF heterostructures is to be studied. This prospect for future work is especially interesting in view of the opposite effects of the weak disorder on superconductivity near half-filling and away from half-filling, what can make the physical picture of the effect of the strong disorder even more rich.

\section{Finite-momentum N\'eel triplets}

\label{finite_momentum}

Triplet pairs, originated from the singlet-triplet conversion in homogeneous superconductors under the action of a Zeeman field, are usually zero-momentum pairs \cite{Sarma1963}, what means that their wave functions are uniform in real space. However, in some narrow regions of parameters an inhomogeneous superconducting state produced by singlet and triplet pairs with finite momentum of the pair, the so-called Fulde-Ferrel-Larkin-Ovchinnikov (FFLO) superconducting state, was predicted \cite{Larkin1964,Fulde1964}. One of the important properties of the triplet pairs generated at F/S interfaces, where the translational invariance is lost, is that the zero-momentum pairs, entering the ferromagnetic region from the superconductor, inevitably acquire a finite momentum of the pair \cite{Buzdin1982,Demler1997} due to the fact that the spin-up and spin-down electrons forming a pair have opposite potential energies in the macroscopic exchange field of the ferromagnet. Thus, the electrons residing at the same energy (at the Fermi-surface)  have not strictly opposite momenta in the ferromagnet (the absolute values of their momenta are different) and the pair as a whole has nonzero total momentum. The finite momentum is acquired both by singlet and triplet pairs, what allows them to be called a mesoscopic analogue of the FFLO superconducting state \cite{Larkin1964,Fulde1964}. The finite momentum, which the Cooper pair acquires in
the exchange field of the ferromagnet, makes the pairing
wave function oscillating. The resulting phase change across
the ferromagnetic layer is responsible for the $\pi$-junction
effects \cite{Buzdin1982,Buzdin2005,Kontos2002,Ryazanov2001,Oboznov2006, Bannykh2009, Robinson2006}, which are widely used now in the superconducting electronics \cite{Yamashita2005,Feofanov2010,Shcherbakova2015}. The interference of the incident and reflected oscillating wave functions determines the oscillatory phenomena of the critical temperature $T_c$ versus the F layer thickness in F/S bilayers and multilayers, which have been widely studied both theoretically \cite{Fominov2002,Radovic1991,Vodopyanov2003,Lazar2000,Buzdin2000,Zareyan2001} and experimentally \cite{Jiang1995,Mercaldo1996,Muhge1996,Zdravkov2006,Zdravkov2010}.

Naively, one does not expect that a Cooper pair penetrating into the antiferromagnet from the superconductor possesses a finite total momentum because the average value of the exchange field in the antiferromagnet is zero, the quasiparticles spectrum is spin-degenerate and, therefore, spin-up and spin-down electrons, forming the pair, should have opposite momenta with equal absolute values $\bm p_\uparrow = -\bm p_\downarrow$. In its turn, that means zero total momentum of the pair and, as a result, absence of the oscillations of the pair amplitude. However, in Refs.~\onlinecite{Bobkov2023_oscillations,Bobkov2024_singleimp} it was shown that in the absence of the translational invariance in S/AF heterostructures, that is at S/AF interfaces or at single impurities the finite-momentum N\'eel triplet pairing  occurs. In Ref.~\onlinecite{Bobkov2023_oscillations} it was demonstrated theoretically that the finite-momentum N\'eel triplet correlations at AF/S interfaces result in the oscillating dependence of the critical temperature on the AF thickness. There are a number of experimental works, where the critical temperature of AF/S bilayers with metallic antiferromagnets has been measured as a function of the AF thickness and the oscillating behavior was observed \cite{Bell2003,Hubener2002,Wu2013}. At the same time, in the regime, when the N\'eel triplets can be disregarded, this dependence has been calculated and no oscillations were reported \cite{Fyhn2022_1}.  Thus, oscillations of the critical temperature of AF/S bilayers can be viewed as a signature of the presence of finite-momentum N\'eel-type triplet correlations in the heterostructure. In two following subsections we discuss the physical nature of the finite-momentum N\'eel triplet pairing and how it manifests itself in the critical temperature of S/AF bilayers with metallic antiferromagnets.

\subsection{Physical mechanism of the finite-momentum N\'eel triplet pairing}

\begin{figure}[tb]
	\begin{center}
		\includegraphics[width=75mm]{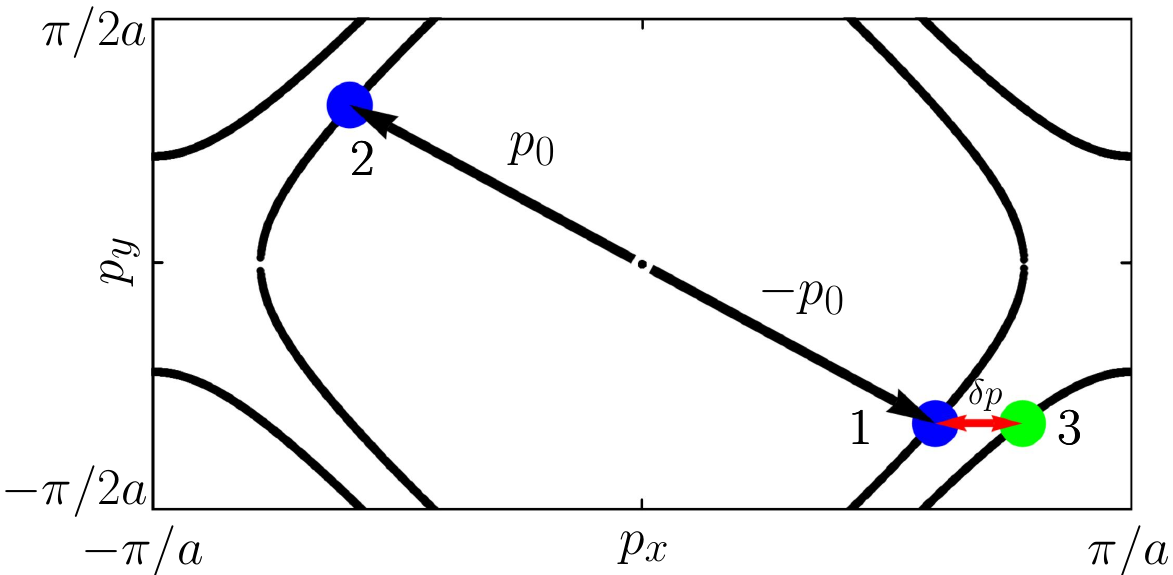}
		\caption{Brillouin zone and Fermi surface (black curves) of the AF layer. Zero-momentum Cooper pair between electrons 1 and 2 is schematically shown by black arrows. There is also N\'eel-type finite-momentum triplet pairing between electrons 2 and 3, which is produced from electron 1 due to the Umklapp reflection process from the AF/S interface. The total momentum of the pair (2,3) $\delta p$ is shown by the red arrow. The picture is adopted from Ref.~\onlinecite{Bobkov2023_oscillations}.}
  \label{fig:BZ}
	\end{center}
\end{figure}

In Sec.~\ref{Neel_origin} we discussed the qualitative mechanism of the N\'eel triplet pairing. It was shown that the N\'eel  pairing is pairing of electrons having the momenta $\bm p$ and $-\bm p + \bm Q$, where $\bm Q$ is the reciprocal lattice vector due to the periodicity enforced by the AF. In the 2D case it is $\bm Q = (\pi/a,\pi/a)$. In real space this pairing manifests itself as the atomic oscillation of the pair amplitude: it flips its sign from a site to its nearest neighbor. But if we only monitor on-site N\'eel triplet pairs on one of the sublattices, we see that their amplitude is homogeneous. Now let us consider a situation where we have breaking of the translational invariance in the system. The simplest case is a plane AF/S interface. Previously we discussed mainly effectively homogeneous systems: thin-film superconductors proximitized by AF insulators, which can be considered as homogeneous superconductors under the action of the N\'eel effective exchange field $h_{eff}$. The spatial dependence of the proximity effect in the direction perpendicular to the interface was neglected due to the small thickness of the S film in this direction. The only exception is Fig.~\ref{fig:trip} of this review, where this simplification was not used. It is clearly seen that the perfect N\'eel structure, which we see along the interface direction, is violated in the direction perpendicular to the interface. Even if we monitor the on-site N\'eel pairing at one of the sublattices, the corresponding amplitude has some oscillations. It is a signature of the formation of finite-momentum N\'eel triplet pairs due to scattering events at the interface. Now let us explain the mechanism of such pairing in more details. 

If we consider a plane interface, which breaks the translational invariance along the $x$-direction, then it is convenient to choose the Brillouin zone (BZ) as shown in Fig.~\ref{fig:BZ}. Due to the doubling of the unit cell BZ is compressed twice in the interface direction $y$. As a result, additional branches of the Fermi surface appear in the reduced BZ. Please note that such additional branches of the Fermi surface do not occur in the 1D case, considered in Sec.~\ref{Neel_origin}. Let us consider an electron $(p_{x1}, p_y)$ incoming to the AF/S interface from the AF side (marked by 1 in Fig.~\ref{fig:BZ}). Because of Umklapp scattering at the the AF/S interface this electron can be reflected as electron 3 from another branch, corresponding to the momentum $(p_{x3}, p_y)$ (for the plane interface the component of the electron momentum along the interface $p_y$ is conserved). That is why the electron 2 with momentum $(-p_{x1}, p_y)$ can form not only a singlet zero-momentum Cooper pair with the electron 1, but also a N\'eel-type triplet pair with the electron 3, which has a finite total momentum $\delta p = |p_{x3}-p_{x1}|$.  The normal state electron dispersion in the reduced BZ takes the form $\varepsilon = -\mu_{AF} + \sqrt{h^2 + 4t^2(\cos p_x a + \cos p_y a + \cos p_z a)^2}$, From this dispersion relation and the condition that $\varepsilon = 0$ at the Fermi surface we obtain $\delta p = \sqrt{\mu_{AF}^2 - h^2}/(ta \sin [p_x a])$. The last expression can be rewritten in terms of the electron Fermi velocity $v_{F,x} \equiv v = \partial \varepsilon/\partial p_x = 2ta \sin [p_x a]$ at $\mu_{AF}=h=0$ as $\delta p = 2\sqrt{\mu_{AF}^2 - h^2}/v$. The main contribution to the oscillations of the critical temperature is given by the normal trajectories with $v_{F,x} \approx v_F$ and then the oscillation period takes the form
\begin{eqnarray}
L_{osc} = \frac{\pi v_F}{\sqrt{\mu_{AF}^2 - h^2}}.
\label{eq:period}
\end{eqnarray}  

\begin{figure}[tb]
	\begin{center}
		\includegraphics[width=85mm]{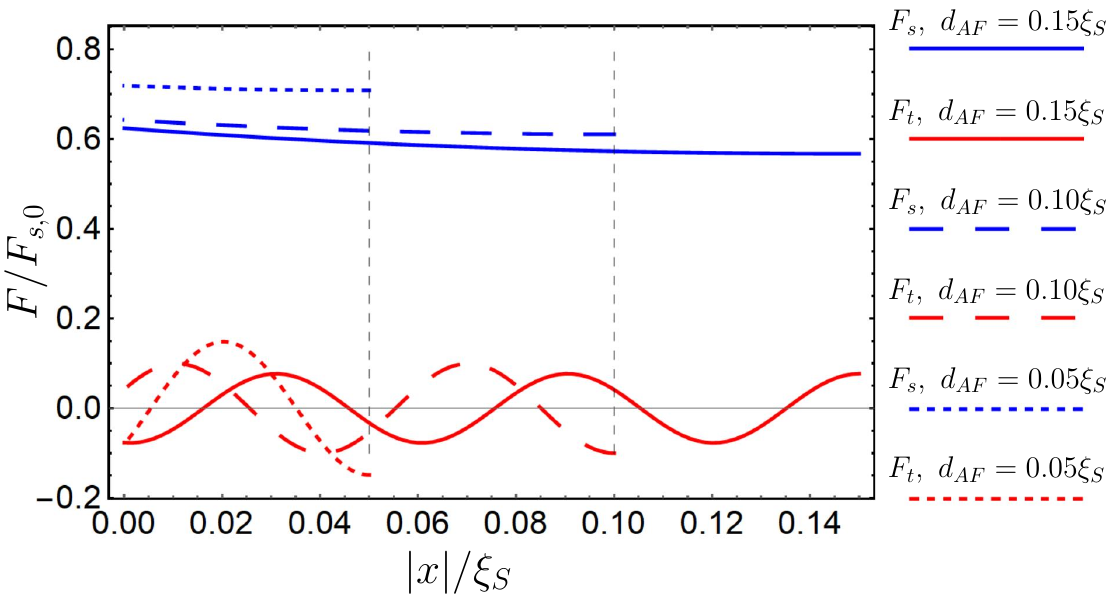}
\caption{Dependence of the $A$-sublattice triplet correlations amplitude $F_t$ (red) and the singlet amplitude $F_s$ (blue) for the normal to the AF/S interface trajectory $v > 0$ on the distance from the S/AF interface inside the AF layer. Different curves correspond to different thicknesses $d_{AF}$ of the AF layer. Each of the curves ends at the distance corresponding to the impenetrable edge of the AF layer. $F_{s,0}$ is the singlet amplitude in the absence of the AF layer. The picture is adopted from Ref.~\onlinecite{Bobkov2023_oscillations}.}
 \label{fig:oscillations_anomalous}
	\end{center}
\end{figure}

Figure \ref{fig:oscillations_anomalous} shows some typical examples of the spatial distribution of the on-site singlet and triplet correlations inside the AF layer at the $A$-sublattice. We see that the triplet correlations oscillate inside the antiferromagnet (period of these oscillations is in agreement with Eq. (\ref{eq:period})), while the singlet correlations just decay without oscillations, unlike the case of S/F heterostructures, where both singlet and triplet correlations manifest oscillations with the same period inside the ferromagnet.  The reason is that according to our qualitative consideration only N\'eel pairs can have finite momentum of the described physical origin. 

\subsection{Oscillations of the critical temperature of S/AF bilayers}

\begin{figure}[tb]
	\begin{center}
		\includegraphics[width=65mm]{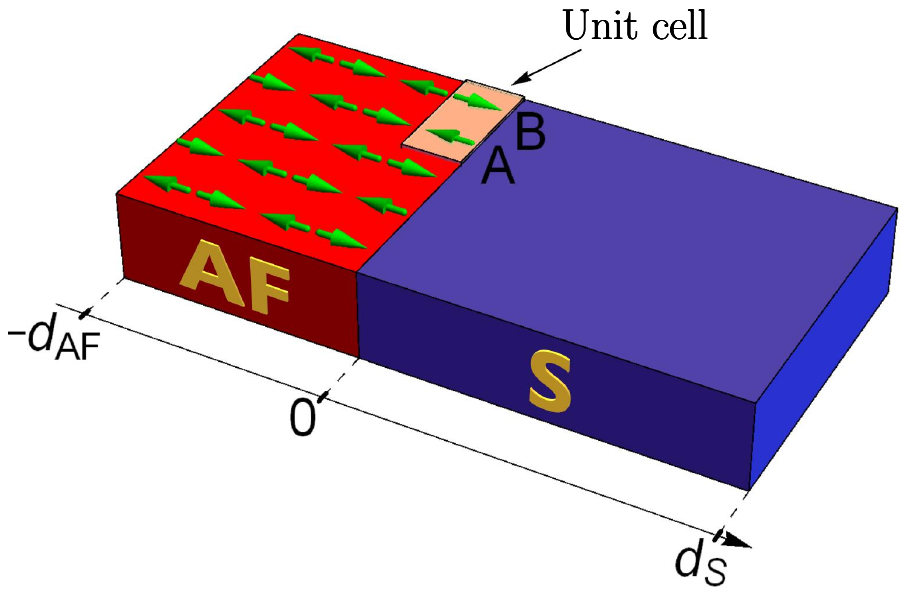}
		\caption{AF/S bilayer with a finite-width metallic AF.  Staggered magnetization of the AF layer is schematically depicted by arrows. The unit cell containing two sites belonging to A and B sublattices is also shown. The picture is adopted from Ref.~\onlinecite{Bobkov2023_oscillations}.}
  \label{fig:sketch_osc}
	\end{center}
 \end{figure}

Now we will discuss the effect which oscillations of the N\'eel triplet correlations have on the critical temperature of S/AF bilayers with metallic antiferromagnets. The sketch of the system is presented in Fig.~\ref{fig:sketch_osc}. In systems with finite-width layers these oscillating correlations can experience constructive or destructive interference due to the reflections from the impenetrable edge of the AF layer. This leads to the oscillating dependence of the N\'eel triplet correlations amplitude on the AF layer width $d_{AF}$, what, in its turn, makes the critical temperature of the bilayer also an oscillating function of $d_{AF}$. 

\begin{figure}[tb]
	\begin{center}
		\includegraphics[width=85mm]{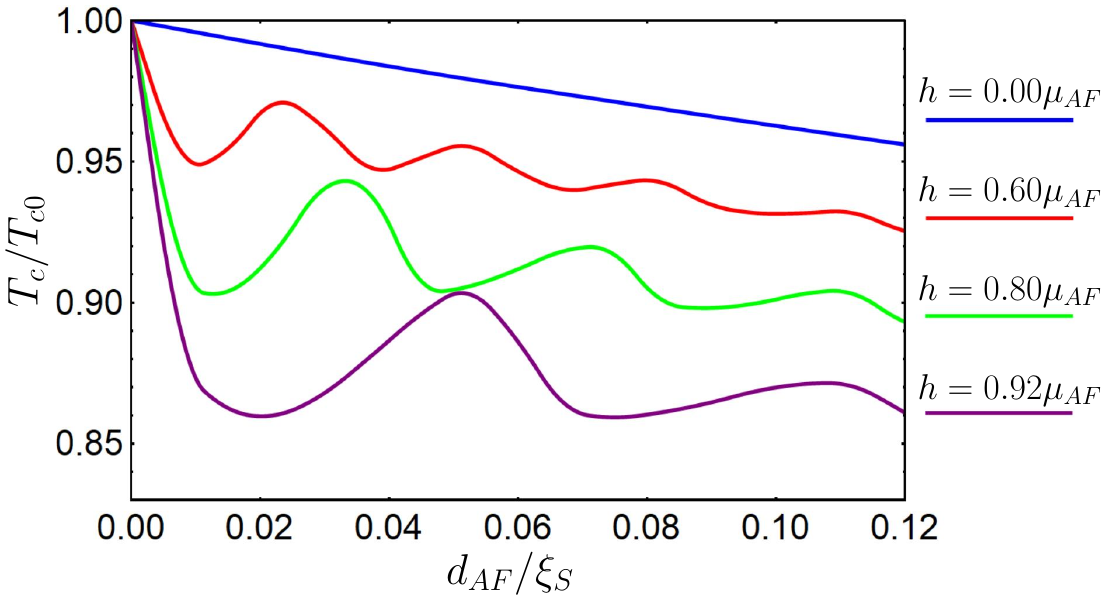}	
  \caption{Critical temperature of the AF/S bilayer as a function of the AF layer thickness $d_{AF}$. $d_S=1.5\xi_S$. The picture is adopted from Ref.~\onlinecite{Bobkov2023_oscillations}.}
  \label{fig:Tc1}
	\end{center}
\end{figure}

\begin{figure}[tb]
	\begin{center}
		\includegraphics[width=85mm]{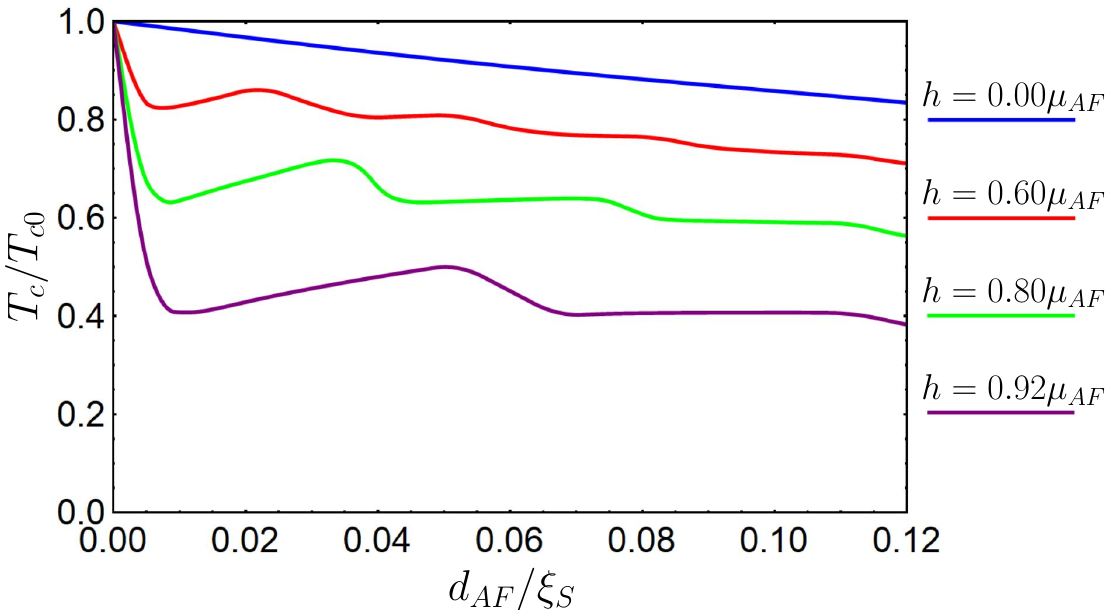}
		\caption{The same as in Fig.~\ref{fig:Tc1} but for $d_S=0.75\xi_S$. The picture is adopted from Ref.~\onlinecite{Bobkov2023_oscillations}.}
  \label{fig:Tc2}
	\end{center}
\end{figure}

\begin{figure}[tb]
	\begin{center}
		\includegraphics[width=85mm]{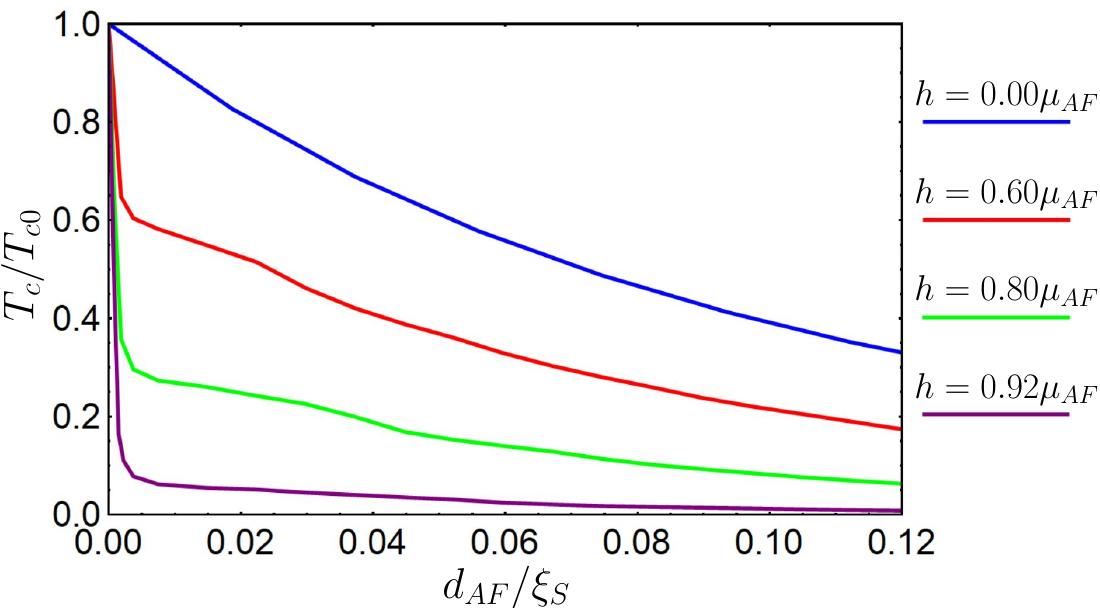}
		\caption{The same as in Fig.~\ref{fig:Tc1} but for $d_S=0.225\xi_S$. The picture is adopted from Ref.~\onlinecite{Bobkov2023_oscillations}. }
  \label{fig:Tc3}
	\end{center}
\end{figure}

Such behavior of the critical temperature was investigated in Ref.~\onlinecite{Bobkov2023_oscillations}. Figs.~\ref{fig:Tc1}-\ref{fig:Tc3} show the critical temperature as a function of $d_{AF}$ for different values of the S layer thickness $d_S$. Different curves in each figure correspond to different exchange fields $h$ of the AF layer. In all three figures we can see the superconductivity suppression accompanied by  oscillations of the critical temperature. The amplitude of the oscillations grows with the value of the exchange field, and their period is described well by Eq.~(\ref{eq:period}) regardless of the thickness of the superconductor. As the S layer gets thinner, the influence of the proximity effect on the superconductor increases, which leads to a stronger suppression of the critical temperature and a higher amplitude of the oscillations To see this we can compare Figs.~\ref{fig:Tc1} and \ref{fig:Tc2}. In Fig.~\ref{fig:Tc2} corresponding to smaller $d_{S}$ the amplitude of the oscillations is higher. This is because of larger amplitude of the triplet wave function reflected from the impenetrable edge of the AF. However, for the thinnest S layer (Fig.~\ref{fig:Tc3}) $T_c$ oscillations are weakly pronounced. This is explained by the fact that the amplitude of the oscillating N\'eel triplets inside the AF layer (and, consequently, the amplitude of $T_c$ oscillations) is greatly suppressed in this case together with superconductivity.

\section{Spin-valve effect in AF/S/AF heterostructures}

In this section we will discuss how N\'eel triplet correlations manifest themselves in AF/S/AF trilayers, following Ref.~\onlinecite{Bobkov2024_spinvalve}. First of all, we need to define what the spin-valve effect is. Let us consider a heterostructure constructed of a superconductor and two magnetic layers. If the superconducting critical temperature of the structure is sensitive to the mutual orientation of the magnetic layers magnetizations, we call it the spin-valve effect and the structure -- a spin valve. Switching between the superconducting and the normal states by changing the mutual orientation of the magnetizations is called the absolute spin-valve effect. 

Spin-valve effect in heterostructures with ferromagnets has been widely studied theoretically \cite{DEGENNES1966,Oh1997,Tagirov1999,Fominov2003,Fominov2010,Wu2012} and experimentally \cite{Li2013,Moraru2006,Singh2007,Zhu2010,Leksin2012,Banerjee2014,Jara2014,Singh2015,Kamashev2019,Westerholt2005,Deutscher1969,Gu2002,Gu2015}, and its origin is intuitively clear. Indeed, let us introduce the misorientation angle $\phi$ between F magnetizations. Then in the parallel (P) configuration, corresponding to $\phi=0$, the exchange fields of the two Fs strengthen each other, while in the antiparallel (AP) case with $\phi=\pi$ they compensate each other. Therefore, we can expect the critical temperature to be lower in the P case, than in the AP (which, however, is not always true because the interference of superconducting correlations makes the situation more complicated). 

What happens to the spin-valve effect in an AF/S/AF structure with fully compensated S/AF interfaces (that is, with zero interface magnetization)? It may seem that such a system is invariant towards reversing the direction of the N\'eel vectors in one of the AFs. Then there would be no physical difference between parallel and antiparallel configurations and, consequently, no spin-valve effect. However, in Ref.~\onlinecite{Johnsen_Kamra_2023} it was theoretically shown that the superconducting spin-valve effect can be realized in AF/S/AF structures with insulating antiferromagnets and fully compensated S/AF interfaces despite the absence of macroscopic magnetization in the AF layers. The explanation is connected with the N\'eel triplet correlations generated by the two S/AF interfaces. Reversing the direction of the N\'eel vectors in one of the AFs means reversing  signs of the amplitudes of the N\'eel correlations generated by the corresponding S/AF interface. Therefore, these correlations, analogously to exchange fields in spin valves with Fs, can be added (subtracted) inside the superconducting layer depending on the misorientation angle between the N\'eel vectors, thus suppressing superconductivity more (less) strongly.

\begin{figure}[tb]
	\begin{center}
		\includegraphics[width=80mm]{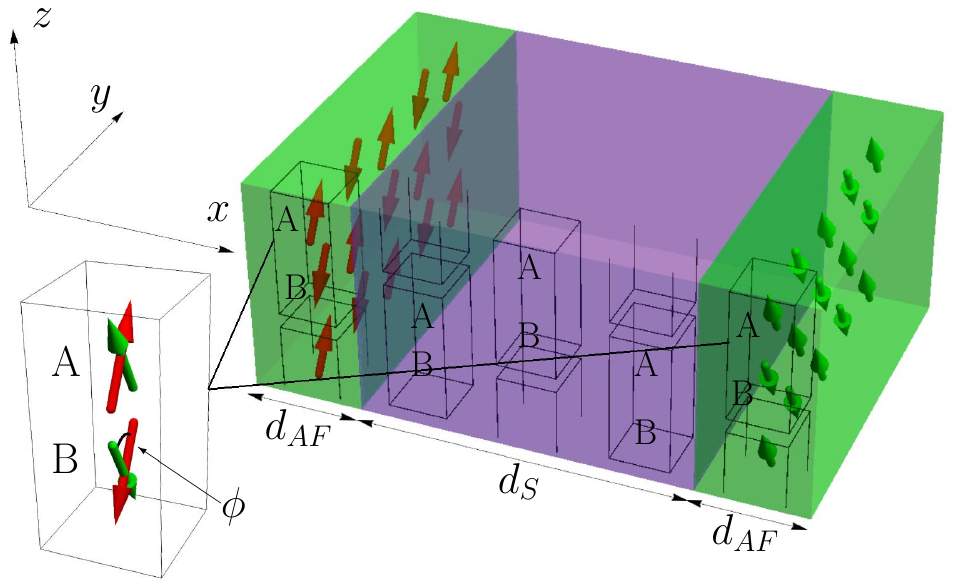}
		\caption{Sketch of the AF/S/AF system. Red and green arrows show N\'eel-type magnetizations of the AFs. The unified division into two sublattices with unit cells containing two sites belonging to $A$ and $B$ sublattices is also shown. The misorientation angle $\phi$ is defined as the angle between the magnetizations of two antiferromagnets at the same sublattice. The picture is adopted from Ref.~\onlinecite{Bobkov2024_spinvalve}.}
        \label{fig:valvesetup}
	\end{center}
\end{figure}

For describing the spin-valve effect in AF/S/AF systems with fully compensated interfaces it is convenient to define the misorientation angle $\phi$ as it is shown in Fig.~\ref{fig:valvesetup}. We perform a unified division of the entire AF/S/AF structure into two sublattices and  define the misorientation angle as the angle between the
magnetizations of two antiferromagnets at the same sublattice. The following two subsections are devoted to the influence of the chemical potential and impurities on the spin-valve effect in AF/S/AF structures with insulating antiferromagnets.

\subsection{Dependence of spin-valve effect in AF/S/AF heterostructures on chemical potential}

In this subsection we discuss the influence of  the chemical potential $\mu_S$ in the superconducting layer on  the dependence $T_c(\phi)$ in the clean case with no impurities in S. According to our definition of the misorientation angle, it seems reasonable to expect the relation $T_c^{P}<T_c^{AP}$ ($T_c(\phi=0) \equiv T_c^P$, $T_c(\phi=\pi) \equiv T_c^{AP}$), as in the parallel case the N\'eel triplets generated by the both S/AF interfaces are effectively summed up and strengthen each other inside the S layer. However, as it will be clear from the following, away from half-filling $\mu_S=0$ the opposite result $T_c^{P}>T_c^{AP}$ can be realized depending on the S width $d_S$.

\begin{figure}[tb]
	\begin{center}
		\includegraphics[width=85mm]{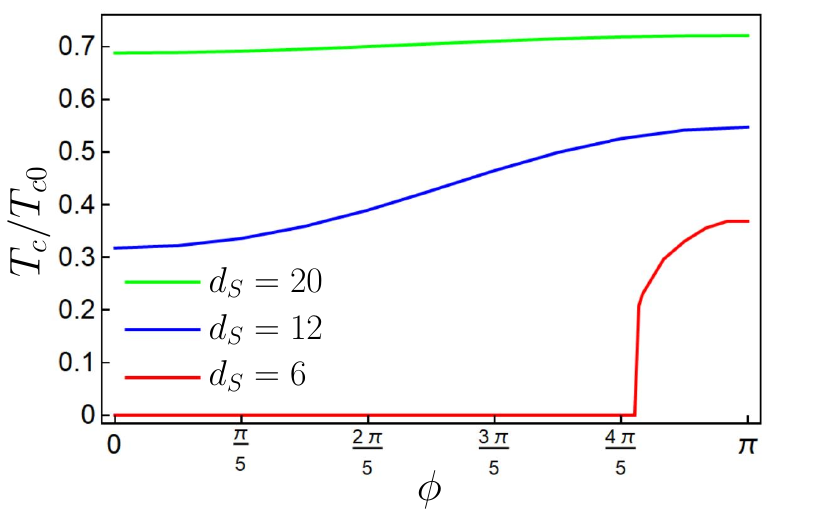}
		\caption{$T_c(\phi)$ for the AF/S/AF structure at half-filling $\mu_S=0$. Different curves correspond to different $d_S$, $d_{AF}=4$ (all widths are measured in the number of monolayers), $\mu_{AF}=0$, $h=0.5t$, $T_{c0}=0.07t$. The picture is adopted from Ref.~\onlinecite{Bobkov2024_spinvalve}.}
        \label{fig:Tcphi_0}
	\end{center}
\end{figure}

At first we consider the $T_c(\phi)$ dependence calculated in the case $\mu_S=0$, which is presented in Fig.~\ref{fig:Tcphi_0}. We can observe that the spin-valve effect is well-pronounced and the relation $T_c^P<T_c^{AP}$ is fulfilled for all considered values of $d_S$. For larger superconducting width the valve effect is reduced, which follows from physical considerations: in the limit $d_S \gg \xi_S$ ($\xi_S\approx 6$ monolayers for the data presented in this subsection) the valve effect should disappear because the two S/AF interfaces do not feel each other and the superconductivity suppression at each of them does not depend on the direction of the N\'eel vector. The curve corresponding to $d_S=6$ monolayers demonstrates that for the system under consideration the absolute spin-valve effect, that is the full suppression of superconducting state for a range of misorientation angles, is also possible.

\begin{figure}[tb]
	\begin{center}
		\includegraphics[width=85mm]{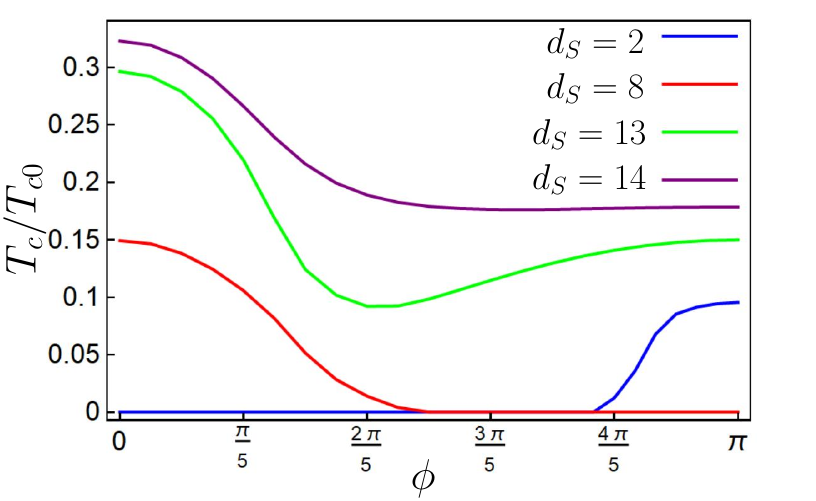}
		\caption{The same as in Fig.~\ref{fig:Tcphi_0} but for $\mu_S=0.2 t$. The picture is adopted from Ref.~\onlinecite{Bobkov2024_spinvalve}.}
        \label{fig:Tcphi_l}
 \end{center}
\end{figure}

\begin{figure}[tb]
	\begin{center}
		\includegraphics[width=85mm]{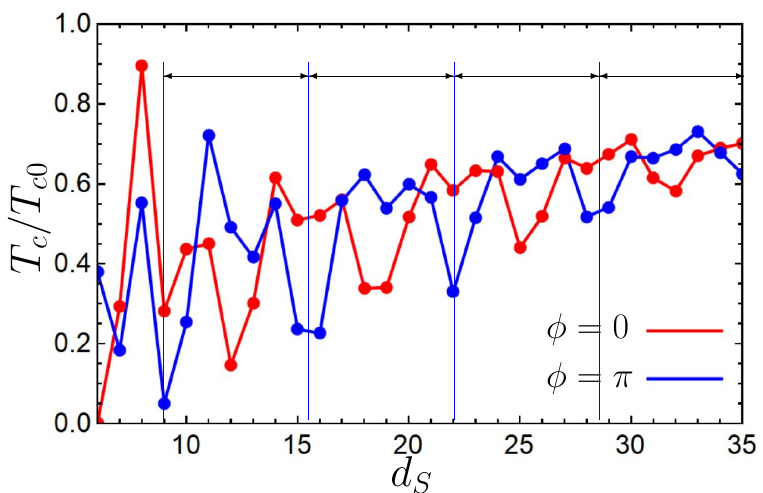}
		\caption{$T_c(0)$ and $T_c(\pi)$ as functions of $d_S$ at $\mu_S=0.9 t$.  $d_{AF}=4$, $\mu_{AF}=0$, $h=t$, $T_{c0}=0.03t$. $L_{\mathrm{ocs}} =  \pi v_F/\mu_S \approx 7$. Four periods (minima $T_c(d_S)$ for $\phi=\pi$) are shown on the plot by vertical blue lines. The picture is adopted from Ref.~\onlinecite{Bobkov2024_spinvalve}.}
        \label{fig:Tc_osc}
	\end{center}
\end{figure}

Fig.~\ref{fig:Tcphi_l} corresponds to  $\mu_S=0.2 t$ and demonstrates that in this case the relation between $T_c^P$ and $T_c^{AP}$ depends on the value of $d_S$. The reason is the finite-momentum N\'eel triplet correlations, discussed in Sec.~\ref{finite_momentum}. Their amplitude oscillates in the S layer with the period $L_{\mathrm{osc}} = \pi v_F/|\mu_S|$. Depending on the width of the S layer the N\'eel triplets generated by the opposite S/AF interfaces can interfere constructively or destructively in the superconductor, which manifests itself in the oscillating behavior of the resulting N\'eel triplet amplitude for a given $\phi$ upon varying $d_S$. This physical picture is further supported by the demonstration of the dependence of $T_c(0,\pi)$ on $d_S$ presented in Fig.~\ref{fig:Tc_osc}. The oscillations of the difference $T_c(\pi)-T_c(0)$ with the period $L_{\mathrm{osc}}$, accompanied by changing sign of the spin-valve effect (that is, the sign of $T_c(\pi)-T_c(0)$), are clearly seen. 

The other interesting feature is the nonmonotonicity of the curves in Fig.~\ref{fig:Tcphi_l}. The dip in the critical temperature at $\phi$ close to $\pi/2$ can be explained by generating of so-called cross product correlations with their amplitude maximal at $\phi \approx \pi/2$. These are triplet correlations determined by $\bm h_l \times \bm h_r$, where $\bm h_{l,r}$ are the N\'eel vectors of the  sublattice $A$ of the left and right AFs, respectively, $|\bm h_l|=|\bm h_r|=h$. These correlations are not of sign-changing N\'eel type and are usual equal-spin triplet correlations. Their amplitude is equal to zero at $\mu_S = 0$, that is why no dips are observed in the curves in Fig.~\ref{fig:Tcphi_0}. In Fig.~\ref{fig:Tcphi_l} the dip can be clearly seen only for $d_S=13$ monolayers. For lower values of the S width the cross product correlations are too weak to result in the pronounced dip feature, as their amplitude is proportional to $d_S/\xi_S$. For higher values of $d_S$ the influence of the cross product correlations is weakened due to smaller overlap of the N\'eel triplet correlations generated by the opposite S/AF interfaces.

\subsection{Dependence of spin-valve effect in AF/S/AF heterostructures on impurities}

In Refs.~\onlinecite{Johnsen_Kamra_2023,Bobkov2024_spinvalve} it was shown that at $h_{\mathrm{eff}}\equiv ha/d_S \ll T_c$ ($a$ is the lattice constant of the superconductor in the $x$-direction) the dependence $T_c(\phi)$ takes the form $T_c=T_{c,\parallel}+ \Delta T_{c,\parallel} \cos \phi + \Delta T_{c,\perp} \sin^2 \phi$, where $T_{c,\parallel} = (T_c(0)+ T_c(\pi))/2$, $\Delta T_{c,\parallel} = (T_c(0)- T_c(\pi))/2$ is the "$0-\pi$" spin-valve effect and $\Delta T_{c,\perp}=T_c(\pi/2)-T_{c,\parallel}$ is the  "perpendicular" spin-valve effect, corresponding to the dips at $T_c(\phi)$ curves at $\phi=\pi/2$. In this subsection we discuss the influence of impurities in the S layer on both "$0-\pi$" and "perpendicular" spin-valve effects. The impurities are modelled as random changes of the chemical potential $\mu_S$ at each site of the superconductor:
$\mu_i=\mu_S+\delta \mu_i, ~\delta \mu_i \in [-\delta\mu,\delta\mu],$ therefore the impurity strength is defined as $\delta\mu$. 

\begin{figure}[tb]
	\begin{center}
		\includegraphics[width=85mm]{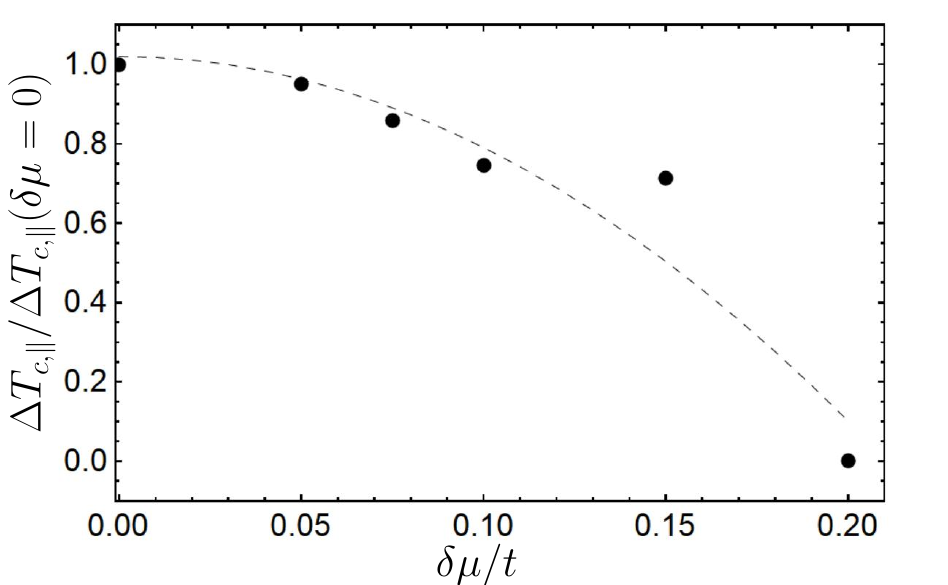}
		\caption{Suppression of the spin-valve effect by impurities. The difference $\Delta T_{c,\parallel}$ is plotted as a function of the impurity strength $\delta \mu$. The difference is normalized to its value at $\delta \mu = 0$. The dashed line is just a fit to provide a guide to eye. $\mu_S=0.9t$, $\mu_{AF}=0$, $h=t$, $d_{AF}=4$, $d_S=20$, $T_{c0}=0.03t$. The picture is adopted from Ref.~\onlinecite{Bobkov2024_spinvalve}.}
         \label{fig:Tcphiimpurities2}
	\end{center}
\end{figure}

Fig.~\ref{fig:Tcphiimpurities2} demonstrates the gradual disappearing of the "$0-\pi$" valve  effect $\Delta T_{c,\parallel} $ under the influence of impurities. This is explained by the fact that the spin-valve effect of this type is produced by the N\'eel triplets, which appear due to interband electron pairing \cite{Bobkov2022} and therefore are suppressed by impurities.

\begin{figure}[tb]
	\begin{center}
		\includegraphics[width=85mm]{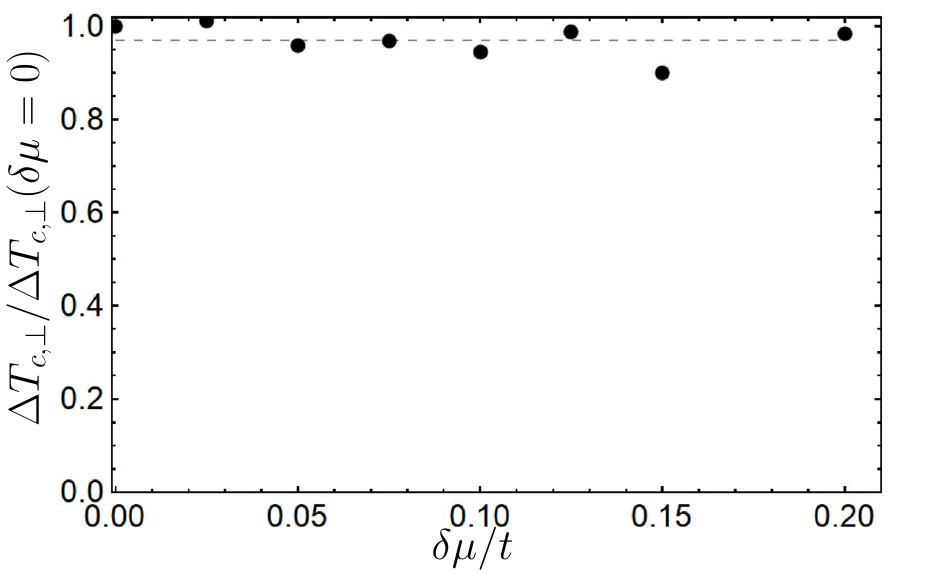}
		\caption{$\Delta T_{c,\perp}$, normalized to its value at $\delta \mu = 0$, plotted as a function of the impurity strength $\delta \mu$. All parameters are the same as in Fig.~\ref{fig:Tcphiimpurities2}. The picture is adopted from Ref.~\onlinecite{Bobkov2024_spinvalve}.}
         \label{fig:Tcphiimpurities3}
	\end{center}
\end{figure}

Fig.~\ref{fig:Tcphiimpurities3} presents the dependence of the "perpendicular" spin-valve effect $\Delta T_{c,\perp}$ on the impurity strength. We see that this effect tends to be insensitive to the presence of impurities. This can be considered as a proof of its origin from  equal-spin cross product triplet correlations, which should be insensitive to impurities as they are conventional (not N\'eel) triplets and correspond to intraband $s$-wave odd-frequency triplet electron pairing, which is not suppressed by nonmagnetic impurities.

\section{Andreev bound states at single impurities in AF/S heterostructures}

It is well-known that the Andreev bound states can occur at single impurities in superconductors  if the impurities suppress superconductivity for a given system \cite{Balatsky2006}. In particular,  magnetic impurities break the time-reversal symmetry and for this reason they are pair-breaking even for conventional $s$-wave superconductors. Well-known spin-split Yu-Shiba-Rusinov states occur at magnetic impurities \cite{Luh1965,Shiba1968,Rusinov1969}. They attracted much attention over the last several decades \cite{Balatsky2006}, primarily due to the fact that on chains of magnetic impurities they can form topological bands due to overlapping of bound states at separate impurities \cite{Nadj-Perge2014,Pawlak2016,Schneider2021}.

\begin{figure}[tb]
	\begin{center}
		\includegraphics[width=85mm]{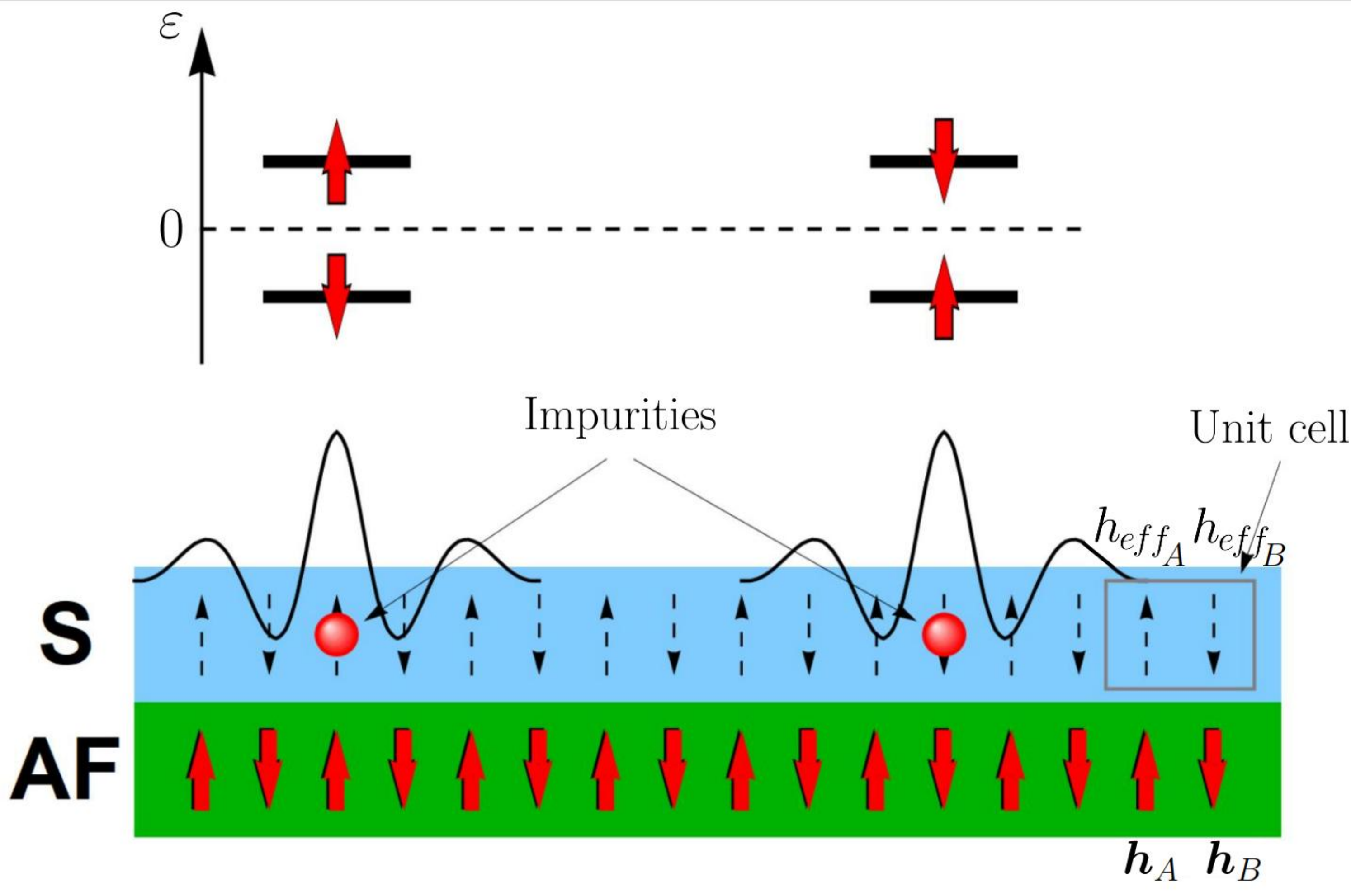}
\caption{Sketch of the S/AF bilayer with non-interacting impurities. Insulating two-sublattice antiferromagnet (AF) with staggered exchange field $\bm h_A = -\bm h_B$ induces an effective staggered exchange field $\bm h_{eff,A} = -\bm h_{eff,B} \equiv \bm h_{eff}$ via the proximity effect in the adjacent thin superconductor (S).  An impurity can occupy site $A$ or $B$ in the S layer. The both possible variants are shown by red balls. The LDOS of Andreev bound states localized at the corresponding impurity is shown schematically. The energy spectrum of the bound states with the appropriate spin structure (red arrows) is also shown above the corresponding impurity. The picture is adopted from Ref.~\onlinecite{Bobkov2024_singleimp}.}
 \label{fig:sketch}
	\end{center}
\end{figure}

In Ref.~\onlinecite{Bobkov2024_singleimp} it was demonstrated that {\it spin-splitted} impurity-induced Andreev bound states can also occur in S/AF heterostructures with conventional intraband $s$-wave pairing at {\it nonmagnetic} impurities. 
The system considered in Ref.~\onlinecite{Bobkov2024_singleimp} is sketched in Fig.~\ref{fig:sketch} and represents a thin-film bilayer composed of a superconductor and a two-sublattice antiferromagnet. The general physical argument allowing for the bound state at a nonmagnetic impurity in such a system is that the impurity can be viewed as effectively magnetic, as it was already discussed in Sec.~\ref{depairing_imp}.  The spin of a particular bound state is determined by the sublattice to which the impurity belongs, see Fig.~\ref{fig:sketch} for illustration. 

\begin{figure}[tb]
	\begin{center}
		\includegraphics[width=85mm]{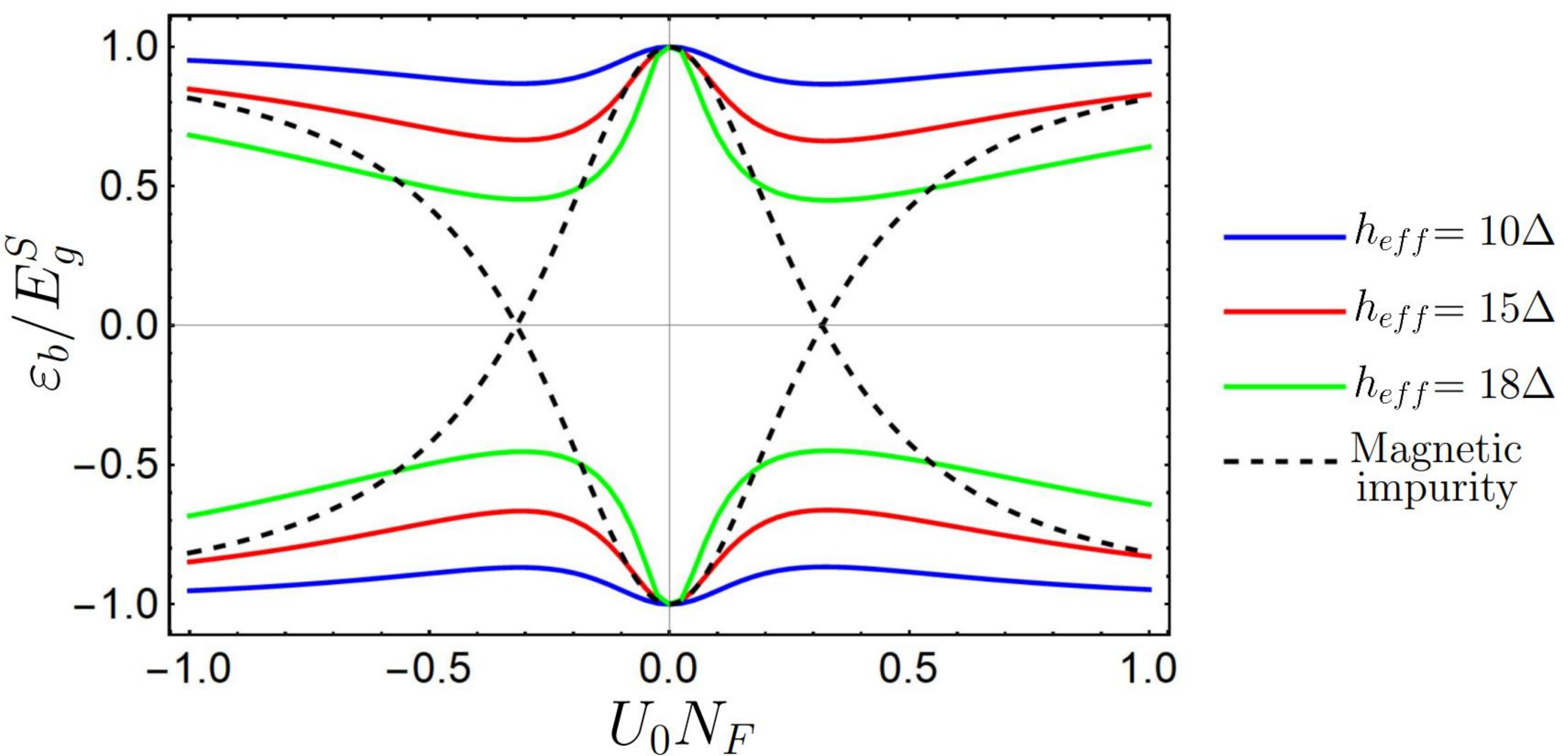}
\caption{Bound state energies as functions of the impurity strength. $\mu=20\Delta$, where $\Delta$ is the value of the order parameter of the superconductor. The energy of the bound state $\varepsilon_b$ is normalized to the value of the superconducting gap $E_g^S$. Different colors correspond to different $h_{eff}$. Dashed lines represent bound state energies at a magnetic impurity with the same strength in a conventional $s$-wave superconductor. The picture is adopted from Ref.~\onlinecite{Bobkov2024_singleimp}.}
 \label{fig:energies}
	\end{center}
\end{figure}

As it was shown in Ref.~\onlinecite{Bobkov2024_singleimp}, the presence or absence of the Andreev bound states at single nonmagnetic impurities in S/AF bilayers is also very sensitive to the value of the chemical potential $\mu$ of the superconductor. It is the third example of the remarkable sensitivity of the physics of S/AF heterostructures to the value of the chemical potential. For large $\mu \gg T_{c0}$, when the impurities can be viewed as effectively magnetic, the bound states exist. Energies of the bound states as functions of the impurity strength are plotted in Fig.~\ref{fig:energies}. It is seen that at stronger staggered effective exchange field $h_{eff}$ the bound state is shifted deeper inside the superconducting gap region. For comparison the energies of the Yu-Shiba-Rusinov bound states at a magnetic impurity with the same strength in a conventional $s$-wave superconducting host are plotted by the dashed lines. Unlike the case of magnetic impurity the nonmagnetic impurities in S/AF bilayers are not able to support low and zero-energy bound states. In this sense one can say that they are weaker pair-breakers as compared to the magnetic impurities. For small $\mu \lesssim T_{c0}$ the bound states do not appear because the impurities are not effectively magnetic \cite{Bobkov2024_singleimp}.

\begin{figure}[tb]
	\begin{center}
		\includegraphics[width=90mm]{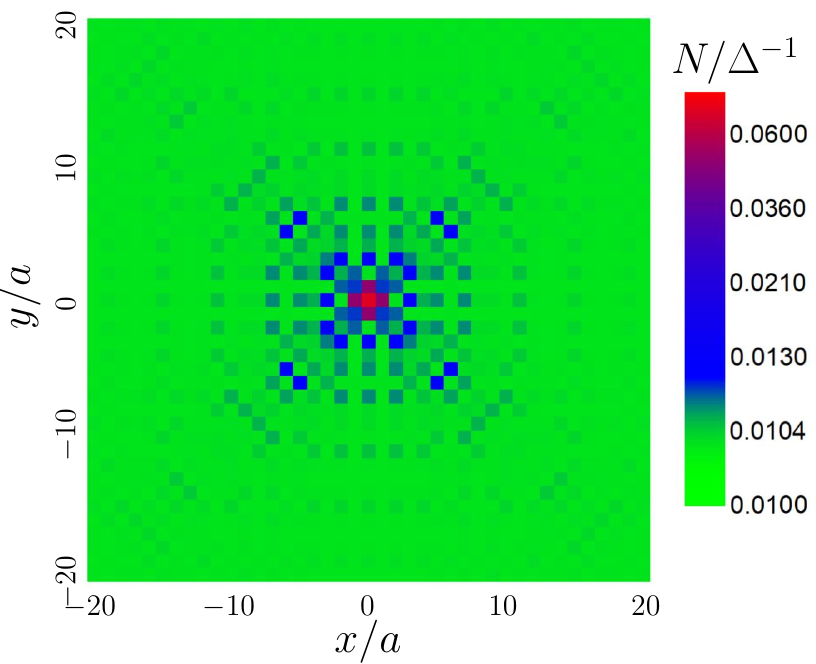}
\caption{LDOS at the energy corresponding to one of the bound states $\varepsilon = -|\varepsilon_b|$ as a function of coordinates. The impurity is at $A$-site in the centre of the presented spatial region. The picture is adopted from Ref.~\onlinecite{Bobkov2024_singleimp}. }
 \label{fig:LDOS_spatial}
	\end{center}
\end{figure}

The spatial region occupied by the bound state has a scale of the order of the superconducting coherence length $\xi_S$. The exponential decay is superimposed by a power-law suppression analogously to the case of magnetic impurities in conventional superconductors \cite{Balatsky2006}. However, unlike the magnetic impurities in conventional superconductors here the local density of states (LDOS) has a "staggered" component, which oscillates between the sublattices. If the impurity is localized at $A$-site, the bound state LDOS is mainly concentrated at the $B$-sublattice everywhere except for the atomic-scale region near the impurity site. The spatial structure of the LDOS is shown in Fig.~\ref{fig:LDOS_spatial}. 

An interesting feature of the spatial structure of the bound state LDOS is that the overall decay of the LDOS and "staggered" oscillations associated with the sublattice structure are also superimposed by  oscillations of a larger spatial scale compared to the atomic one, which is nevertheless significantly smaller than the superconducting coherence length scale. These oscillations are due to the generation of finite-momentum N\'eel-type triplet correlations, which were discussed in Sec.~\ref{finite_momentum}. Here they are produced due to the Umklapp electron scattering processes at the impurities. The period of oscillations is $L_{osc} = \pi v_F/ \sqrt{\mu^2 - h^2}$. Data presented in Fig.~\ref{fig:LDOS_spatial} are calculated at $h=1.5 t$ and $\mu = 2 t$. Then $L_{osc} \approx 4 a$, which is in agreement with the additional oscillation period seen in the figure. 

The presence of Andreev bound states at single nonmagnetic impurities in S/AF bilayers is in agreement with the behavior of the superconducting critical temperature of such systems in the presence of random disorder, which has already been studied \cite{Bobkov2023_impurities}  and is discussed in Sec.~\ref{impurities}. At $\mu \lesssim T_{c0}$  the nonmagnetic impurities are not pair-breaking and enhance superconductivity of S/AF bilayers due to the suppression of the N\'eel triplet correlations \cite{Bobkov2022,Bobkov2023_impurities}. On the contrary, if $\mu \gg T_{c0}$ the superconductivity is suppressed by random nonmagnetic disorder\cite{Buzdin1986,Bobkov2023_impurities,Fyhn2022,Fyhn2022_1}. The same sensitivity to the value of the chemical potential occurs in the problem of a single impurity: the bound states only exist at $\mu \gg T_{c0}$, when the impurities are suppressing for superconductivity.

\section{N\'eel triplets in the presence of spin-orbit interaction}

Many studies are devoted to the interplay of conventional correlations and spin-orbit coupling (SOC) in S/F hybrid structures \cite{Gorkov2001,Annunziata2012,Bergeret2013,Bergeret2014,Edelstein2003,Edelstein2003_JETPLett,Jacobsen2015,Linder_review}. It was predicted and observed that SOC in S/F bilayers can produce an anisotropic depairing effect on triplets. One of the manifestations of the anisotropic depairing is that the critical temperature $T_c$ of the bilayer depends on the orientation of the F layer magnetization with respect to the S/F interface\cite{Jacobsen2015,Ouassou2016,Simensen2018,Banerjee2018}. This behavior is interesting not only from a fundamental point of view, but also for spintronics applications because there is a possibility for a reciprocal effect i.e., a reorientation of the F layer magnetization due to superconductivity\cite{Jonsen2019,Gonzalez-Ruano2020,Gonzalez-Ruano2021}. The possibility to control magnetic anisotropies using superconductivity is a key step
in designing future cryogenic magnetic memories and spintronics applications. 

In this section we discuss anisotropic effect of Rashba SOC on the N\'eel triplets in S/AF thin film bilayers. The key feature is that in addition to the anisotropic depairing of triplet correlations known in S/F hybrids, a unique effect of anisotropic enhancement of the triplets by the SOC occurs in the S/AF case. We discuss the physical mechanism of the effect and demonstrate that it can manifest itself in opposite trend in the anisotropy of the superconducting transition as compared to S/F heterostructures. The SOC results in the depairing or enhancement of the N\'eel triplets depending on the value of the chemical potential of the superconductor and it is the fourth physical manifestation of the strong sensitivity of the physics of S/AF hybrid structures to the value of the chemical potential.

\subsection{Anisotropy of the N\'eel triplets and $T_c$}

\begin{figure}[tb]
	\begin{center}
		\includegraphics[width=75mm]{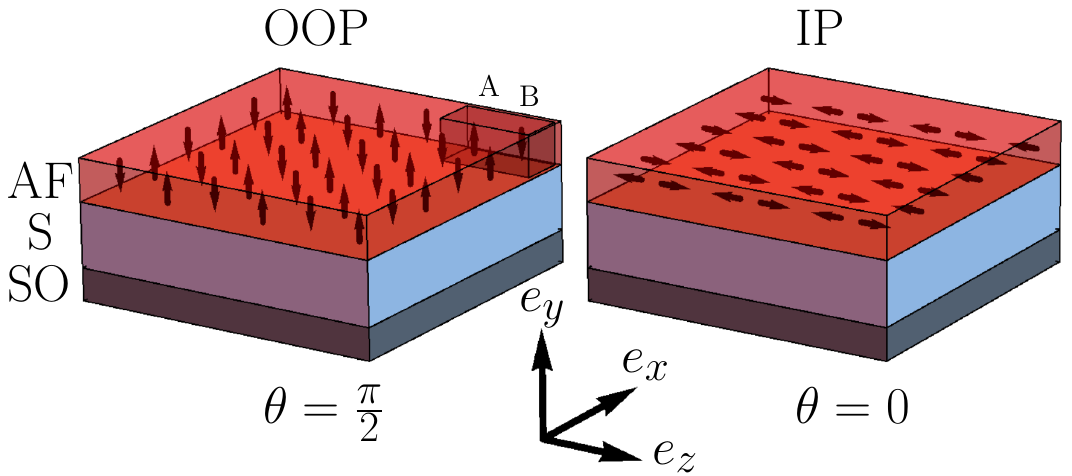}
		\caption{Sketch of the thin-film AF/S bilayer with SOC.  The N\'eel vector of the AF makes angle $\theta$ with the plane of the structure. $\theta=0$ corresponds to the in-plane (IP) and $\theta=\pi/2$ accounts for the  out-of-plane (OOP) orientations. Unit cell with two sites A and B is also shown. The picture is adopted from Ref.~\onlinecite{Bobkov2023_anisotropy}.}
        \label{fig:setup}
	\end{center}
\end{figure}

The anisotropic effect of Rashba SOC was considered in Ref.~\onlinecite{Bobkov2023_anisotropy} by an example of a thin-film S/AF bilayer, where the antiferromagnet is assumed to be an insulator, see Fig.~\ref{fig:setup}. The SOC is induced in the S layer by proximity to a heavy metal layer like Pt (shown as the SO layer in Fig.~\ref{fig:setup}). The SOC can also be due to inversion symmetry breaking in the S film by itself. The magnetism is staggered. The S/AF interfaces were assumed  fully compensated, that is the interface magnetization had zero average value. The influence of the antiferromagnetic insulator on the superconductor was described by the  exchange field $\bm h_{eff,\bm i} = \bm (-1)^{i_x+i_z} \bm h_{eff} $.  

\begin{figure}[tb]
	\begin{center}
		\includegraphics[width=65mm]{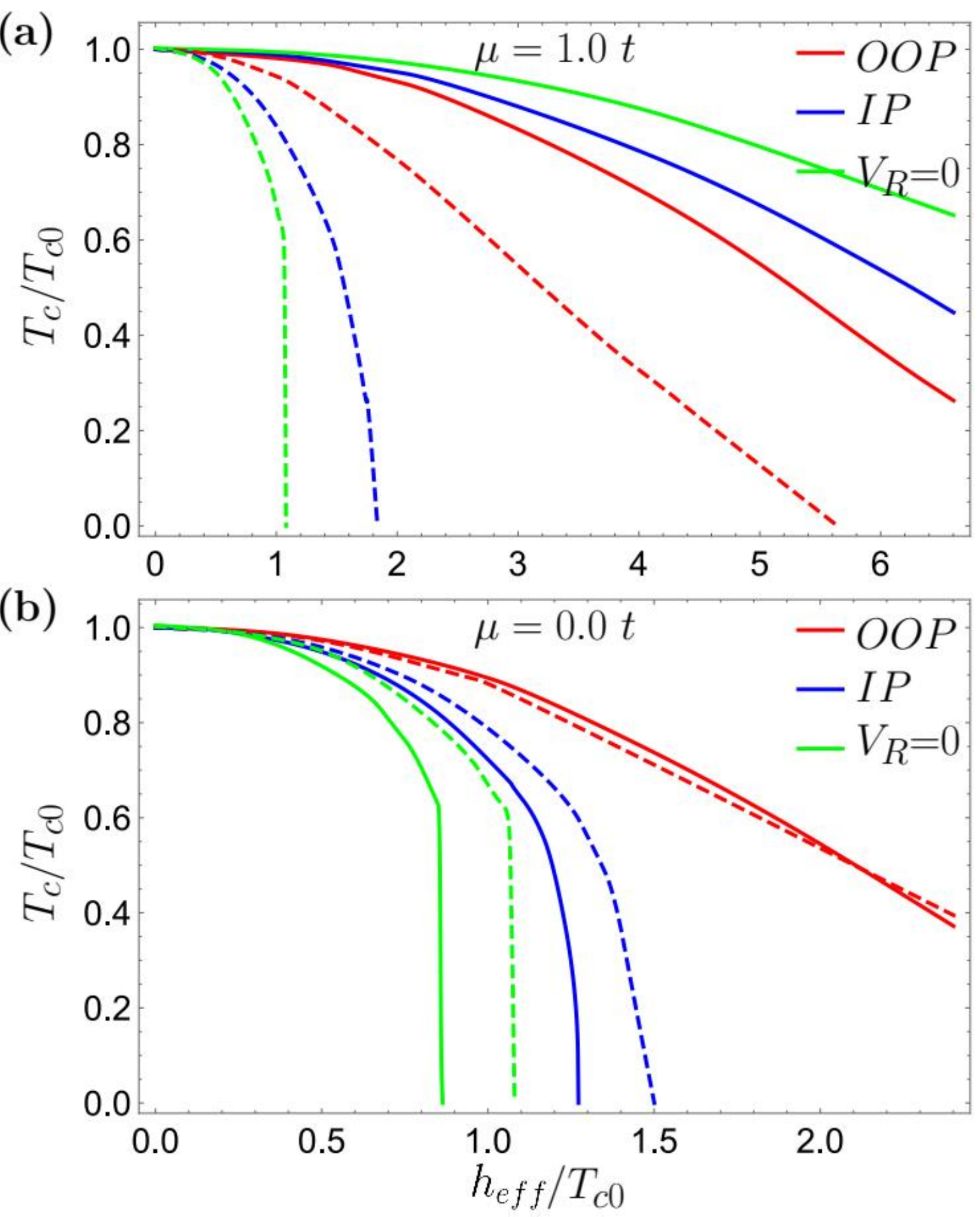}
		\caption{Critical temperature of S/AF (solid curves) and S/F bilayers (dashed) as a function of the effective exchange field $h_{eff}$. Panels (a) and (b) correspond to $\mu=12 T_{c0}$ and $\mu=0$, respectively. Green curves represent the results with no SOC, red and blue curves are for out-of-plane (OOP) and in-plane (IP) orientations, respectively and $V_R = 0.4t$ is the Rashba SOC strength. The picture is adopted from Ref.~\onlinecite{Bobkov2023_anisotropy}.}
        \label{fig:Tc}
	\end{center}
\end{figure}

Fig.~\ref{fig:Tc} demonstrates the dependencies $T_c(h_{eff})$ for S/AF structures with in-plane (IP) and out-of-plane (OOP) orientations of the N\'eel vector. They have been compared to $T_c(h_{eff})$ of the S/F system with the same absolute value of the effective exchange field $h_{eff}$ (conventional, not staggered). First, it is seen that while for S/F heterostructures $T_c$ is always higher in the presence of SOC (dashed curves), for AF/S heterostructures the trends are opposite for large $\mu \gg T_{c0}$ and for small $\mu \lesssim T_{c0}$. At small $\mu$ the behavior of $T_c$ is qualitatively similar to the case of S/F bilayers, and at large $\mu$ it is opposite - the presence of SOC {\it suppresses} $T_c$.

Furthermore, in the presence of SOC $T_c$ is anisotropic depending on the angle $\theta$ between the magnetization and the interface plane. For S/F heterostructures $T_c$ is always higher for OOP orientation (dashed curves) \cite{Jacobsen2015,Ouassou2016,Simensen2018,Banerjee2018}. At the same time for AF/S heterostructures the ratio between the values of $T_c$ for IP and OOP orientations is again opposite for large $\mu \gg T_{c0}$ and for small $\mu \lesssim T_{c0}$.
At $\mu \lesssim T_{c0}$ for S/AF heterostructures the ratio between $T_c$ for IP and OPP orientations is the same as for the S/F case. At $\mu \gg T_{c0}$ the anisotropy of the critical temperature, that is the difference between the IP and OOP $T_c$ is opposite. 

\subsection{Physical mechanism of the $T_c$ anisotropy}

Here we discuss the physical reasons of the numerical findings described in the previous subsection. First of all, as it was already discussed in Sec.~\ref{canted}, if $\mu \lesssim \pi T_{c0}$ the most important contribution to the pairing correlations is given by the electronic states corresponding to $\xi_\pm \approx 0$, what means that the electrons are at the edge of the antiferromagnetic gap and, therefore, practically fully localized at one of the sublattices. Consequently, they only feel the magnetization of the corresponding sublattice and behave in the same way as in the ferromagnet. For this reason our results at $\mu \lesssim \pi T_c$ demonstrate the same trends as the corresponding results for S/F structures. It is well-known that in S/F heterostructures the critical temperature is higher for OOP magnetization orientation than for the IP magnetization\cite{Banerjee2018} due to the fact that SOC suppresses triplets oriented OOP more than triplets oriented IP. The same is observed in Fig.~\ref{fig:Tc}(b) for S/AF bilayers with $\mu = 0$. Moreover, the SOC always suppresses triplets and, correspondingly, enhances $T_c$. The same is seen in Fig.~\ref{fig:Tc}(b) for S/AF heterostructures.  

The opposite trend at $\mu \gg T_{c0}$ is due to the existence of a unique mechanism of generation of the N\'eel triplet correlations in S/AF heterostructures, which differs from the mechanism of the direct singlet-triplet conversion known in S/F heterostructures. The N\'eel triplets are generated via the normal state N\'eel-type spin polarization of the DOS \cite{Bobkov2023_anisotropy}.  Up to the leading order in $h_{eff}/|\mu|$ and $h_R/|\mu|$ the N\'eel-type polarization of the normal state DOS along $\bm h_{eff}$ and at the Fermi surface takes the form 
\begin{eqnarray}
P_{\bm h}^A  = - P_{\bm h}^B  = 2N_F \left[ \frac{h_{eff}}{\mu} + \frac{h_{eff} h_R^2 \sin^2 \phi}{\mu^3} \right],
\label{eq:dos_polarization_normal}
\end{eqnarray}
where $N_F$ is the normal state DOS at the Fermi surface of the isolated superconductor, $\bm h_R \propto V_R (\bm e_y \times \bm v_F)$ is the effective Rashba pseudomagnetic field seen by an electron moving along the trajectory determined by the Fermi velocity $\bm v_F$ and $\phi$ is the angle between $\bm h_{eff}$ and $\bm h_R$. It is seen that (i) the absolute value of the polarization is always enhanced by the SOC and (ii) the enhancement is anisotropic. It reaches maximal possible value for all the trajectories for OOP orientation of $\bm h_{eff}$ because $\bm h_R$ is always in-plane. The reason for the described above enhancement is the specific reconstruction of normal state electron spectra under the influence of the SOC \cite{Bobkov2023_anisotropy}.

In S/F heterostructures the effective exchange field $\bm h_{eff}$ is also generated in the superconductor due to proximity with a ferromagnetic insulator. It results in the occurrence of the normal state polarization of the DOS of conventional type $ P \sim N_F (h_{eff}/\varepsilon_F) \ll N_F$. However, this polarization is always very small, because the effective exchange induced in the superconductor cannot be higher than the superconducting order parameter \cite{Sarma1963}, otherwise it completely suppresses superconductivity. Therefore, in S/F bilayers this quantity does not play a significant role in the generation of triplets. At the same time,  the N\'eel type polarization, described by Eq.~(\ref{eq:dos_polarization_normal}) is not necessary small and provides a generator for the N\'eel triplets. The enhancement of the N\'eel type polarization of the normal state by SOC leads to the {\it enhancement} of the N\'eel-type triplet correlations in S/AF structures. It was obtained in Ref.~\onlinecite{Bobkov2023_anisotropy} that at $\mu \gg T_{c0}$ up to the leading order in $h_{eff}/|\mu|$ and $h_R/|\mu|$ the anomalous Green's function of the N\'eel-type triplet correlations takes the form:
\begin{eqnarray}
\bm f^y = {\rm sgn}\omega \left[\frac{i\Delta \bm h_{eff}}{\mu^2} + \frac{3i\Delta [\bm h_R \times (\bm h_{eff} \times \bm h_R)]}{\mu^4}\right],~
\label{eq:triplet_sol_AF}
\end{eqnarray}
where $\omega$ is the Matsubara frequency. The component of $\bm f^y$ along the effective exchange field $\bm h_{eff}$, which accounts for the suppression of the critical temperature by the triplets, takes the form:
\begin{eqnarray}
\bm f_{\bm h}^y = {\rm sgn}\omega \frac{\bm h_{eff}}{\mu} \left[\frac{i\Delta }{\mu} + \frac{3i\Delta  h_R^2 \sin^2 \phi}{\mu^3}\right].~
\label{eq:triplet_sol_AF}
\end{eqnarray}
Thus, the amplitude of the N\'eel triplet correlations qualitatively follows the normal state N\'eel polarization and it is also enhanced by the SOC. The effect of the enhancement is the strongest for the OOP orientation corresponding to $\phi = \pi/2$.

\section{Conclusions}

In this review we discussed the physics of N\'eel triplet proximity effect in S/AF heterostructures, which was studied in a number of recent papers. The main findings can be summarized as follows:
\begin{itemize}
    \item 
 {At S/AF interfaces unconventional triplet correlations are generated. The amplitude of the corresponding pair wave function flips sign from one site of the materials to the nearest one following the N\'eel structure of the antiferromagnetic order parameter. The correlations can occur in superconductors due to proximity to the antiferromagnetic insulators and metals and penetrate into antiferromagnetic metals from superconductors. They are generated even at fully compensated S/AF interfaces with zero net magnetization of the interface. These triplet correlations were called the N\'eel triplet correlations.} 

 \item 
 {
 The N\'eel triplets suppress superconductivity. The efficiency of the suppression is of the same order and frequently a bit stronger than the suppression by the conventional proximity-induced triplet correlations in S/F heterostructures.
 }

 \item
 {
 The mechanism of the N\'eel triplets generation differs from the well-known direct singlet-triplet conversion in S/F heterostructures. They are produced via  the effect caused on the singlet superconducting correlations by  the N\'eel-type normal state electron polarization, induced in the superconductor by proximity to the AF. This mechanism results in the opposite as compared to S/F heterostructures trends in the anisotropy of the critical temperature of S/AF heterostructures in the presence of SOC. 
 }

 \item
 {The higher the N\'eel exchange field of the AF or proximity-induced effective exchange field in the S the stronger the amplitude of the N\'eel triplets, is. At the same time, unlike S/F heterostuctures the amplitude of the N\'eel triplet correlations is very sensitive to the value of the chemical potential in the material, where they are induced. Deviation from half filling suppresses the amplitude of the N\'eel triplets for a given value of the N\'eel exchange field.}

\item
{Near half-filling the N\'eel triplets are interband, and far from half-filling they are intraband. It results in very different response of the N\'eel triplet correlations on the nonmagnetic impurities: near half-filling the N\'eel triplets are suppressed by impurities and far from half-filling they are immune to impurities.}

\item
{The above behavior of the N\'eel triplets in the presence of nonmagnetic impurities leads to very different dependencies of the critical temperature of S/AF bilayers on the impurity strength: near half-filling the impurities enhance the critical temperature, and far from half-filling they suppress it. The second tendency is caused by an additional mechanism: far from half-filling the nonmagnetic impurities in S/AF heterostructures behave like effectively magnetic and suppress singlet superconductivity by themselves.}

\item
{The effective magnetic character of the nonmagnetic impurities in S/AF heterostructures leads to the occurrence of the spin-split Andreev bound states at single impurities, which are reminiscent of well-known Yu-Shiba-Rusinov bound states at magnetic impurities in superconducting materials, but occur at {\it nonmagnetic} impurities.}

\item 
{The N\'eel triplet correlations provide a possibility to realize a spin-valve effect in AF/S/AF trilayers even with fully compensated interfaces.}

\item
{Due to the lost of the translational invariance at S/AF interfaces and at single nonmagnetic impurities a finite-momentum N\'eel triplet pairing can occur. It results in the oscillations of the critical temperature of S/AF bilayers with metallic antiferromagnets as a function of the thickness of the AF layer, specific oscillations of the impurity-induced DOS around nonmagnetic impurities and changing sign of the spin-valve effect in AF/S/AF trilayers upon varying the width of the S layer.}

\item 
{Unlike S/F heterostructures the physics of S/AF heterostructures is crucially sensitive to the value of the chemical potential. Two physically very different regimes can be identified near half-filling and far from half-filling. In particular, the dependence of the critical temperature of S/AF heterostructures on impurity concentration is opposite in these regimes, the dependence of the critical temperature of heterostructures with canted AFs on the canting angle is also opposite, the same applies to the magnetic anisotropy in S/AF heterostructures with SOC and to the formation of Andreev bound states at single nonmagnetic impurities.}
\end{itemize}

It could be especially interesting to study the described effects in heterostructures composed of antiferromagnets and 2D superconductors because of possibility of external control of the chemical potential.

\begin{acknowledgments}

We are grateful to A. A. Golubov for the discussion of the
obtained results. The work was supported by Grant from
the Ministry of Science and Higher Education of the Russian
Federation No. 075-15-2024-632.

\end{acknowledgments}

\bibliography{Neel.bib}

\begin{thebibliography}{119}%
\makeatletter
\providecommand \@ifxundefined [1]{%
 \@ifx{#1\undefined}
}%
\providecommand \@ifnum [1]{%
 \ifnum #1\expandafter \@firstoftwo
 \else \expandafter \@secondoftwo
 \fi
}%
\providecommand \@ifx [1]{%
 \ifx #1\expandafter \@firstoftwo
 \else \expandafter \@secondoftwo
 \fi
}%
\providecommand \natexlab [1]{#1}%
\providecommand \enquote  [1]{``#1''}%
\providecommand \bibnamefont  [1]{#1}%
\providecommand \bibfnamefont [1]{#1}%
\providecommand \citenamefont [1]{#1}%
\providecommand \href@noop [0]{\@secondoftwo}%
\providecommand \href [0]{\begingroup \@sanitize@url \@href}%
\providecommand \@href[1]{\@@startlink{#1}\@@href}%
\providecommand \@@href[1]{\endgroup#1\@@endlink}%
\providecommand \@sanitize@url [0]{\catcode `\\12\catcode `\$12\catcode `\&12\catcode `\#12\catcode `\^12\catcode `\_12\catcode `\%12\relax}%
\providecommand \@@startlink[1]{}%
\providecommand \@@endlink[0]{}%
\providecommand \url  [0]{\begingroup\@sanitize@url \@url }%
\providecommand \@url [1]{\endgroup\@href {#1}{\urlprefix }}%
\providecommand \urlprefix  [0]{URL }%
\providecommand \Eprint [0]{\href }%
\providecommand \doibase [0]{http://dx.doi.org/}%
\providecommand \selectlanguage [0]{\@gobble}%
\providecommand \bibinfo  [0]{\@secondoftwo}%
\providecommand \bibfield  [0]{\@secondoftwo}%
\providecommand \translation [1]{[#1]}%
\providecommand \BibitemOpen [0]{}%
\providecommand \bibitemStop [0]{}%
\providecommand \bibitemNoStop [0]{.\EOS\space}%
\providecommand \EOS [0]{\spacefactor3000\relax}%
\providecommand \BibitemShut  [1]{\csname bibitem#1\endcsname}%
\let\auto@bib@innerbib\@empty
\bibitem [{\citenamefont {Cooper}(1956)}]{Cooper1956}%
  \BibitemOpen
  \bibfield  {author} {\bibinfo {author} {\bibfnamefont {Leon~N.}\ \bibnamefont {Cooper}},\ }\bibfield  {title} {\enquote {\bibinfo {title} {Bound electron pairs in a degenerate fermi gas},}\ }\href {\doibase 10.1103/PhysRev.104.1189} {\bibfield  {journal} {\bibinfo  {journal} {Phys. Rev.}\ }\textbf {\bibinfo {volume} {104}},\ \bibinfo {pages} {1189--1190} (\bibinfo {year} {1956})}\BibitemShut {NoStop}%
\bibitem [{\citenamefont {Bardeen}\ \emph {et~al.}(1957)\citenamefont {Bardeen}, \citenamefont {Cooper},\ and\ \citenamefont {Schrieffer}}]{Bardeen1957}%
  \BibitemOpen
  \bibfield  {author} {\bibinfo {author} {\bibfnamefont {J.}~\bibnamefont {Bardeen}}, \bibinfo {author} {\bibfnamefont {L.~N.}\ \bibnamefont {Cooper}}, \ and\ \bibinfo {author} {\bibfnamefont {J.~R.}\ \bibnamefont {Schrieffer}},\ }\bibfield  {title} {\enquote {\bibinfo {title} {Microscopic theory of superconductivity},}\ }\href {\doibase 10.1103/PhysRev.106.162} {\bibfield  {journal} {\bibinfo  {journal} {Phys. Rev.}\ }\textbf {\bibinfo {volume} {106}},\ \bibinfo {pages} {162--164} (\bibinfo {year} {1957})}\BibitemShut {NoStop}%
\bibitem [{\citenamefont {Bergeret}\ \emph {et~al.}(2005)\citenamefont {Bergeret}, \citenamefont {Volkov},\ and\ \citenamefont {Efetov}}]{Bergeret2005}%
  \BibitemOpen
  \bibfield  {author} {\bibinfo {author} {\bibfnamefont {F.~S.}\ \bibnamefont {Bergeret}}, \bibinfo {author} {\bibfnamefont {A.~F.}\ \bibnamefont {Volkov}}, \ and\ \bibinfo {author} {\bibfnamefont {K.~B.}\ \bibnamefont {Efetov}},\ }\bibfield  {title} {\enquote {\bibinfo {title} {Odd triplet superconductivity and related phenomena in superconductor-ferromagnet structures},}\ }\href {\doibase 10.1103/RevModPhys.77.1321} {\bibfield  {journal} {\bibinfo  {journal} {Rev. Mod. Phys.}\ }\textbf {\bibinfo {volume} {77}},\ \bibinfo {pages} {1321--1373} (\bibinfo {year} {2005})}\BibitemShut {NoStop}%
\bibitem [{\citenamefont {Buzdin}(2005)}]{Buzdin2005}%
  \BibitemOpen
  \bibfield  {author} {\bibinfo {author} {\bibfnamefont {A.~I.}\ \bibnamefont {Buzdin}},\ }\bibfield  {title} {\enquote {\bibinfo {title} {Proximity effects in superconductor-ferromagnet heterostructures},}\ }\href {\doibase 10.1103/RevModPhys.77.935} {\bibfield  {journal} {\bibinfo  {journal} {Rev. Mod. Phys.}\ }\textbf {\bibinfo {volume} {77}},\ \bibinfo {pages} {935--976} (\bibinfo {year} {2005})}\BibitemShut {NoStop}%
\bibitem [{\citenamefont {Bergeret}\ \emph {et~al.}(2018)\citenamefont {Bergeret}, \citenamefont {Silaev}, \citenamefont {Virtanen},\ and\ \citenamefont {Heikkil\"a}}]{Bergeret2018}%
  \BibitemOpen
  \bibfield  {author} {\bibinfo {author} {\bibfnamefont {F.~Sebastian}\ \bibnamefont {Bergeret}}, \bibinfo {author} {\bibfnamefont {Mikhail}\ \bibnamefont {Silaev}}, \bibinfo {author} {\bibfnamefont {Pauli}\ \bibnamefont {Virtanen}}, \ and\ \bibinfo {author} {\bibfnamefont {Tero~T.}\ \bibnamefont {Heikkil\"a}},\ }\bibfield  {title} {\enquote {\bibinfo {title} {Colloquium: Nonequilibrium effects in superconductors with a spin-splitting field},}\ }\href {\doibase 10.1103/RevModPhys.90.041001} {\bibfield  {journal} {\bibinfo  {journal} {Rev. Mod. Phys.}\ }\textbf {\bibinfo {volume} {90}},\ \bibinfo {pages} {041001} (\bibinfo {year} {2018})}\BibitemShut {NoStop}%
\bibitem [{\citenamefont {Eschrig}\ \emph {et~al.}(2015)\citenamefont {Eschrig}, \citenamefont {Cottet}, \citenamefont {Belzig},\ and\ \citenamefont {Linder}}]{Eschrig2015}%
  \BibitemOpen
  \bibfield  {author} {\bibinfo {author} {\bibfnamefont {M.}~\bibnamefont {Eschrig}}, \bibinfo {author} {\bibfnamefont {A.}~\bibnamefont {Cottet}}, \bibinfo {author} {\bibfnamefont {W.}~\bibnamefont {Belzig}}, \ and\ \bibinfo {author} {\bibfnamefont {J.}~\bibnamefont {Linder}},\ }\bibfield  {title} {\enquote {\bibinfo {title} {General boundary conditions for quasiclassical theory of superconductivity in the diffusive limit: application to strongly spin-polarized systems},}\ }\href {\doibase 10.1088/1367-2630/17/8/083037} {\bibfield  {journal} {\bibinfo  {journal} {New Journal of Physics}\ }\textbf {\bibinfo {volume} {17}},\ \bibinfo {pages} {083037} (\bibinfo {year} {2015})}\BibitemShut {NoStop}%
\bibitem [{\citenamefont {Chandrasekhar}(1962)}]{Chandrasekhar1962}%
  \BibitemOpen
  \bibfield  {author} {\bibinfo {author} {\bibfnamefont {B.~S.}\ \bibnamefont {Chandrasekhar}},\ }\bibfield  {title} {\enquote {\bibinfo {title} {A note on the maximum critical field of high‐field superconductors},}\ }\href {\doibase 10.1063/1.1777362} {\bibfield  {journal} {\bibinfo  {journal} {Applied Physics Letters}\ }\textbf {\bibinfo {volume} {1}},\ \bibinfo {pages} {7--8} (\bibinfo {year} {1962})}\BibitemShut {NoStop}%
\bibitem [{\citenamefont {Clogston}(1962)}]{Clogston1962}%
  \BibitemOpen
  \bibfield  {author} {\bibinfo {author} {\bibfnamefont {A.~M.}\ \bibnamefont {Clogston}},\ }\bibfield  {title} {\enquote {\bibinfo {title} {Upper limit for the critical field in hard superconductors},}\ }\href {\doibase 10.1103/PhysRevLett.9.266} {\bibfield  {journal} {\bibinfo  {journal} {Phys. Rev. Lett.}\ }\textbf {\bibinfo {volume} {9}},\ \bibinfo {pages} {266--267} (\bibinfo {year} {1962})}\BibitemShut {NoStop}%
\bibitem [{\citenamefont {Sarma}(1963)}]{Sarma1963}%
  \BibitemOpen
  \bibfield  {author} {\bibinfo {author} {\bibfnamefont {G.}~\bibnamefont {Sarma}},\ }\bibfield  {title} {\enquote {\bibinfo {title} {On the influence of a uniform exchange field acting on the spins of the conduction electrons in a superconductor},}\ }\href {https://www.sciencedirect.com/science/article/pii/0022369763900076} {\bibfield  {journal} {\bibinfo  {journal} {Journal of Physics and Chemistry of Solids}\ }\textbf {\bibinfo {volume} {24}},\ \bibinfo {pages} {1029--1032} (\bibinfo {year} {1963})}\BibitemShut {NoStop}%
\bibitem [{\citenamefont {Linder}\ and\ \citenamefont {Robinson}(2015)}]{Linder_review}%
  \BibitemOpen
  \bibfield  {author} {\bibinfo {author} {\bibfnamefont {Jacob}\ \bibnamefont {Linder}}\ and\ \bibinfo {author} {\bibfnamefont {Jason W.~A.}\ \bibnamefont {Robinson}},\ }\bibfield  {title} {\enquote {\bibinfo {title} {Superconducting spintronics},}\ }\href {\doibase 10.1038/nphys3242} {\bibfield  {journal} {\bibinfo  {journal} {Nature Physics}\ }\textbf {\bibinfo {volume} {11}},\ \bibinfo {pages} {307--315} (\bibinfo {year} {2015})}\BibitemShut {NoStop}%
\bibitem [{\citenamefont {Eschrig}(2015)}]{Eschrig_review}%
  \BibitemOpen
  \bibfield  {author} {\bibinfo {author} {\bibfnamefont {Matthias}\ \bibnamefont {Eschrig}},\ }\bibfield  {title} {\enquote {\bibinfo {title} {Spin-polarized supercurrents for spintronics: a review of current progress},}\ }\href {\doibase 10.1088/0034-4885/78/10/104501} {\bibfield  {journal} {\bibinfo  {journal} {Reports on Progress in Physics}\ }\textbf {\bibinfo {volume} {78}},\ \bibinfo {pages} {104501} (\bibinfo {year} {2015})}\BibitemShut {NoStop}%
\bibitem [{\citenamefont {Hauser}\ \emph {et~al.}(1966)\citenamefont {Hauser}, \citenamefont {Theuerer},\ and\ \citenamefont {Werthamer}}]{Werthammer1966}%
  \BibitemOpen
  \bibfield  {author} {\bibinfo {author} {\bibfnamefont {J.~J.}\ \bibnamefont {Hauser}}, \bibinfo {author} {\bibfnamefont {H.~C.}\ \bibnamefont {Theuerer}}, \ and\ \bibinfo {author} {\bibfnamefont {N.~R.}\ \bibnamefont {Werthamer}},\ }\bibfield  {title} {\enquote {\bibinfo {title} {Proximity effects between superconducting and magnetic films},}\ }\href {\doibase 10.1103/PhysRev.142.118} {\bibfield  {journal} {\bibinfo  {journal} {Phys. Rev.}\ }\textbf {\bibinfo {volume} {142}},\ \bibinfo {pages} {118--126} (\bibinfo {year} {1966})}\BibitemShut {NoStop}%
\bibitem [{\citenamefont {Kamra}\ \emph {et~al.}(2018)\citenamefont {Kamra}, \citenamefont {Rezaei},\ and\ \citenamefont {Belzig}}]{Kamra2018}%
  \BibitemOpen
  \bibfield  {author} {\bibinfo {author} {\bibfnamefont {Akashdeep}\ \bibnamefont {Kamra}}, \bibinfo {author} {\bibfnamefont {Ali}\ \bibnamefont {Rezaei}}, \ and\ \bibinfo {author} {\bibfnamefont {Wolfgang}\ \bibnamefont {Belzig}},\ }\bibfield  {title} {\enquote {\bibinfo {title} {Spin splitting induced in a superconductor by an antiferromagnetic insulator},}\ }\href {\doibase 10.1103/PhysRevLett.121.247702} {\bibfield  {journal} {\bibinfo  {journal} {Phys. Rev. Lett.}\ }\textbf {\bibinfo {volume} {121}},\ \bibinfo {pages} {247702} (\bibinfo {year} {2018})}\BibitemShut {NoStop}%
\bibitem [{\citenamefont {Bobkov}\ \emph {et~al.}(2021)\citenamefont {Bobkov}, \citenamefont {Bobkova}, \citenamefont {Bobkov},\ and\ \citenamefont {Kamra}}]{Bobkov2021}%
  \BibitemOpen
  \bibfield  {author} {\bibinfo {author} {\bibfnamefont {G.~A.}\ \bibnamefont {Bobkov}}, \bibinfo {author} {\bibfnamefont {I.~V.}\ \bibnamefont {Bobkova}}, \bibinfo {author} {\bibfnamefont {A.~M.}\ \bibnamefont {Bobkov}}, \ and\ \bibinfo {author} {\bibfnamefont {Akashdeep}\ \bibnamefont {Kamra}},\ }\bibfield  {title} {\enquote {\bibinfo {title} {Thermally induced spin torque and domain-wall motion in superconductor/antiferromagnetic-insulator bilayers},}\ }\href {\doibase 10.1103/PhysRevB.103.094506} {\bibfield  {journal} {\bibinfo  {journal} {Phys. Rev. B}\ }\textbf {\bibinfo {volume} {103}},\ \bibinfo {pages} {094506} (\bibinfo {year} {2021})}\BibitemShut {NoStop}%
\bibitem [{\citenamefont {Rabinovich}\ \emph {et~al.}(2019)\citenamefont {Rabinovich}, \citenamefont {Bobkova},\ and\ \citenamefont {Bobkov}}]{Rabinovich2019}%
  \BibitemOpen
  \bibfield  {author} {\bibinfo {author} {\bibfnamefont {D.~S.}\ \bibnamefont {Rabinovich}}, \bibinfo {author} {\bibfnamefont {I.~V.}\ \bibnamefont {Bobkova}}, \ and\ \bibinfo {author} {\bibfnamefont {A.~M.}\ \bibnamefont {Bobkov}},\ }\bibfield  {title} {\enquote {\bibinfo {title} {Anomalous phase shift in a josephson junction via an antiferromagnetic interlayer},}\ }\href {\doibase 10.1103/PhysRevResearch.1.033095} {\bibfield  {journal} {\bibinfo  {journal} {Phys. Rev. Research}\ }\textbf {\bibinfo {volume} {1}},\ \bibinfo {pages} {033095} (\bibinfo {year} {2019})}\BibitemShut {NoStop}%
\bibitem [{\citenamefont {Falch}\ and\ \citenamefont {Linder}(2022)}]{Falch2022}%
  \BibitemOpen
  \bibfield  {author} {\bibinfo {author} {\bibfnamefont {Vemund}\ \bibnamefont {Falch}}\ and\ \bibinfo {author} {\bibfnamefont {Jacob}\ \bibnamefont {Linder}},\ }\bibfield  {title} {\enquote {\bibinfo {title} {Giant magnetoanisotropy in the josephson effect and switching of staggered order in antiferromagnets},}\ }\href {\doibase 10.1103/PhysRevB.106.214511} {\bibfield  {journal} {\bibinfo  {journal} {Phys. Rev. B}\ }\textbf {\bibinfo {volume} {106}},\ \bibinfo {pages} {214511} (\bibinfo {year} {2022})}\BibitemShut {NoStop}%
\bibitem [{\citenamefont {Buzdin}\ and\ \citenamefont {Bulaevski{\u{\i}}}(1986)}]{Buzdin1986}%
  \BibitemOpen
  \bibfield  {author} {\bibinfo {author} {\bibfnamefont {Aleksandr~I}\ \bibnamefont {Buzdin}}\ and\ \bibinfo {author} {\bibfnamefont {L~N}\ \bibnamefont {Bulaevski{\u{\i}}}},\ }\bibfield  {title} {\enquote {\bibinfo {title} {Antiferromagnetic superconductors},}\ }\href {\doibase 10.1070/pu1986v029n05abeh003375} {\bibfield  {journal} {\bibinfo  {journal} {Soviet Physics Uspekhi}\ }\textbf {\bibinfo {volume} {29}},\ \bibinfo {pages} {412--425} (\bibinfo {year} {1986})}\BibitemShut {NoStop}%
\bibitem [{\citenamefont {Fyhn}\ \emph {et~al.}(2023{\natexlab{a}})\citenamefont {Fyhn}, \citenamefont {Brataas}, \citenamefont {Qaiumzadeh},\ and\ \citenamefont {Linder}}]{Fyhn2022}%
  \BibitemOpen
  \bibfield  {author} {\bibinfo {author} {\bibfnamefont {Eirik~Holm}\ \bibnamefont {Fyhn}}, \bibinfo {author} {\bibfnamefont {Arne}\ \bibnamefont {Brataas}}, \bibinfo {author} {\bibfnamefont {Alireza}\ \bibnamefont {Qaiumzadeh}}, \ and\ \bibinfo {author} {\bibfnamefont {Jacob}\ \bibnamefont {Linder}},\ }\bibfield  {title} {\enquote {\bibinfo {title} {Quasiclassical theory for antiferromagnetic metals},}\ }\href {\doibase 10.1103/PhysRevB.107.174503} {\bibfield  {journal} {\bibinfo  {journal} {Phys. Rev. B}\ }\textbf {\bibinfo {volume} {107}},\ \bibinfo {pages} {174503} (\bibinfo {year} {2023}{\natexlab{a}})}\BibitemShut {NoStop}%
\bibitem [{\citenamefont {Fyhn}\ \emph {et~al.}(2023{\natexlab{b}})\citenamefont {Fyhn}, \citenamefont {Brataas}, \citenamefont {Qaiumzadeh},\ and\ \citenamefont {Linder}}]{Fyhn2022_1}%
  \BibitemOpen
  \bibfield  {author} {\bibinfo {author} {\bibfnamefont {Eirik~Holm}\ \bibnamefont {Fyhn}}, \bibinfo {author} {\bibfnamefont {Arne}\ \bibnamefont {Brataas}}, \bibinfo {author} {\bibfnamefont {Alireza}\ \bibnamefont {Qaiumzadeh}}, \ and\ \bibinfo {author} {\bibfnamefont {Jacob}\ \bibnamefont {Linder}},\ }\bibfield  {title} {\enquote {\bibinfo {title} {Superconducting proximity effect and long-ranged triplets in dirty metallic antiferromagnets},}\ }\href {\doibase 10.1103/PhysRevLett.131.076001} {\bibfield  {journal} {\bibinfo  {journal} {Phys. Rev. Lett.}\ }\textbf {\bibinfo {volume} {131}},\ \bibinfo {pages} {076001} (\bibinfo {year} {2023}{\natexlab{b}})}\BibitemShut {NoStop}%
\bibitem [{\citenamefont {Bell}\ \emph {et~al.}(2003)\citenamefont {Bell}, \citenamefont {Tarte}, \citenamefont {Burnell}, \citenamefont {Leung}, \citenamefont {Kang},\ and\ \citenamefont {Blamire}}]{Bell2003}%
  \BibitemOpen
  \bibfield  {author} {\bibinfo {author} {\bibfnamefont {C.}~\bibnamefont {Bell}}, \bibinfo {author} {\bibfnamefont {E.~J.}\ \bibnamefont {Tarte}}, \bibinfo {author} {\bibfnamefont {G.}~\bibnamefont {Burnell}}, \bibinfo {author} {\bibfnamefont {C.~W.}\ \bibnamefont {Leung}}, \bibinfo {author} {\bibfnamefont {D.-J.}\ \bibnamefont {Kang}}, \ and\ \bibinfo {author} {\bibfnamefont {M.~G.}\ \bibnamefont {Blamire}},\ }\bibfield  {title} {\enquote {\bibinfo {title} {Proximity and josephson effects in superconductor/antiferromagnetic $\mathrm{Nb}/\ensuremath{\gamma}\ensuremath{-}{\mathrm{fe}}_{50}{\mathrm{mn}}_{50}$ heterostructures},}\ }\href {\doibase 10.1103/PhysRevB.68.144517} {\bibfield  {journal} {\bibinfo  {journal} {Phys. Rev. B}\ }\textbf {\bibinfo {volume} {68}},\ \bibinfo {pages} {144517} (\bibinfo {year} {2003})}\BibitemShut {NoStop}%
\bibitem [{\citenamefont {H{\"u}bener}\ \emph {et~al.}(2002)\citenamefont {H{\"u}bener}, \citenamefont {Tikhonov}, \citenamefont {Garifullin}, \citenamefont {Westerholt},\ and\ \citenamefont {Zabel}}]{Hubener2002}%
  \BibitemOpen
  \bibfield  {author} {\bibinfo {author} {\bibfnamefont {M}~\bibnamefont {H{\"u}bener}}, \bibinfo {author} {\bibfnamefont {D}~\bibnamefont {Tikhonov}}, \bibinfo {author} {\bibfnamefont {I~A}\ \bibnamefont {Garifullin}}, \bibinfo {author} {\bibfnamefont {K}~\bibnamefont {Westerholt}}, \ and\ \bibinfo {author} {\bibfnamefont {H}~\bibnamefont {Zabel}},\ }\bibfield  {title} {\enquote {\bibinfo {title} {{The antiferromagnet/superconductor proximity effect in Cr/V/Cr trilayers}},}\ }\href@noop {} {\bibfield  {journal} {\bibinfo  {journal} {Journal Of Physics-Condensed Matter}\ }\textbf {\bibinfo {volume} {14}},\ \bibinfo {pages} {8687--8696} (\bibinfo {year} {2002})}\BibitemShut {NoStop}%
\bibitem [{\citenamefont {Wu}\ \emph {et~al.}(2013)\citenamefont {Wu}, \citenamefont {Yang}, \citenamefont {Guo}, \citenamefont {Wu},\ and\ \citenamefont {Qiu}}]{Wu2013}%
  \BibitemOpen
  \bibfield  {author} {\bibinfo {author} {\bibfnamefont {B.~L.}\ \bibnamefont {Wu}}, \bibinfo {author} {\bibfnamefont {Y.~M.}\ \bibnamefont {Yang}}, \bibinfo {author} {\bibfnamefont {Z.~B.}\ \bibnamefont {Guo}}, \bibinfo {author} {\bibfnamefont {Y.~H.}\ \bibnamefont {Wu}}, \ and\ \bibinfo {author} {\bibfnamefont {J.~J.}\ \bibnamefont {Qiu}},\ }\bibfield  {title} {\enquote {\bibinfo {title} {Suppression of superconductivity in nb by irmn in irmn/nb bilayers},}\ }\href {\doibase 10.1063/1.4824891} {\bibfield  {journal} {\bibinfo  {journal} {Applied Physics Letters}\ }\textbf {\bibinfo {volume} {103}},\ \bibinfo {pages} {152602} (\bibinfo {year} {2013})}\BibitemShut {NoStop}%
\bibitem [{\citenamefont {Seeger}\ \emph {et~al.}(2021)\citenamefont {Seeger}, \citenamefont {Forestier}, \citenamefont {Gladii}, \citenamefont {Leivisk\"a}, \citenamefont {Auffret}, \citenamefont {Joumard}, \citenamefont {Gomez}, \citenamefont {Rubio-Roy}, \citenamefont {Buzdin}, \citenamefont {Houzet},\ and\ \citenamefont {Baltz}}]{Seeger2021}%
  \BibitemOpen
  \bibfield  {author} {\bibinfo {author} {\bibfnamefont {R.~L.}\ \bibnamefont {Seeger}}, \bibinfo {author} {\bibfnamefont {G.}~\bibnamefont {Forestier}}, \bibinfo {author} {\bibfnamefont {O.}~\bibnamefont {Gladii}}, \bibinfo {author} {\bibfnamefont {M.}~\bibnamefont {Leivisk\"a}}, \bibinfo {author} {\bibfnamefont {S.}~\bibnamefont {Auffret}}, \bibinfo {author} {\bibfnamefont {I.}~\bibnamefont {Joumard}}, \bibinfo {author} {\bibfnamefont {C.}~\bibnamefont {Gomez}}, \bibinfo {author} {\bibfnamefont {M.}~\bibnamefont {Rubio-Roy}}, \bibinfo {author} {\bibfnamefont {A.~I.}\ \bibnamefont {Buzdin}}, \bibinfo {author} {\bibfnamefont {M.}~\bibnamefont {Houzet}}, \ and\ \bibinfo {author} {\bibfnamefont {V.}~\bibnamefont {Baltz}},\ }\bibfield  {title} {\enquote {\bibinfo {title} {Penetration depth of cooper pairs in the irmn antiferromagnet},}\ }\href {\doibase 10.1103/PhysRevB.104.054413} {\bibfield  {journal} {\bibinfo  {journal} {Phys. Rev. B}\ }\textbf {\bibinfo {volume} {104}},\ \bibinfo {pages} {054413} (\bibinfo
  {year} {2021})}\BibitemShut {NoStop}%
\bibitem [{\citenamefont {Krivoruchko}(1996)}]{Krivoruchko1996}%
  \BibitemOpen
  \bibfield  {author} {\bibinfo {author} {\bibfnamefont {V.~N.}\ \bibnamefont {Krivoruchko}},\ }\bibfield  {title} {\enquote {\bibinfo {title} {Upper critical fields of the superconducting state of a superconductor-antiferromagnetic metal superlattice},}\ }\href@noop {} {\bibfield  {journal} {\bibinfo  {journal} {Zh. Eksp.Teor. Fiz.}\ }\textbf {\bibinfo {volume} {109}},\ \bibinfo {pages} {649} (\bibinfo {year} {1996})}\BibitemShut {NoStop}%
\bibitem [{\citenamefont {Belashchenko}(2010)}]{Belashchenko2010}%
  \BibitemOpen
  \bibfield  {author} {\bibinfo {author} {\bibfnamefont {K.~D.}\ \bibnamefont {Belashchenko}},\ }\bibfield  {title} {\enquote {\bibinfo {title} {Equilibrium magnetization at the boundary of a magnetoelectric antiferromagnet},}\ }\href {\doibase 10.1103/PhysRevLett.105.147204} {\bibfield  {journal} {\bibinfo  {journal} {Phys. Rev. Lett.}\ }\textbf {\bibinfo {volume} {105}},\ \bibinfo {pages} {147204} (\bibinfo {year} {2010})}\BibitemShut {NoStop}%
\bibitem [{\citenamefont {He}\ \emph {et~al.}(2010)\citenamefont {He}, \citenamefont {Wang}, \citenamefont {Wu}, \citenamefont {Caruso}, \citenamefont {Vescovo}, \citenamefont {Belashchenko}, \citenamefont {Dowben},\ and\ \citenamefont {Binek}}]{He2010}%
  \BibitemOpen
  \bibfield  {author} {\bibinfo {author} {\bibfnamefont {Xi}~\bibnamefont {He}}, \bibinfo {author} {\bibfnamefont {Yi}~\bibnamefont {Wang}}, \bibinfo {author} {\bibfnamefont {Ning}\ \bibnamefont {Wu}}, \bibinfo {author} {\bibfnamefont {Anthony~N.}\ \bibnamefont {Caruso}}, \bibinfo {author} {\bibfnamefont {Elio}\ \bibnamefont {Vescovo}}, \bibinfo {author} {\bibfnamefont {Kirill~D.}\ \bibnamefont {Belashchenko}}, \bibinfo {author} {\bibfnamefont {Peter~A.}\ \bibnamefont {Dowben}}, \ and\ \bibinfo {author} {\bibfnamefont {Christian}\ \bibnamefont {Binek}},\ }\bibfield  {title} {\enquote {\bibinfo {title} {Robust isothermal electric control of exchange bias at room temperature},}\ }\href {\doibase 10.1038/nmat2785} {\bibfield  {journal} {\bibinfo  {journal} {Nature Materials}\ }\textbf {\bibinfo {volume} {9}},\ \bibinfo {pages} {579} (\bibinfo {year} {2010})}\BibitemShut {NoStop}%
\bibitem [{\citenamefont {Manna}\ and\ \citenamefont {Yusuf}(2014)}]{Manna2014}%
  \BibitemOpen
  \bibfield  {author} {\bibinfo {author} {\bibfnamefont {P.K.}\ \bibnamefont {Manna}}\ and\ \bibinfo {author} {\bibfnamefont {S.M.}\ \bibnamefont {Yusuf}},\ }\bibfield  {title} {\enquote {\bibinfo {title} {Two interface effects: Exchange bias and magnetic proximity},}\ }\href {\doibase https://doi.org/10.1016/j.physrep.2013.10.002} {\bibfield  {journal} {\bibinfo  {journal} {Physics Reports}\ }\textbf {\bibinfo {volume} {535}},\ \bibinfo {pages} {61 -- 99} (\bibinfo {year} {2014})}\BibitemShut {NoStop}%
\bibitem [{\citenamefont {Bobkova}\ \emph {et~al.}(2005)\citenamefont {Bobkova}, \citenamefont {Hirschfeld},\ and\ \citenamefont {Barash}}]{Bobkova2005}%
  \BibitemOpen
  \bibfield  {author} {\bibinfo {author} {\bibfnamefont {I.~V.}\ \bibnamefont {Bobkova}}, \bibinfo {author} {\bibfnamefont {P.~J.}\ \bibnamefont {Hirschfeld}}, \ and\ \bibinfo {author} {\bibfnamefont {Yu.~S.}\ \bibnamefont {Barash}},\ }\bibfield  {title} {\enquote {\bibinfo {title} {Spin-dependent quasiparticle reflection and bound states at interfaces with itinerant antiferromagnets},}\ }\href {\doibase 10.1103/PhysRevLett.94.037005} {\bibfield  {journal} {\bibinfo  {journal} {Phys. Rev. Lett.}\ }\textbf {\bibinfo {volume} {94}},\ \bibinfo {pages} {037005} (\bibinfo {year} {2005})}\BibitemShut {NoStop}%
\bibitem [{\citenamefont {Andersen}\ \emph {et~al.}(2005)\citenamefont {Andersen}, \citenamefont {Bobkova}, \citenamefont {Hirschfeld},\ and\ \citenamefont {Barash}}]{Andersen2005}%
  \BibitemOpen
  \bibfield  {author} {\bibinfo {author} {\bibfnamefont {Brian~M.}\ \bibnamefont {Andersen}}, \bibinfo {author} {\bibfnamefont {I.~V.}\ \bibnamefont {Bobkova}}, \bibinfo {author} {\bibfnamefont {P.~J.}\ \bibnamefont {Hirschfeld}}, \ and\ \bibinfo {author} {\bibfnamefont {Yu.~S.}\ \bibnamefont {Barash}},\ }\bibfield  {title} {\enquote {\bibinfo {title} {Bound states at the interface between antiferromagnets and superconductors},}\ }\href {\doibase 10.1103/PhysRevB.72.184510} {\bibfield  {journal} {\bibinfo  {journal} {Phys. Rev. B}\ }\textbf {\bibinfo {volume} {72}},\ \bibinfo {pages} {184510} (\bibinfo {year} {2005})}\BibitemShut {NoStop}%
\bibitem [{\citenamefont {Andersen}\ \emph {et~al.}(2006)\citenamefont {Andersen}, \citenamefont {Bobkova}, \citenamefont {Hirschfeld},\ and\ \citenamefont {Barash}}]{Andersen2006}%
  \BibitemOpen
  \bibfield  {author} {\bibinfo {author} {\bibfnamefont {Brian~M.}\ \bibnamefont {Andersen}}, \bibinfo {author} {\bibfnamefont {I.~V.}\ \bibnamefont {Bobkova}}, \bibinfo {author} {\bibfnamefont {P.~J.}\ \bibnamefont {Hirschfeld}}, \ and\ \bibinfo {author} {\bibfnamefont {Yu.~S.}\ \bibnamefont {Barash}},\ }\bibfield  {title} {\enquote {\bibinfo {title} {$0\ensuremath{-}\ensuremath{\pi}$ transitions in josephson junctions with antiferromagnetic interlayers},}\ }\href {\doibase 10.1103/PhysRevLett.96.117005} {\bibfield  {journal} {\bibinfo  {journal} {Phys. Rev. Lett.}\ }\textbf {\bibinfo {volume} {96}},\ \bibinfo {pages} {117005} (\bibinfo {year} {2006})}\BibitemShut {NoStop}%
\bibitem [{\citenamefont {Enoksen}\ \emph {et~al.}(2013)\citenamefont {Enoksen}, \citenamefont {Linder},\ and\ \citenamefont {Sudb\o{}}}]{Enoksen2013}%
  \BibitemOpen
  \bibfield  {author} {\bibinfo {author} {\bibfnamefont {Henrik}\ \bibnamefont {Enoksen}}, \bibinfo {author} {\bibfnamefont {Jacob}\ \bibnamefont {Linder}}, \ and\ \bibinfo {author} {\bibfnamefont {Asle}\ \bibnamefont {Sudb\o{}}},\ }\bibfield  {title} {\enquote {\bibinfo {title} {Pressure-induced 0-$\ensuremath{\pi}$ transitions and supercurrent crossover in antiferromagnetic weak links},}\ }\href {\doibase 10.1103/PhysRevB.88.214512} {\bibfield  {journal} {\bibinfo  {journal} {Phys. Rev. B}\ }\textbf {\bibinfo {volume} {88}},\ \bibinfo {pages} {214512} (\bibinfo {year} {2013})}\BibitemShut {NoStop}%
\bibitem [{\citenamefont {Bulaevskii}\ \emph {et~al.}(2017)\citenamefont {Bulaevskii}, \citenamefont {Eneias},\ and\ \citenamefont {Ferraz}}]{Bulaevskii2017}%
  \BibitemOpen
  \bibfield  {author} {\bibinfo {author} {\bibfnamefont {Lev}\ \bibnamefont {Bulaevskii}}, \bibinfo {author} {\bibfnamefont {Ronivon}\ \bibnamefont {Eneias}}, \ and\ \bibinfo {author} {\bibfnamefont {Alvaro}\ \bibnamefont {Ferraz}},\ }\bibfield  {title} {\enquote {\bibinfo {title} {Superconductor-antiferromagnet-superconductor $\ensuremath{\pi}$ josephson junction based on an antiferromagnetic barrier},}\ }\href {\doibase 10.1103/PhysRevB.95.104513} {\bibfield  {journal} {\bibinfo  {journal} {Phys. Rev. B}\ }\textbf {\bibinfo {volume} {95}},\ \bibinfo {pages} {104513} (\bibinfo {year} {2017})}\BibitemShut {NoStop}%
\bibitem [{\citenamefont {Johnsen}\ \emph {et~al.}(2021)\citenamefont {Johnsen}, \citenamefont {Jacobsen},\ and\ \citenamefont {Linder}}]{Johnsen2021}%
  \BibitemOpen
  \bibfield  {author} {\bibinfo {author} {\bibfnamefont {Lina~G.}\ \bibnamefont {Johnsen}}, \bibinfo {author} {\bibfnamefont {Sol~H.}\ \bibnamefont {Jacobsen}}, \ and\ \bibinfo {author} {\bibfnamefont {Jacob}\ \bibnamefont {Linder}},\ }\bibfield  {title} {\enquote {\bibinfo {title} {Magnetic control of superconducting heterostructures using compensated antiferromagnets},}\ }\href {\doibase 10.1103/PhysRevB.103.L060505} {\bibfield  {journal} {\bibinfo  {journal} {Phys. Rev. B}\ }\textbf {\bibinfo {volume} {103}},\ \bibinfo {pages} {L060505} (\bibinfo {year} {2021})}\BibitemShut {NoStop}%
\bibitem [{\citenamefont {Bobkov}\ \emph {et~al.}(2022)\citenamefont {Bobkov}, \citenamefont {Bobkova}, \citenamefont {Bobkov},\ and\ \citenamefont {Kamra}}]{Bobkov2022}%
  \BibitemOpen
  \bibfield  {author} {\bibinfo {author} {\bibfnamefont {G.~A.}\ \bibnamefont {Bobkov}}, \bibinfo {author} {\bibfnamefont {I.~V.}\ \bibnamefont {Bobkova}}, \bibinfo {author} {\bibfnamefont {A.~M.}\ \bibnamefont {Bobkov}}, \ and\ \bibinfo {author} {\bibfnamefont {Akashdeep}\ \bibnamefont {Kamra}},\ }\bibfield  {title} {\enquote {\bibinfo {title} {N\'eel proximity effect at antiferromagnet/superconductor interfaces},}\ }\href {\doibase 10.1103/PhysRevB.106.144512} {\bibfield  {journal} {\bibinfo  {journal} {Phys. Rev. B}\ }\textbf {\bibinfo {volume} {106}},\ \bibinfo {pages} {144512} (\bibinfo {year} {2022})}\BibitemShut {NoStop}%
\bibitem [{\citenamefont {Cheng}\ \emph {et~al.}(2014)\citenamefont {Cheng}, \citenamefont {Xiao}, \citenamefont {Niu},\ and\ \citenamefont {Brataas}}]{Cheng2014}%
  \BibitemOpen
  \bibfield  {author} {\bibinfo {author} {\bibfnamefont {Ran}\ \bibnamefont {Cheng}}, \bibinfo {author} {\bibfnamefont {Jiang}\ \bibnamefont {Xiao}}, \bibinfo {author} {\bibfnamefont {Qian}\ \bibnamefont {Niu}}, \ and\ \bibinfo {author} {\bibfnamefont {Arne}\ \bibnamefont {Brataas}},\ }\bibfield  {title} {\enquote {\bibinfo {title} {Spin pumping and spin-transfer torques in antiferromagnets},}\ }\href {\doibase 10.1103/PhysRevLett.113.057601} {\bibfield  {journal} {\bibinfo  {journal} {Phys. Rev. Lett.}\ }\textbf {\bibinfo {volume} {113}},\ \bibinfo {pages} {057601} (\bibinfo {year} {2014})}\BibitemShut {NoStop}%
\bibitem [{\citenamefont {Takei}\ \emph {et~al.}(2014)\citenamefont {Takei}, \citenamefont {Halperin}, \citenamefont {Yacoby},\ and\ \citenamefont {Tserkovnyak}}]{Takei2014}%
  \BibitemOpen
  \bibfield  {author} {\bibinfo {author} {\bibfnamefont {So}~\bibnamefont {Takei}}, \bibinfo {author} {\bibfnamefont {Bertrand~I.}\ \bibnamefont {Halperin}}, \bibinfo {author} {\bibfnamefont {Amir}\ \bibnamefont {Yacoby}}, \ and\ \bibinfo {author} {\bibfnamefont {Yaroslav}\ \bibnamefont {Tserkovnyak}},\ }\bibfield  {title} {\enquote {\bibinfo {title} {Superfluid spin transport through antiferromagnetic insulators},}\ }\href {\doibase 10.1103/PhysRevB.90.094408} {\bibfield  {journal} {\bibinfo  {journal} {Phys. Rev. B}\ }\textbf {\bibinfo {volume} {90}},\ \bibinfo {pages} {094408} (\bibinfo {year} {2014})}\BibitemShut {NoStop}%
\bibitem [{\citenamefont {Baltz}\ \emph {et~al.}(2018)\citenamefont {Baltz}, \citenamefont {Manchon}, \citenamefont {Tsoi}, \citenamefont {Moriyama}, \citenamefont {Ono},\ and\ \citenamefont {Tserkovnyak}}]{Baltz2016}%
  \BibitemOpen
  \bibfield  {author} {\bibinfo {author} {\bibfnamefont {V.}~\bibnamefont {Baltz}}, \bibinfo {author} {\bibfnamefont {A.}~\bibnamefont {Manchon}}, \bibinfo {author} {\bibfnamefont {M.}~\bibnamefont {Tsoi}}, \bibinfo {author} {\bibfnamefont {T.}~\bibnamefont {Moriyama}}, \bibinfo {author} {\bibfnamefont {T.}~\bibnamefont {Ono}}, \ and\ \bibinfo {author} {\bibfnamefont {Y.}~\bibnamefont {Tserkovnyak}},\ }\bibfield  {title} {\enquote {\bibinfo {title} {Antiferromagnetic spintronics},}\ }\href {\doibase 10.1103/RevModPhys.90.015005} {\bibfield  {journal} {\bibinfo  {journal} {Rev. Mod. Phys.}\ }\textbf {\bibinfo {volume} {90}},\ \bibinfo {pages} {015005} (\bibinfo {year} {2018})}\BibitemShut {NoStop}%
\bibitem [{\citenamefont {Bobkov}\ \emph {et~al.}(2023{\natexlab{a}})\citenamefont {Bobkov}, \citenamefont {Bobkova},\ and\ \citenamefont {Bobkov}}]{Bobkov2023_impurities}%
  \BibitemOpen
  \bibfield  {author} {\bibinfo {author} {\bibfnamefont {G.~A.}\ \bibnamefont {Bobkov}}, \bibinfo {author} {\bibfnamefont {I.~V.}\ \bibnamefont {Bobkova}}, \ and\ \bibinfo {author} {\bibfnamefont {A.~M.}\ \bibnamefont {Bobkov}},\ }\bibfield  {title} {\enquote {\bibinfo {title} {Proximity effect in superconductor/antiferromagnet hybrids: N\'eel triplets and impurity suppression of superconductivity},}\ }\href {\doibase 10.1103/PhysRevB.108.054510} {\bibfield  {journal} {\bibinfo  {journal} {Phys. Rev. B}\ }\textbf {\bibinfo {volume} {108}},\ \bibinfo {pages} {054510} (\bibinfo {year} {2023}{\natexlab{a}})}\BibitemShut {NoStop}%
\bibitem [{\citenamefont {Chourasia}\ \emph {et~al.}(2023)\citenamefont {Chourasia}, \citenamefont {Kamra}, \citenamefont {Bobkova},\ and\ \citenamefont {Kamra}}]{Chourasia2023}%
  \BibitemOpen
  \bibfield  {author} {\bibinfo {author} {\bibfnamefont {Simran}\ \bibnamefont {Chourasia}}, \bibinfo {author} {\bibfnamefont {Lina~Johnsen}\ \bibnamefont {Kamra}}, \bibinfo {author} {\bibfnamefont {Irina~V.}\ \bibnamefont {Bobkova}}, \ and\ \bibinfo {author} {\bibfnamefont {Akashdeep}\ \bibnamefont {Kamra}},\ }\bibfield  {title} {\enquote {\bibinfo {title} {Generation of spin-triplet cooper pairs via a canted antiferromagnet},}\ }\href {\doibase 10.1103/PhysRevB.108.064515} {\bibfield  {journal} {\bibinfo  {journal} {Phys. Rev. B}\ }\textbf {\bibinfo {volume} {108}},\ \bibinfo {pages} {064515} (\bibinfo {year} {2023})}\BibitemShut {NoStop}%
\bibitem [{\citenamefont {Abrikosov}\ and\ \citenamefont {Gor’kov}(1961)}]{Abrikosov1961}%
  \BibitemOpen
  \bibfield  {author} {\bibinfo {author} {\bibfnamefont {A.A.}\ \bibnamefont {Abrikosov}}\ and\ \bibinfo {author} {\bibfnamefont {L.P.}\ \bibnamefont {Gor’kov}},\ }\bibfield  {title} {\enquote {\bibinfo {title} {Contribution to the theory of superconducting alloys with paramagnetic impurities},}\ }\href@noop {} {\bibfield  {journal} {\bibinfo  {journal} {Sov. Phys. JETP}\ }\textbf {\bibinfo {volume} {12}},\ \bibinfo {pages} {1243} (\bibinfo {year} {1961})}\BibitemShut {NoStop}%
\bibitem [{\citenamefont {Goldman}\ and\ \citenamefont {Markovi{\'{c}}}(1998)}]{Goldman1998}%
  \BibitemOpen
  \bibfield  {author} {\bibinfo {author} {\bibfnamefont {Allen~M.}\ \bibnamefont {Goldman}}\ and\ \bibinfo {author} {\bibfnamefont {Nina}\ \bibnamefont {Markovi{\'{c}}}},\ }\bibfield  {title} {\enquote {\bibinfo {title} {Superconductor‐insulator transitions in the two‐dimensional limit},}\ }\href {\doibase 10.1063/1.882069} {\bibfield  {journal} {\bibinfo  {journal} {Physics Today}\ }\textbf {\bibinfo {volume} {51}},\ \bibinfo {pages} {39--44} (\bibinfo {year} {1998})}\BibitemShut {NoStop}%
\bibitem [{\citenamefont {Sadovskii}(1997)}]{Sadovskii1997}%
  \BibitemOpen
  \bibfield  {author} {\bibinfo {author} {\bibfnamefont {Michael~V.}\ \bibnamefont {Sadovskii}},\ }\bibfield  {title} {\enquote {\bibinfo {title} {Superconductivity and localization},}\ }\href {https://www.sciencedirect.com/science/article/pii/S0370157396000361} {\bibfield  {journal} {\bibinfo  {journal} {Physics Reports}\ }\textbf {\bibinfo {volume} {282}},\ \bibinfo {pages} {225--348} (\bibinfo {year} {1997})}\BibitemShut {NoStop}%
\bibitem [{\citenamefont {Gantmakher}\ and\ \citenamefont {Dolgopolov}(2010)}]{Gantmakher2010}%
  \BibitemOpen
  \bibfield  {author} {\bibinfo {author} {\bibfnamefont {Vsevolod~F.}\ \bibnamefont {Gantmakher}}\ and\ \bibinfo {author} {\bibfnamefont {Valery~T.}\ \bibnamefont {Dolgopolov}},\ }\bibfield  {title} {\enquote {\bibinfo {title} {Superconductor--insulator quantum phase transition},}\ }\href {\doibase 10.3367/UFNe.0180.201001a.0003} {\bibfield  {journal} {\bibinfo  {journal} {Physics-Uspekhi}\ }\textbf {\bibinfo {volume} {53}},\ \bibinfo {pages} {1} (\bibinfo {year} {2010})}\BibitemShut {NoStop}%
\bibitem [{\citenamefont {Sac\'ep\'e}\ \emph {et~al.}(2008)\citenamefont {Sac\'ep\'e}, \citenamefont {Chapelier}, \citenamefont {Baturina}, \citenamefont {Vinokur}, \citenamefont {Baklanov},\ and\ \citenamefont {Sanquer}}]{Sacepe2008}%
  \BibitemOpen
  \bibfield  {author} {\bibinfo {author} {\bibfnamefont {B.}~\bibnamefont {Sac\'ep\'e}}, \bibinfo {author} {\bibfnamefont {C.}~\bibnamefont {Chapelier}}, \bibinfo {author} {\bibfnamefont {T.~I.}\ \bibnamefont {Baturina}}, \bibinfo {author} {\bibfnamefont {V.~M.}\ \bibnamefont {Vinokur}}, \bibinfo {author} {\bibfnamefont {M.~R.}\ \bibnamefont {Baklanov}}, \ and\ \bibinfo {author} {\bibfnamefont {M.}~\bibnamefont {Sanquer}},\ }\bibfield  {title} {\enquote {\bibinfo {title} {Disorder-induced inhomogeneities of the superconducting state close to the superconductor-insulator transition},}\ }\href {\doibase 10.1103/PhysRevLett.101.157006} {\bibfield  {journal} {\bibinfo  {journal} {Phys. Rev. Lett.}\ }\textbf {\bibinfo {volume} {101}},\ \bibinfo {pages} {157006} (\bibinfo {year} {2008})}\BibitemShut {NoStop}%
\bibitem [{\citenamefont {Sac{\'e}p{\'e}}\ \emph {et~al.}(2010)\citenamefont {Sac{\'e}p{\'e}}, \citenamefont {Chapelier}, \citenamefont {Baturina}, \citenamefont {Vinokur}, \citenamefont {Baklanov},\ and\ \citenamefont {Sanquer}}]{Sacepe2010}%
  \BibitemOpen
  \bibfield  {author} {\bibinfo {author} {\bibfnamefont {Benjamin}\ \bibnamefont {Sac{\'e}p{\'e}}}, \bibinfo {author} {\bibfnamefont {Claude}\ \bibnamefont {Chapelier}}, \bibinfo {author} {\bibfnamefont {Tatyana~I.}\ \bibnamefont {Baturina}}, \bibinfo {author} {\bibfnamefont {Valerii~M.}\ \bibnamefont {Vinokur}}, \bibinfo {author} {\bibfnamefont {Mikhail~R.}\ \bibnamefont {Baklanov}}, \ and\ \bibinfo {author} {\bibfnamefont {Marc}\ \bibnamefont {Sanquer}},\ }\bibfield  {title} {\enquote {\bibinfo {title} {Pseudogap in a thin film of a conventional superconductor},}\ }\href {\doibase 10.1038/ncomms1140} {\bibfield  {journal} {\bibinfo  {journal} {Nature Communications}\ }\textbf {\bibinfo {volume} {1}},\ \bibinfo {pages} {140} (\bibinfo {year} {2010})}\BibitemShut {NoStop}%
\bibitem [{\citenamefont {Arrigoni}\ and\ \citenamefont {Kivelson}(2003)}]{Arrigoni2003}%
  \BibitemOpen
  \bibfield  {author} {\bibinfo {author} {\bibfnamefont {E.}~\bibnamefont {Arrigoni}}\ and\ \bibinfo {author} {\bibfnamefont {S.~A.}\ \bibnamefont {Kivelson}},\ }\bibfield  {title} {\enquote {\bibinfo {title} {Optimal inhomogeneity for superconductivity},}\ }\href {\doibase 10.1103/PhysRevB.68.180503} {\bibfield  {journal} {\bibinfo  {journal} {Phys. Rev. B}\ }\textbf {\bibinfo {volume} {68}},\ \bibinfo {pages} {180503} (\bibinfo {year} {2003})}\BibitemShut {NoStop}%
\bibitem [{\citenamefont {Gastiasoro}\ and\ \citenamefont {Andersen}(2018)}]{Gastiasoro2018}%
  \BibitemOpen
  \bibfield  {author} {\bibinfo {author} {\bibfnamefont {Maria~N.}\ \bibnamefont {Gastiasoro}}\ and\ \bibinfo {author} {\bibfnamefont {Brian~M.}\ \bibnamefont {Andersen}},\ }\bibfield  {title} {\enquote {\bibinfo {title} {Enhancing superconductivity by disorder},}\ }\href {\doibase 10.1103/PhysRevB.98.184510} {\bibfield  {journal} {\bibinfo  {journal} {Phys. Rev. B}\ }\textbf {\bibinfo {volume} {98}},\ \bibinfo {pages} {184510} (\bibinfo {year} {2018})}\BibitemShut {NoStop}%
\bibitem [{\citenamefont {Martin}\ \emph {et~al.}(2005)\citenamefont {Martin}, \citenamefont {Podolsky},\ and\ \citenamefont {Kivelson}}]{Martin2005}%
  \BibitemOpen
  \bibfield  {author} {\bibinfo {author} {\bibfnamefont {Ivar}\ \bibnamefont {Martin}}, \bibinfo {author} {\bibfnamefont {Daniel}\ \bibnamefont {Podolsky}}, \ and\ \bibinfo {author} {\bibfnamefont {Steven~A.}\ \bibnamefont {Kivelson}},\ }\bibfield  {title} {\enquote {\bibinfo {title} {Enhancement of superconductivity by local inhomogeneities},}\ }\href {\doibase 10.1103/PhysRevB.72.060502} {\bibfield  {journal} {\bibinfo  {journal} {Phys. Rev. B}\ }\textbf {\bibinfo {volume} {72}},\ \bibinfo {pages} {060502} (\bibinfo {year} {2005})}\BibitemShut {NoStop}%
\bibitem [{\citenamefont {Zhao}\ \emph {et~al.}(2019)\citenamefont {Zhao}, \citenamefont {Lin}, \citenamefont {Xiao}, \citenamefont {Huang}, \citenamefont {Yao}, \citenamefont {Yan}, \citenamefont {Xing}, \citenamefont {Zhang}, \citenamefont {Li}, \citenamefont {Hoshino}, \citenamefont {Wang}, \citenamefont {Zhou}, \citenamefont {Gu}, \citenamefont {Bahramy}, \citenamefont {Yao}, \citenamefont {Nagaosa}, \citenamefont {Xue}, \citenamefont {Law}, \citenamefont {Chen},\ and\ \citenamefont {Ji}}]{Zhao2019}%
  \BibitemOpen
  \bibfield  {author} {\bibinfo {author} {\bibfnamefont {Kun}\ \bibnamefont {Zhao}}, \bibinfo {author} {\bibfnamefont {Haicheng}\ \bibnamefont {Lin}}, \bibinfo {author} {\bibfnamefont {Xiao}\ \bibnamefont {Xiao}}, \bibinfo {author} {\bibfnamefont {Wantong}\ \bibnamefont {Huang}}, \bibinfo {author} {\bibfnamefont {Wei}\ \bibnamefont {Yao}}, \bibinfo {author} {\bibfnamefont {Mingzhe}\ \bibnamefont {Yan}}, \bibinfo {author} {\bibfnamefont {Ying}\ \bibnamefont {Xing}}, \bibinfo {author} {\bibfnamefont {Qinghua}\ \bibnamefont {Zhang}}, \bibinfo {author} {\bibfnamefont {Zi-Xiang}\ \bibnamefont {Li}}, \bibinfo {author} {\bibfnamefont {Shintaro}\ \bibnamefont {Hoshino}}, \bibinfo {author} {\bibfnamefont {Jian}\ \bibnamefont {Wang}}, \bibinfo {author} {\bibfnamefont {Shuyun}\ \bibnamefont {Zhou}}, \bibinfo {author} {\bibfnamefont {Lin}\ \bibnamefont {Gu}}, \bibinfo {author} {\bibfnamefont {Mohammad~Saeed}\ \bibnamefont {Bahramy}}, \bibinfo {author} {\bibfnamefont {Hong}\ \bibnamefont {Yao}}, \bibinfo {author}
  {\bibfnamefont {Naoto}\ \bibnamefont {Nagaosa}}, \bibinfo {author} {\bibfnamefont {Qi-Kun}\ \bibnamefont {Xue}}, \bibinfo {author} {\bibfnamefont {Kam~Tuen}\ \bibnamefont {Law}}, \bibinfo {author} {\bibfnamefont {Xi}~\bibnamefont {Chen}}, \ and\ \bibinfo {author} {\bibfnamefont {Shuai-Hua}\ \bibnamefont {Ji}},\ }\bibfield  {title} {\enquote {\bibinfo {title} {Disorder-induced multifractal superconductivity in monolayer niobium dichalcogenides},}\ }\href {\doibase 10.1038/s41567-019-0570-0} {\bibfield  {journal} {\bibinfo  {journal} {Nature Physics}\ }\textbf {\bibinfo {volume} {15}},\ \bibinfo {pages} {904--910} (\bibinfo {year} {2019})}\BibitemShut {NoStop}%
\bibitem [{\citenamefont {Petrovi{\'{c}}}\ \emph {et~al.}(2016)\citenamefont {Petrovi{\'{c}}}, \citenamefont {Ansermet}, \citenamefont {Chernyshov}, \citenamefont {Hoesch}, \citenamefont {Salloum}, \citenamefont {Gougeon}, \citenamefont {Potel}, \citenamefont {Boeri},\ and\ \citenamefont {Panagopoulos}}]{Petrovic2016}%
  \BibitemOpen
  \bibfield  {author} {\bibinfo {author} {\bibfnamefont {A.~P.}\ \bibnamefont {Petrovi{\'{c}}}}, \bibinfo {author} {\bibfnamefont {D.}~\bibnamefont {Ansermet}}, \bibinfo {author} {\bibfnamefont {D.}~\bibnamefont {Chernyshov}}, \bibinfo {author} {\bibfnamefont {M.}~\bibnamefont {Hoesch}}, \bibinfo {author} {\bibfnamefont {D.}~\bibnamefont {Salloum}}, \bibinfo {author} {\bibfnamefont {P.}~\bibnamefont {Gougeon}}, \bibinfo {author} {\bibfnamefont {M.}~\bibnamefont {Potel}}, \bibinfo {author} {\bibfnamefont {L.}~\bibnamefont {Boeri}}, \ and\ \bibinfo {author} {\bibfnamefont {C.}~\bibnamefont {Panagopoulos}},\ }\bibfield  {title} {\enquote {\bibinfo {title} {A disorder-enhanced quasi-one-dimensional superconductor},}\ }\href {\doibase 10.1038/ncomms12262} {\bibfield  {journal} {\bibinfo  {journal} {Nature Communications}\ }\textbf {\bibinfo {volume} {7}},\ \bibinfo {pages} {12262} (\bibinfo {year} {2016})}\BibitemShut {NoStop}%
\bibitem [{\citenamefont {Peng}\ \emph {et~al.}(2018)\citenamefont {Peng}, \citenamefont {Yu}, \citenamefont {Wu}, \citenamefont {Zhou}, \citenamefont {Guo}, \citenamefont {Li}, \citenamefont {Zhao}, \citenamefont {Wu},\ and\ \citenamefont {Xie}}]{Peng2018}%
  \BibitemOpen
  \bibfield  {author} {\bibinfo {author} {\bibfnamefont {Jing}\ \bibnamefont {Peng}}, \bibinfo {author} {\bibfnamefont {Zhi}\ \bibnamefont {Yu}}, \bibinfo {author} {\bibfnamefont {Jiajing}\ \bibnamefont {Wu}}, \bibinfo {author} {\bibfnamefont {Yuan}\ \bibnamefont {Zhou}}, \bibinfo {author} {\bibfnamefont {Yuqiao}\ \bibnamefont {Guo}}, \bibinfo {author} {\bibfnamefont {Zejun}\ \bibnamefont {Li}}, \bibinfo {author} {\bibfnamefont {Jiyin}\ \bibnamefont {Zhao}}, \bibinfo {author} {\bibfnamefont {Changzheng}\ \bibnamefont {Wu}}, \ and\ \bibinfo {author} {\bibfnamefont {Yi}~\bibnamefont {Xie}},\ }\bibfield  {title} {\enquote {\bibinfo {title} {Disorder enhanced superconductivity toward tas2 monolayer},}\ }\href {\doibase 10.1021/acsnano.8b04718} {\bibfield  {journal} {\bibinfo  {journal} {ACS Nano}\ }\textbf {\bibinfo {volume} {12}},\ \bibinfo {pages} {9461--9466} (\bibinfo {year} {2018})}\BibitemShut {NoStop}%
\bibitem [{\citenamefont {Neverov}\ \emph {et~al.}(2022)\citenamefont {Neverov}, \citenamefont {Lukyanov}, \citenamefont {Krasavin}, \citenamefont {Vagov},\ and\ \citenamefont {Croitoru}}]{Neverov2022}%
  \BibitemOpen
  \bibfield  {author} {\bibinfo {author} {\bibfnamefont {Vyacheslav~D.}\ \bibnamefont {Neverov}}, \bibinfo {author} {\bibfnamefont {Alexander~E.}\ \bibnamefont {Lukyanov}}, \bibinfo {author} {\bibfnamefont {Andrey~V.}\ \bibnamefont {Krasavin}}, \bibinfo {author} {\bibfnamefont {Alexei}\ \bibnamefont {Vagov}}, \ and\ \bibinfo {author} {\bibfnamefont {Mihail~D.}\ \bibnamefont {Croitoru}},\ }\bibfield  {title} {\enquote {\bibinfo {title} {Correlated disorder as a way towards robust superconductivity},}\ }\href {\doibase 10.1038/s42005-022-00933-z} {\bibfield  {journal} {\bibinfo  {journal} {Communications Physics}\ }\textbf {\bibinfo {volume} {5}},\ \bibinfo {pages} {177} (\bibinfo {year} {2022})}\BibitemShut {NoStop}%
\bibitem [{\citenamefont {Larkin}\ and\ \citenamefont {Ovchinnikov}(1964)}]{Larkin1964}%
  \BibitemOpen
  \bibfield  {author} {\bibinfo {author} {\bibfnamefont {A.~I.}\ \bibnamefont {Larkin}}\ and\ \bibinfo {author} {\bibfnamefont {Y.~N.}\ \bibnamefont {Ovchinnikov}},\ }\bibfield  {title} {\enquote {\bibinfo {title} {{Nonuniform state of superconductors}},}\ }\href@noop {} {\bibfield  {journal} {\bibinfo  {journal} {Zh. Eksp. Teor. Fiz.}\ }\textbf {\bibinfo {volume} {47}},\ \bibinfo {pages} {1136--1146} (\bibinfo {year} {1964})}\BibitemShut {NoStop}%
\bibitem [{\citenamefont {Fulde}\ and\ \citenamefont {Ferrell}(1964)}]{Fulde1964}%
  \BibitemOpen
  \bibfield  {author} {\bibinfo {author} {\bibfnamefont {Peter}\ \bibnamefont {Fulde}}\ and\ \bibinfo {author} {\bibfnamefont {Richard~A.}\ \bibnamefont {Ferrell}},\ }\bibfield  {title} {\enquote {\bibinfo {title} {Superconductivity in a strong spin-exchange field},}\ }\href {\doibase 10.1103/PhysRev.135.A550} {\bibfield  {journal} {\bibinfo  {journal} {Phys. Rev.}\ }\textbf {\bibinfo {volume} {135}},\ \bibinfo {pages} {A550--A563} (\bibinfo {year} {1964})}\BibitemShut {NoStop}%
\bibitem [{\citenamefont {Buzdin}\ \emph {et~al.}(Feb 1982)\citenamefont {Buzdin}, \citenamefont {Bulaevskii},\ and\ \citenamefont {Panyukov}}]{Buzdin1982}%
  \BibitemOpen
  \bibfield  {author} {\bibinfo {author} {\bibfnamefont {A.~I.}\ \bibnamefont {Buzdin}}, \bibinfo {author} {\bibfnamefont {L.~N.}\ \bibnamefont {Bulaevskii}}, \ and\ \bibinfo {author} {\bibfnamefont {S.~V.}\ \bibnamefont {Panyukov}},\ }\bibfield  {title} {\enquote {\bibinfo {title} {Critical-current oscillations as a function of the exchange field and thickness of the ferromagnetic metal (f) in an s-f-s josephson junction},}\ }\href {http://inis.iaea.org/search/search.aspx?orig_q=RN:14748338} {\bibfield  {journal} {\bibinfo  {journal} {JETP Letters}\ }\textbf {\bibinfo {volume} {35}},\ \bibinfo {pages} {178--180} (\bibinfo {year} {Feb 1982})}\BibitemShut {NoStop}%
\bibitem [{\citenamefont {Demler}\ \emph {et~al.}(1997)\citenamefont {Demler}, \citenamefont {Arnold},\ and\ \citenamefont {Beasley}}]{Demler1997}%
  \BibitemOpen
  \bibfield  {author} {\bibinfo {author} {\bibfnamefont {E.~A.}\ \bibnamefont {Demler}}, \bibinfo {author} {\bibfnamefont {G.~B.}\ \bibnamefont {Arnold}}, \ and\ \bibinfo {author} {\bibfnamefont {M.~R.}\ \bibnamefont {Beasley}},\ }\bibfield  {title} {\enquote {\bibinfo {title} {Superconducting proximity effects in magnetic metals},}\ }\href {\doibase 10.1103/PhysRevB.55.15174} {\bibfield  {journal} {\bibinfo  {journal} {Phys. Rev. B}\ }\textbf {\bibinfo {volume} {55}},\ \bibinfo {pages} {15174--15182} (\bibinfo {year} {1997})}\BibitemShut {NoStop}%
\bibitem [{\citenamefont {Kontos}\ \emph {et~al.}(2002)\citenamefont {Kontos}, \citenamefont {Aprili}, \citenamefont {Lesueur}, \citenamefont {Gen\^et}, \citenamefont {Stephanidis},\ and\ \citenamefont {Boursier}}]{Kontos2002}%
  \BibitemOpen
  \bibfield  {author} {\bibinfo {author} {\bibfnamefont {T.}~\bibnamefont {Kontos}}, \bibinfo {author} {\bibfnamefont {M.}~\bibnamefont {Aprili}}, \bibinfo {author} {\bibfnamefont {J.}~\bibnamefont {Lesueur}}, \bibinfo {author} {\bibfnamefont {F.}~\bibnamefont {Gen\^et}}, \bibinfo {author} {\bibfnamefont {B.}~\bibnamefont {Stephanidis}}, \ and\ \bibinfo {author} {\bibfnamefont {R.}~\bibnamefont {Boursier}},\ }\bibfield  {title} {\enquote {\bibinfo {title} {Josephson junction through a thin ferromagnetic layer: Negative coupling},}\ }\href {\doibase 10.1103/PhysRevLett.89.137007} {\bibfield  {journal} {\bibinfo  {journal} {Phys. Rev. Lett.}\ }\textbf {\bibinfo {volume} {89}},\ \bibinfo {pages} {137007} (\bibinfo {year} {2002})}\BibitemShut {NoStop}%
\bibitem [{\citenamefont {Ryazanov}\ \emph {et~al.}(2001)\citenamefont {Ryazanov}, \citenamefont {Oboznov}, \citenamefont {Rusanov}, \citenamefont {Veretennikov}, \citenamefont {Golubov},\ and\ \citenamefont {Aarts}}]{Ryazanov2001}%
  \BibitemOpen
  \bibfield  {author} {\bibinfo {author} {\bibfnamefont {V.~V.}\ \bibnamefont {Ryazanov}}, \bibinfo {author} {\bibfnamefont {V.~A.}\ \bibnamefont {Oboznov}}, \bibinfo {author} {\bibfnamefont {A.~Yu.}\ \bibnamefont {Rusanov}}, \bibinfo {author} {\bibfnamefont {A.~V.}\ \bibnamefont {Veretennikov}}, \bibinfo {author} {\bibfnamefont {A.~A.}\ \bibnamefont {Golubov}}, \ and\ \bibinfo {author} {\bibfnamefont {J.}~\bibnamefont {Aarts}},\ }\bibfield  {title} {\enquote {\bibinfo {title} {Coupling of two superconductors through a ferromagnet: Evidence for a $\ensuremath{\pi}$ junction},}\ }\href {\doibase 10.1103/PhysRevLett.86.2427} {\bibfield  {journal} {\bibinfo  {journal} {Phys. Rev. Lett.}\ }\textbf {\bibinfo {volume} {86}},\ \bibinfo {pages} {2427--2430} (\bibinfo {year} {2001})}\BibitemShut {NoStop}%
\bibitem [{\citenamefont {Oboznov}\ \emph {et~al.}(2006)\citenamefont {Oboznov}, \citenamefont {Bol'ginov}, \citenamefont {Feofanov}, \citenamefont {Ryazanov},\ and\ \citenamefont {Buzdin}}]{Oboznov2006}%
  \BibitemOpen
  \bibfield  {author} {\bibinfo {author} {\bibfnamefont {V.~A.}\ \bibnamefont {Oboznov}}, \bibinfo {author} {\bibfnamefont {V.~V.}\ \bibnamefont {Bol'ginov}}, \bibinfo {author} {\bibfnamefont {A.~K.}\ \bibnamefont {Feofanov}}, \bibinfo {author} {\bibfnamefont {V.~V.}\ \bibnamefont {Ryazanov}}, \ and\ \bibinfo {author} {\bibfnamefont {A.~I.}\ \bibnamefont {Buzdin}},\ }\bibfield  {title} {\enquote {\bibinfo {title} {Thickness dependence of the josephson ground states of superconductor-ferromagnet-superconductor junctions},}\ }\href {\doibase 10.1103/PhysRevLett.96.197003} {\bibfield  {journal} {\bibinfo  {journal} {Phys. Rev. Lett.}\ }\textbf {\bibinfo {volume} {96}},\ \bibinfo {pages} {197003} (\bibinfo {year} {2006})}\BibitemShut {NoStop}%
\bibitem [{\citenamefont {Bannykh}\ \emph {et~al.}(2009)\citenamefont {Bannykh}, \citenamefont {Pfeiffer}, \citenamefont {Stolyarov}, \citenamefont {Batov}, \citenamefont {Ryazanov},\ and\ \citenamefont {Weides}}]{Bannykh2009}%
  \BibitemOpen
  \bibfield  {author} {\bibinfo {author} {\bibfnamefont {A.~A.}\ \bibnamefont {Bannykh}}, \bibinfo {author} {\bibfnamefont {J.}~\bibnamefont {Pfeiffer}}, \bibinfo {author} {\bibfnamefont {V.~S.}\ \bibnamefont {Stolyarov}}, \bibinfo {author} {\bibfnamefont {I.~E.}\ \bibnamefont {Batov}}, \bibinfo {author} {\bibfnamefont {V.~V.}\ \bibnamefont {Ryazanov}}, \ and\ \bibinfo {author} {\bibfnamefont {M.}~\bibnamefont {Weides}},\ }\bibfield  {title} {\enquote {\bibinfo {title} {Josephson tunnel junctions with a strong ferromagnetic interlayer},}\ }\href {https://journals.aps.org/prb/abstract/10.1103/PhysRevB.79.054501} {\bibfield  {journal} {\bibinfo  {journal} {Physical Review B}\ }\textbf {\bibinfo {volume} {79}},\ \bibinfo {pages} {054501} (\bibinfo {year} {2009})}\BibitemShut {NoStop}%
\bibitem [{\citenamefont {Robinson}\ \emph {et~al.}(2006)\citenamefont {Robinson}, \citenamefont {Piano}, \citenamefont {Burnell}, \citenamefont {Bell},\ and\ \citenamefont {Blamire}}]{Robinson2006}%
  \BibitemOpen
  \bibfield  {author} {\bibinfo {author} {\bibfnamefont {J.~W.~A.}\ \bibnamefont {Robinson}}, \bibinfo {author} {\bibfnamefont {S.}~\bibnamefont {Piano}}, \bibinfo {author} {\bibfnamefont {G.}~\bibnamefont {Burnell}}, \bibinfo {author} {\bibfnamefont {C.}~\bibnamefont {Bell}}, \ and\ \bibinfo {author} {\bibfnamefont {M.~G.}\ \bibnamefont {Blamire}},\ }\bibfield  {title} {\enquote {\bibinfo {title} {Critical current oscillations in strong ferromagnetic $\pi$ junctions},}\ }\href {https://journals.aps.org/prl/abstract/10.1103/PhysRevLett.97.177003} {\bibfield  {journal} {\bibinfo  {journal} {Physical Review Letters}\ }\textbf {\bibinfo {volume} {97}},\ \bibinfo {pages} {177003} (\bibinfo {year} {2006})}\BibitemShut {NoStop}%
\bibitem [{\citenamefont {Yamashita}\ \emph {et~al.}(2005)\citenamefont {Yamashita}, \citenamefont {Tanikawa}, \citenamefont {Takahashi},\ and\ \citenamefont {Maekawa}}]{Yamashita2005}%
  \BibitemOpen
  \bibfield  {author} {\bibinfo {author} {\bibfnamefont {T.}~\bibnamefont {Yamashita}}, \bibinfo {author} {\bibfnamefont {K.}~\bibnamefont {Tanikawa}}, \bibinfo {author} {\bibfnamefont {S.}~\bibnamefont {Takahashi}}, \ and\ \bibinfo {author} {\bibfnamefont {S.}~\bibnamefont {Maekawa}},\ }\bibfield  {title} {\enquote {\bibinfo {title} {Superconducting $\ensuremath{\pi}$ qubit with a ferromagnetic josephson junction},}\ }\href {\doibase 10.1103/PhysRevLett.95.097001} {\bibfield  {journal} {\bibinfo  {journal} {Phys. Rev. Lett.}\ }\textbf {\bibinfo {volume} {95}},\ \bibinfo {pages} {097001} (\bibinfo {year} {2005})}\BibitemShut {NoStop}%
\bibitem [{\citenamefont {Feofanov}\ \emph {et~al.}(2010)\citenamefont {Feofanov}, \citenamefont {Oboznov}, \citenamefont {Bol'ginov}, \citenamefont {Lisenfeld}, \citenamefont {Poletto}, \citenamefont {Ryazanov}, \citenamefont {Rossolenko}, \citenamefont {Khabipov}, \citenamefont {Balashov}, \citenamefont {Zorin}, \citenamefont {Dmitriev}, \citenamefont {Koshelets},\ and\ \citenamefont {Ustinov}}]{Feofanov2010}%
  \BibitemOpen
  \bibfield  {author} {\bibinfo {author} {\bibfnamefont {A.~K.}\ \bibnamefont {Feofanov}}, \bibinfo {author} {\bibfnamefont {V.~A.}\ \bibnamefont {Oboznov}}, \bibinfo {author} {\bibfnamefont {V.~V.}\ \bibnamefont {Bol'ginov}}, \bibinfo {author} {\bibfnamefont {J.}~\bibnamefont {Lisenfeld}}, \bibinfo {author} {\bibfnamefont {S.}~\bibnamefont {Poletto}}, \bibinfo {author} {\bibfnamefont {V.~V.}\ \bibnamefont {Ryazanov}}, \bibinfo {author} {\bibfnamefont {A.~N.}\ \bibnamefont {Rossolenko}}, \bibinfo {author} {\bibfnamefont {M.}~\bibnamefont {Khabipov}}, \bibinfo {author} {\bibfnamefont {D.}~\bibnamefont {Balashov}}, \bibinfo {author} {\bibfnamefont {A.~B.}\ \bibnamefont {Zorin}}, \bibinfo {author} {\bibfnamefont {P.~N.}\ \bibnamefont {Dmitriev}}, \bibinfo {author} {\bibfnamefont {V.~P.}\ \bibnamefont {Koshelets}}, \ and\ \bibinfo {author} {\bibfnamefont {A.~V.}\ \bibnamefont {Ustinov}},\ }\bibfield  {title} {\enquote {\bibinfo {title} {Implementation of superconductor/ferromagnet/ superconductor $\pi$-shifters
  in superconducting digital and quantum circuits},}\ }\href {\doibase 10.1038/nphys1700} {\bibfield  {journal} {\bibinfo  {journal} {Nature Physics}\ }\textbf {\bibinfo {volume} {6}},\ \bibinfo {pages} {593--597} (\bibinfo {year} {2010})}\BibitemShut {NoStop}%
\bibitem [{\citenamefont {Shcherbakova}\ \emph {et~al.}(2015)\citenamefont {Shcherbakova}, \citenamefont {Fedorov}, \citenamefont {Shulga}, \citenamefont {Ryazanov}, \citenamefont {Bolginov}, \citenamefont {Oboznov}, \citenamefont {Egorov}, \citenamefont {Shkolnikov}, \citenamefont {Wolf}, \citenamefont {Beckmann},\ and\ \citenamefont {Ustinov}}]{Shcherbakova2015}%
  \BibitemOpen
  \bibfield  {author} {\bibinfo {author} {\bibfnamefont {A.~V.}\ \bibnamefont {Shcherbakova}}, \bibinfo {author} {\bibfnamefont {K.~G.}\ \bibnamefont {Fedorov}}, \bibinfo {author} {\bibfnamefont {K.~V.}\ \bibnamefont {Shulga}}, \bibinfo {author} {\bibfnamefont {V.~V.}\ \bibnamefont {Ryazanov}}, \bibinfo {author} {\bibfnamefont {V.~V.}\ \bibnamefont {Bolginov}}, \bibinfo {author} {\bibfnamefont {V.~A.}\ \bibnamefont {Oboznov}}, \bibinfo {author} {\bibfnamefont {S.~V.}\ \bibnamefont {Egorov}}, \bibinfo {author} {\bibfnamefont {V.~O.}\ \bibnamefont {Shkolnikov}}, \bibinfo {author} {\bibfnamefont {M.~J.}\ \bibnamefont {Wolf}}, \bibinfo {author} {\bibfnamefont {D.}~\bibnamefont {Beckmann}}, \ and\ \bibinfo {author} {\bibfnamefont {A.~V.}\ \bibnamefont {Ustinov}},\ }\bibfield  {title} {\enquote {\bibinfo {title} {Fabrication and measurements of hybrid nb/al josephson junctions and flux qubits with $\pi$-shifters},}\ }\href {\doibase 10.1088/0953-2048/28/2/025009} {\bibfield  {journal} {\bibinfo  {journal}
  {Superconductor Science and Technology}\ }\textbf {\bibinfo {volume} {28}},\ \bibinfo {pages} {025009} (\bibinfo {year} {2015})}\BibitemShut {NoStop}%
\bibitem [{\citenamefont {Fominov}\ \emph {et~al.}(2002)\citenamefont {Fominov}, \citenamefont {Chtchelkatchev},\ and\ \citenamefont {Golubov}}]{Fominov2002}%
  \BibitemOpen
  \bibfield  {author} {\bibinfo {author} {\bibfnamefont {Ya.~V.}\ \bibnamefont {Fominov}}, \bibinfo {author} {\bibfnamefont {N.~M.}\ \bibnamefont {Chtchelkatchev}}, \ and\ \bibinfo {author} {\bibfnamefont {A.~A.}\ \bibnamefont {Golubov}},\ }\bibfield  {title} {\enquote {\bibinfo {title} {Nonmonotonic critical temperature in superconductor/ferromagnet bilayers},}\ }\href {\doibase 10.1103/PhysRevB.66.014507} {\bibfield  {journal} {\bibinfo  {journal} {Phys. Rev. B}\ }\textbf {\bibinfo {volume} {66}},\ \bibinfo {pages} {014507} (\bibinfo {year} {2002})}\BibitemShut {NoStop}%
\bibitem [{\citenamefont {Radovi\ifmmode~\acute{c}\else \'{c}\fi{}}\ \emph {et~al.}(1991)\citenamefont {Radovi\ifmmode~\acute{c}\else \'{c}\fi{}}, \citenamefont {Ledvij}, \citenamefont {Dobrosavljevi\ifmmode \acute{c}\else \'{c}\fi{}-Gruji\ifmmode~\acute{c}\else \'{c}\fi{}}, \citenamefont {Buzdin},\ and\ \citenamefont {Clem}}]{Radovic1991}%
  \BibitemOpen
  \bibfield  {author} {\bibinfo {author} {\bibfnamefont {Zoran}\ \bibnamefont {Radovi\ifmmode~\acute{c}\else \'{c}\fi{}}}, \bibinfo {author} {\bibfnamefont {Marko}\ \bibnamefont {Ledvij}}, \bibinfo {author} {\bibfnamefont {Ljiljana}\ \bibnamefont {Dobrosavljevi\ifmmode \acute{c}\else \'{c}\fi{}-Gruji\ifmmode~\acute{c}\else \'{c}\fi{}}}, \bibinfo {author} {\bibfnamefont {A.~I.}\ \bibnamefont {Buzdin}}, \ and\ \bibinfo {author} {\bibfnamefont {John~R.}\ \bibnamefont {Clem}},\ }\bibfield  {title} {\enquote {\bibinfo {title} {Transition temperatures of superconductor-ferromagnet superlattices},}\ }\href {\doibase 10.1103/PhysRevB.44.759} {\bibfield  {journal} {\bibinfo  {journal} {Phys. Rev. B}\ }\textbf {\bibinfo {volume} {44}},\ \bibinfo {pages} {759--764} (\bibinfo {year} {1991})}\BibitemShut {NoStop}%
\bibitem [{\citenamefont {Vodopyanov}\ and\ \citenamefont {Tagirov}(2003)}]{Vodopyanov2003}%
  \BibitemOpen
  \bibfield  {author} {\bibinfo {author} {\bibfnamefont {B.~P.}\ \bibnamefont {Vodopyanov}}\ and\ \bibinfo {author} {\bibfnamefont {L.~R.}\ \bibnamefont {Tagirov}},\ }\bibfield  {title} {\enquote {\bibinfo {title} {Oscillations of superconducting transition temperature in strong ferromagnet-superconductor bilayers},}\ }\href {\doibase 10.1134/1.1641483} {\bibfield  {journal} {\bibinfo  {journal} {JETP Letters}\ }\textbf {\bibinfo {volume} {78}},\ \bibinfo {pages} {555--559} (\bibinfo {year} {2003})}\BibitemShut {NoStop}%
\bibitem [{\citenamefont {Lazar}\ \emph {et~al.}(2000)\citenamefont {Lazar}, \citenamefont {Westerholt}, \citenamefont {Zabel}, \citenamefont {Tagirov}, \citenamefont {Goryunov}, \citenamefont {Garif'yanov},\ and\ \citenamefont {Garifullin}}]{Lazar2000}%
  \BibitemOpen
  \bibfield  {author} {\bibinfo {author} {\bibfnamefont {L.}~\bibnamefont {Lazar}}, \bibinfo {author} {\bibfnamefont {K.}~\bibnamefont {Westerholt}}, \bibinfo {author} {\bibfnamefont {H.}~\bibnamefont {Zabel}}, \bibinfo {author} {\bibfnamefont {L.~R.}\ \bibnamefont {Tagirov}}, \bibinfo {author} {\bibfnamefont {Yu.~V.}\ \bibnamefont {Goryunov}}, \bibinfo {author} {\bibfnamefont {N.~N.}\ \bibnamefont {Garif'yanov}}, \ and\ \bibinfo {author} {\bibfnamefont {I.~A.}\ \bibnamefont {Garifullin}},\ }\bibfield  {title} {\enquote {\bibinfo {title} {Superconductor/ferromagnet proximity effect in fe/pb/fe trilayers},}\ }\href {\doibase 10.1103/PhysRevB.61.3711} {\bibfield  {journal} {\bibinfo  {journal} {Phys. Rev. B}\ }\textbf {\bibinfo {volume} {61}},\ \bibinfo {pages} {3711--3722} (\bibinfo {year} {2000})}\BibitemShut {NoStop}%
\bibitem [{\citenamefont {Buzdin}(2000)}]{Buzdin2000}%
  \BibitemOpen
  \bibfield  {author} {\bibinfo {author} {\bibfnamefont {A.}~\bibnamefont {Buzdin}},\ }\bibfield  {title} {\enquote {\bibinfo {title} {Density of states oscillations in a ferromagnetic metal in contact with a superconductor},}\ }\href {\doibase 10.1103/PhysRevB.62.11377} {\bibfield  {journal} {\bibinfo  {journal} {Phys. Rev. B}\ }\textbf {\bibinfo {volume} {62}},\ \bibinfo {pages} {11377--11379} (\bibinfo {year} {2000})}\BibitemShut {NoStop}%
\bibitem [{\citenamefont {Zareyan}\ \emph {et~al.}(2001)\citenamefont {Zareyan}, \citenamefont {Belzig},\ and\ \citenamefont {Nazarov}}]{Zareyan2001}%
  \BibitemOpen
  \bibfield  {author} {\bibinfo {author} {\bibfnamefont {M.}~\bibnamefont {Zareyan}}, \bibinfo {author} {\bibfnamefont {W.}~\bibnamefont {Belzig}}, \ and\ \bibinfo {author} {\bibfnamefont {Yu.~V.}\ \bibnamefont {Nazarov}},\ }\bibfield  {title} {\enquote {\bibinfo {title} {Oscillations of andreev states in clean ferromagnetic films},}\ }\href {\doibase 10.1103/PhysRevLett.86.308} {\bibfield  {journal} {\bibinfo  {journal} {Phys. Rev. Lett.}\ }\textbf {\bibinfo {volume} {86}},\ \bibinfo {pages} {308--311} (\bibinfo {year} {2001})}\BibitemShut {NoStop}%
\bibitem [{\citenamefont {Jiang}\ \emph {et~al.}(1995)\citenamefont {Jiang}, \citenamefont {Davidovi\ifmmode~\acute{c}\else \'{c}\fi{}}, \citenamefont {Reich},\ and\ \citenamefont {Chien}}]{Jiang1995}%
  \BibitemOpen
  \bibfield  {author} {\bibinfo {author} {\bibfnamefont {J.~S.}\ \bibnamefont {Jiang}}, \bibinfo {author} {\bibfnamefont {D.}~\bibnamefont {Davidovi\ifmmode~\acute{c}\else \'{c}\fi{}}}, \bibinfo {author} {\bibfnamefont {Daniel~H.}\ \bibnamefont {Reich}}, \ and\ \bibinfo {author} {\bibfnamefont {C.~L.}\ \bibnamefont {Chien}},\ }\bibfield  {title} {\enquote {\bibinfo {title} {Oscillatory superconducting transition temperature in nb/gd multilayers},}\ }\href {\doibase 10.1103/PhysRevLett.74.314} {\bibfield  {journal} {\bibinfo  {journal} {Phys. Rev. Lett.}\ }\textbf {\bibinfo {volume} {74}},\ \bibinfo {pages} {314--317} (\bibinfo {year} {1995})}\BibitemShut {NoStop}%
\bibitem [{\citenamefont {Mercaldo}\ \emph {et~al.}(1996)\citenamefont {Mercaldo}, \citenamefont {Attanasio}, \citenamefont {Coccorese}, \citenamefont {Maritato}, \citenamefont {Prischepa},\ and\ \citenamefont {Salvato}}]{Mercaldo1996}%
  \BibitemOpen
  \bibfield  {author} {\bibinfo {author} {\bibfnamefont {L.~V.}\ \bibnamefont {Mercaldo}}, \bibinfo {author} {\bibfnamefont {C.}~\bibnamefont {Attanasio}}, \bibinfo {author} {\bibfnamefont {C.}~\bibnamefont {Coccorese}}, \bibinfo {author} {\bibfnamefont {L.}~\bibnamefont {Maritato}}, \bibinfo {author} {\bibfnamefont {S.~L.}\ \bibnamefont {Prischepa}}, \ and\ \bibinfo {author} {\bibfnamefont {M.}~\bibnamefont {Salvato}},\ }\bibfield  {title} {\enquote {\bibinfo {title} {Superconducting-critical-temperature oscillations in nb/cumn multilayers},}\ }\href {\doibase 10.1103/PhysRevB.53.14040} {\bibfield  {journal} {\bibinfo  {journal} {Phys. Rev. B}\ }\textbf {\bibinfo {volume} {53}},\ \bibinfo {pages} {14040--14042} (\bibinfo {year} {1996})}\BibitemShut {NoStop}%
\bibitem [{\citenamefont {M\"uhge}\ \emph {et~al.}(1996)\citenamefont {M\"uhge}, \citenamefont {Garif'yanov}, \citenamefont {Goryunov}, \citenamefont {Khaliullin}, \citenamefont {Tagirov}, \citenamefont {Westerholt}, \citenamefont {Garifullin},\ and\ \citenamefont {Zabel}}]{Muhge1996}%
  \BibitemOpen
  \bibfield  {author} {\bibinfo {author} {\bibfnamefont {Th.}\ \bibnamefont {M\"uhge}}, \bibinfo {author} {\bibfnamefont {N.~N.}\ \bibnamefont {Garif'yanov}}, \bibinfo {author} {\bibfnamefont {Yu.~V.}\ \bibnamefont {Goryunov}}, \bibinfo {author} {\bibfnamefont {G.~G.}\ \bibnamefont {Khaliullin}}, \bibinfo {author} {\bibfnamefont {L.~R.}\ \bibnamefont {Tagirov}}, \bibinfo {author} {\bibfnamefont {K.}~\bibnamefont {Westerholt}}, \bibinfo {author} {\bibfnamefont {I.~A.}\ \bibnamefont {Garifullin}}, \ and\ \bibinfo {author} {\bibfnamefont {H.}~\bibnamefont {Zabel}},\ }\bibfield  {title} {\enquote {\bibinfo {title} {Possible origin for oscillatory superconducting transition temperature in superconductor/ferromagnet multilayers},}\ }\href {\doibase 10.1103/PhysRevLett.77.1857} {\bibfield  {journal} {\bibinfo  {journal} {Phys. Rev. Lett.}\ }\textbf {\bibinfo {volume} {77}},\ \bibinfo {pages} {1857--1860} (\bibinfo {year} {1996})}\BibitemShut {NoStop}%
\bibitem [{\citenamefont {Zdravkov}\ \emph {et~al.}(2006)\citenamefont {Zdravkov}, \citenamefont {Sidorenko}, \citenamefont {Obermeier}, \citenamefont {Gsell}, \citenamefont {Schreck}, \citenamefont {M\"uller}, \citenamefont {Horn}, \citenamefont {Tidecks},\ and\ \citenamefont {Tagirov}}]{Zdravkov2006}%
  \BibitemOpen
  \bibfield  {author} {\bibinfo {author} {\bibfnamefont {V.}~\bibnamefont {Zdravkov}}, \bibinfo {author} {\bibfnamefont {A.}~\bibnamefont {Sidorenko}}, \bibinfo {author} {\bibfnamefont {G.}~\bibnamefont {Obermeier}}, \bibinfo {author} {\bibfnamefont {S.}~\bibnamefont {Gsell}}, \bibinfo {author} {\bibfnamefont {M.}~\bibnamefont {Schreck}}, \bibinfo {author} {\bibfnamefont {C.}~\bibnamefont {M\"uller}}, \bibinfo {author} {\bibfnamefont {S.}~\bibnamefont {Horn}}, \bibinfo {author} {\bibfnamefont {R.}~\bibnamefont {Tidecks}}, \ and\ \bibinfo {author} {\bibfnamefont {L.~R.}\ \bibnamefont {Tagirov}},\ }\bibfield  {title} {\enquote {\bibinfo {title} {Reentrant superconductivity in $\mathrm{Nb}/{\mathrm{cu}}_{1\ensuremath{-}x}{\mathrm{ni}}_{x}$ bilayers},}\ }\href {\doibase 10.1103/PhysRevLett.97.057004} {\bibfield  {journal} {\bibinfo  {journal} {Phys. Rev. Lett.}\ }\textbf {\bibinfo {volume} {97}},\ \bibinfo {pages} {057004} (\bibinfo {year} {2006})}\BibitemShut {NoStop}%
\bibitem [{\citenamefont {Zdravkov}\ \emph {et~al.}(2010)\citenamefont {Zdravkov}, \citenamefont {Kehrle}, \citenamefont {Obermeier}, \citenamefont {Gsell}, \citenamefont {Schreck}, \citenamefont {M\"uller}, \citenamefont {Krug~von Nidda}, \citenamefont {Lindner}, \citenamefont {Moosburger-Will}, \citenamefont {Nold}, \citenamefont {Morari}, \citenamefont {Ryazanov}, \citenamefont {Sidorenko}, \citenamefont {Horn}, \citenamefont {Tidecks},\ and\ \citenamefont {Tagirov}}]{Zdravkov2010}%
  \BibitemOpen
  \bibfield  {author} {\bibinfo {author} {\bibfnamefont {V.~I.}\ \bibnamefont {Zdravkov}}, \bibinfo {author} {\bibfnamefont {J.}~\bibnamefont {Kehrle}}, \bibinfo {author} {\bibfnamefont {G.}~\bibnamefont {Obermeier}}, \bibinfo {author} {\bibfnamefont {S.}~\bibnamefont {Gsell}}, \bibinfo {author} {\bibfnamefont {M.}~\bibnamefont {Schreck}}, \bibinfo {author} {\bibfnamefont {C.}~\bibnamefont {M\"uller}}, \bibinfo {author} {\bibfnamefont {H.-A.}\ \bibnamefont {Krug~von Nidda}}, \bibinfo {author} {\bibfnamefont {J.}~\bibnamefont {Lindner}}, \bibinfo {author} {\bibfnamefont {J.}~\bibnamefont {Moosburger-Will}}, \bibinfo {author} {\bibfnamefont {E.}~\bibnamefont {Nold}}, \bibinfo {author} {\bibfnamefont {R.}~\bibnamefont {Morari}}, \bibinfo {author} {\bibfnamefont {V.~V.}\ \bibnamefont {Ryazanov}}, \bibinfo {author} {\bibfnamefont {A.~S.}\ \bibnamefont {Sidorenko}}, \bibinfo {author} {\bibfnamefont {S.}~\bibnamefont {Horn}}, \bibinfo {author} {\bibfnamefont {R.}~\bibnamefont {Tidecks}}, \ and\ \bibinfo {author}
  {\bibfnamefont {L.~R.}\ \bibnamefont {Tagirov}},\ }\bibfield  {title} {\enquote {\bibinfo {title} {Reentrant superconductivity in superconductor/ferromagnetic-alloy bilayers},}\ }\href {\doibase 10.1103/PhysRevB.82.054517} {\bibfield  {journal} {\bibinfo  {journal} {Phys. Rev. B}\ }\textbf {\bibinfo {volume} {82}},\ \bibinfo {pages} {054517} (\bibinfo {year} {2010})}\BibitemShut {NoStop}%
\bibitem [{\citenamefont {Bobkov}\ \emph {et~al.}(2023{\natexlab{b}})\citenamefont {Bobkov}, \citenamefont {Gordeeva}, \citenamefont {Bobkov},\ and\ \citenamefont {Bobkova}}]{Bobkov2023_oscillations}%
  \BibitemOpen
  \bibfield  {author} {\bibinfo {author} {\bibfnamefont {G.~A.}\ \bibnamefont {Bobkov}}, \bibinfo {author} {\bibfnamefont {V.~M.}\ \bibnamefont {Gordeeva}}, \bibinfo {author} {\bibfnamefont {A.~M.}\ \bibnamefont {Bobkov}}, \ and\ \bibinfo {author} {\bibfnamefont {I.~V.}\ \bibnamefont {Bobkova}},\ }\bibfield  {title} {\enquote {\bibinfo {title} {Oscillatory superconducting transition temperature in superconductor/antiferromagnet heterostructures},}\ }\href {\doibase 10.1103/PhysRevB.108.184509} {\bibfield  {journal} {\bibinfo  {journal} {Phys. Rev. B}\ }\textbf {\bibinfo {volume} {108}},\ \bibinfo {pages} {184509} (\bibinfo {year} {2023}{\natexlab{b}})}\BibitemShut {NoStop}%
\bibitem [{\citenamefont {Bobkov}\ \emph {et~al.}(2024{\natexlab{a}})\citenamefont {Bobkov}, \citenamefont {Bobkova},\ and\ \citenamefont {Bobkov}}]{Bobkov2024_singleimp}%
  \BibitemOpen
  \bibfield  {author} {\bibinfo {author} {\bibfnamefont {G.~A.}\ \bibnamefont {Bobkov}}, \bibinfo {author} {\bibfnamefont {I.~V.}\ \bibnamefont {Bobkova}}, \ and\ \bibinfo {author} {\bibfnamefont {A.~M.}\ \bibnamefont {Bobkov}},\ }\bibfield  {title} {\enquote {\bibinfo {title} {Andreev bound states at nonmagnetic impurities in superconductor/antiferromagnet heterostructures},}\ }\href {\doibase 10.1103/PhysRevB.109.214508} {\bibfield  {journal} {\bibinfo  {journal} {Phys. Rev. B}\ }\textbf {\bibinfo {volume} {109}},\ \bibinfo {pages} {214508} (\bibinfo {year} {2024}{\natexlab{a}})}\BibitemShut {NoStop}%
\bibitem [{\citenamefont {Bobkov}\ \emph {et~al.}(2024{\natexlab{b}})\citenamefont {Bobkov}, \citenamefont {Gordeeva}, \citenamefont {Johnsen~Kamra}, \citenamefont {Chourasia}, \citenamefont {Bobkov}, \citenamefont {Kamra},\ and\ \citenamefont {Bobkova}}]{Bobkov2024_spinvalve}%
  \BibitemOpen
  \bibfield  {author} {\bibinfo {author} {\bibfnamefont {G.~A.}\ \bibnamefont {Bobkov}}, \bibinfo {author} {\bibfnamefont {V.~M.}\ \bibnamefont {Gordeeva}}, \bibinfo {author} {\bibfnamefont {Lina}\ \bibnamefont {Johnsen~Kamra}}, \bibinfo {author} {\bibfnamefont {Simran}\ \bibnamefont {Chourasia}}, \bibinfo {author} {\bibfnamefont {A.~M.}\ \bibnamefont {Bobkov}}, \bibinfo {author} {\bibfnamefont {Akashdeep}\ \bibnamefont {Kamra}}, \ and\ \bibinfo {author} {\bibfnamefont {I.~V.}\ \bibnamefont {Bobkova}},\ }\bibfield  {title} {\enquote {\bibinfo {title} {Superconducting spin valves based on antiferromagnet/superconductor/antiferromagnet heterostructures},}\ }\href {\doibase 10.1103/PhysRevB.109.184504} {\bibfield  {journal} {\bibinfo  {journal} {Phys. Rev. B}\ }\textbf {\bibinfo {volume} {109}},\ \bibinfo {pages} {184504} (\bibinfo {year} {2024}{\natexlab{b}})}\BibitemShut {NoStop}%
\bibitem [{\citenamefont {{De Gennes}}(1966)}]{DEGENNES1966}%
  \BibitemOpen
  \bibfield  {author} {\bibinfo {author} {\bibfnamefont {P.G.}\ \bibnamefont {{De Gennes}}},\ }\bibfield  {title} {\enquote {\bibinfo {title} {Coupling between ferromagnets through a superconducting layer},}\ }\href {\doibase https://doi.org/10.1016/0031-9163(66)90229-0} {\bibfield  {journal} {\bibinfo  {journal} {Physics Letters}\ }\textbf {\bibinfo {volume} {23}},\ \bibinfo {pages} {10--11} (\bibinfo {year} {1966})}\BibitemShut {NoStop}%
\bibitem [{\citenamefont {Oh}\ \emph {et~al.}(1997)\citenamefont {Oh}, \citenamefont {Youm},\ and\ \citenamefont {Beasley}}]{Oh1997}%
  \BibitemOpen
  \bibfield  {author} {\bibinfo {author} {\bibfnamefont {Sangjun}\ \bibnamefont {Oh}}, \bibinfo {author} {\bibfnamefont {D.}~\bibnamefont {Youm}}, \ and\ \bibinfo {author} {\bibfnamefont {M.~R.}\ \bibnamefont {Beasley}},\ }\bibfield  {title} {\enquote {\bibinfo {title} {A superconductive magnetoresistive memory element using controlled exchange interaction},}\ }\href {\doibase 10.1063/1.120032} {\bibfield  {journal} {\bibinfo  {journal} {Appl. Phys. Lett.}\ }\textbf {\bibinfo {volume} {71}},\ \bibinfo {pages} {2376} (\bibinfo {year} {1997})}\BibitemShut {NoStop}%
\bibitem [{\citenamefont {Tagirov}(1999)}]{Tagirov1999}%
  \BibitemOpen
  \bibfield  {author} {\bibinfo {author} {\bibfnamefont {L.~R.}\ \bibnamefont {Tagirov}},\ }\bibfield  {title} {\enquote {\bibinfo {title} {Low-field superconducting spin switch based on a superconductor $/$ferromagnet multilayer},}\ }\href {\doibase 10.1103/PhysRevLett.83.2058} {\bibfield  {journal} {\bibinfo  {journal} {Phys. Rev. Lett.}\ }\textbf {\bibinfo {volume} {83}},\ \bibinfo {pages} {2058--2061} (\bibinfo {year} {1999})}\BibitemShut {NoStop}%
\bibitem [{\citenamefont {Fominov}\ \emph {et~al.}(2003)\citenamefont {Fominov}, \citenamefont {Golubov},\ and\ \citenamefont {Kupriyanov}}]{Fominov2003}%
  \BibitemOpen
  \bibfield  {author} {\bibinfo {author} {\bibfnamefont {Ya.~V.}\ \bibnamefont {Fominov}}, \bibinfo {author} {\bibfnamefont {A.~A.}\ \bibnamefont {Golubov}}, \ and\ \bibinfo {author} {\bibfnamefont {M.~Yu.}\ \bibnamefont {Kupriyanov}},\ }\bibfield  {title} {\enquote {\bibinfo {title} {Triplet proximity effect in fsf trilayers},}\ }\href {\doibase 10.1134/1.1591981} {\bibfield  {journal} {\bibinfo  {journal} {JETP Letters}\ }\textbf {\bibinfo {volume} {77}},\ \bibinfo {pages} {510} (\bibinfo {year} {2003})}\BibitemShut {NoStop}%
\bibitem [{\citenamefont {Fominov}\ \emph {et~al.}(2010)\citenamefont {Fominov}, \citenamefont {Golubov}, \citenamefont {Karminskaya}, \citenamefont {Kupriyanov}, \citenamefont {Deminov},\ and\ \citenamefont {Tagirov}}]{Fominov2010}%
  \BibitemOpen
  \bibfield  {author} {\bibinfo {author} {\bibfnamefont {Ya.~V.}\ \bibnamefont {Fominov}}, \bibinfo {author} {\bibfnamefont {A.~A.}\ \bibnamefont {Golubov}}, \bibinfo {author} {\bibfnamefont {T.~Yu.}\ \bibnamefont {Karminskaya}}, \bibinfo {author} {\bibfnamefont {M.~Yu.}\ \bibnamefont {Kupriyanov}}, \bibinfo {author} {\bibfnamefont {R.~G.}\ \bibnamefont {Deminov}}, \ and\ \bibinfo {author} {\bibfnamefont {L.~R.}\ \bibnamefont {Tagirov}},\ }\bibfield  {title} {\enquote {\bibinfo {title} {Superconducting triplet spin valve},}\ }\href {\doibase 10.1134/S002136401006010X} {\bibfield  {journal} {\bibinfo  {journal} {JETP Letters}\ }\textbf {\bibinfo {volume} {91}},\ \bibinfo {pages} {308} (\bibinfo {year} {2010})}\BibitemShut {NoStop}%
\bibitem [{\citenamefont {Wu}\ and\ \citenamefont {Valls}(2012)}]{Wu2012}%
  \BibitemOpen
  \bibfield  {author} {\bibinfo {author} {\bibfnamefont {Chien-Te}\ \bibnamefont {Wu}}\ and\ \bibinfo {author} {\bibfnamefont {Oriol~T.}\ \bibnamefont {Valls}},\ }\bibfield  {title} {\enquote {\bibinfo {title} {Superconducting proximity effects in ferromagnet/superconductor heterostructures},}\ }\href {\doibase 10.1007/s10948-012-1645-7} {\bibfield  {journal} {\bibinfo  {journal} {J. Supercond. Nov. Magn.}\ }\textbf {\bibinfo {volume} {25}},\ \bibinfo {pages} {2173} (\bibinfo {year} {2012})}\BibitemShut {NoStop}%
\bibitem [{\citenamefont {Li}\ \emph {et~al.}(2013)\citenamefont {Li}, \citenamefont {Roschewsky}, \citenamefont {Assaf}, \citenamefont {Eich}, \citenamefont {Epstein-Martin}, \citenamefont {Heiman}, \citenamefont {M\"unzenberg},\ and\ \citenamefont {Moodera}}]{Li2013}%
  \BibitemOpen
  \bibfield  {author} {\bibinfo {author} {\bibfnamefont {Bin}\ \bibnamefont {Li}}, \bibinfo {author} {\bibfnamefont {Niklas}\ \bibnamefont {Roschewsky}}, \bibinfo {author} {\bibfnamefont {Badih~A.}\ \bibnamefont {Assaf}}, \bibinfo {author} {\bibfnamefont {Marius}\ \bibnamefont {Eich}}, \bibinfo {author} {\bibfnamefont {Marguerite}\ \bibnamefont {Epstein-Martin}}, \bibinfo {author} {\bibfnamefont {Don}\ \bibnamefont {Heiman}}, \bibinfo {author} {\bibfnamefont {Markus}\ \bibnamefont {M\"unzenberg}}, \ and\ \bibinfo {author} {\bibfnamefont {Jagadeesh~S.}\ \bibnamefont {Moodera}},\ }\bibfield  {title} {\enquote {\bibinfo {title} {Superconducting spin switch with infinite magnetoresistance induced by an internal exchange field},}\ }\href {\doibase 10.1103/PhysRevLett.110.097001} {\bibfield  {journal} {\bibinfo  {journal} {Phys. Rev. Lett.}\ }\textbf {\bibinfo {volume} {110}},\ \bibinfo {pages} {097001} (\bibinfo {year} {2013})}\BibitemShut {NoStop}%
\bibitem [{\citenamefont {Moraru}\ \emph {et~al.}(2006)\citenamefont {Moraru}, \citenamefont {Pratt},\ and\ \citenamefont {Birge}}]{Moraru2006}%
  \BibitemOpen
  \bibfield  {author} {\bibinfo {author} {\bibfnamefont {Ion~C.}\ \bibnamefont {Moraru}}, \bibinfo {author} {\bibfnamefont {W.~P.}\ \bibnamefont {Pratt}}, \ and\ \bibinfo {author} {\bibfnamefont {Norman~O.}\ \bibnamefont {Birge}},\ }\bibfield  {title} {\enquote {\bibinfo {title} {Magnetization-dependent ${T}_{c}$ shift in ferromagnet/superconductor/ferromagnet trilayers with a strong ferromagnet},}\ }\href {\doibase 10.1103/PhysRevLett.96.037004} {\bibfield  {journal} {\bibinfo  {journal} {Phys. Rev. Lett.}\ }\textbf {\bibinfo {volume} {96}},\ \bibinfo {pages} {037004} (\bibinfo {year} {2006})}\BibitemShut {NoStop}%
\bibitem [{\citenamefont {Singh}\ \emph {et~al.}(2007)\citenamefont {Singh}, \citenamefont {S\"urgers},\ and\ \citenamefont {L\"ohneysen}}]{Singh2007}%
  \BibitemOpen
  \bibfield  {author} {\bibinfo {author} {\bibfnamefont {A.}~\bibnamefont {Singh}}, \bibinfo {author} {\bibfnamefont {C.}~\bibnamefont {S\"urgers}}, \ and\ \bibinfo {author} {\bibfnamefont {H.~v.}\ \bibnamefont {L\"ohneysen}},\ }\bibfield  {title} {\enquote {\bibinfo {title} {Superconducting spin switch with perpendicular magnetic anisotropy},}\ }\href {\doibase 10.1103/PhysRevB.75.024513} {\bibfield  {journal} {\bibinfo  {journal} {Phys. Rev. B}\ }\textbf {\bibinfo {volume} {75}},\ \bibinfo {pages} {024513} (\bibinfo {year} {2007})}\BibitemShut {NoStop}%
\bibitem [{\citenamefont {Zhu}\ \emph {et~al.}(2010)\citenamefont {Zhu}, \citenamefont {Krivorotov}, \citenamefont {Halterman},\ and\ \citenamefont {Valls}}]{Zhu2010}%
  \BibitemOpen
  \bibfield  {author} {\bibinfo {author} {\bibfnamefont {Jian}\ \bibnamefont {Zhu}}, \bibinfo {author} {\bibfnamefont {Ilya~N.}\ \bibnamefont {Krivorotov}}, \bibinfo {author} {\bibfnamefont {Klaus}\ \bibnamefont {Halterman}}, \ and\ \bibinfo {author} {\bibfnamefont {Oriol~T.}\ \bibnamefont {Valls}},\ }\bibfield  {title} {\enquote {\bibinfo {title} {Angular dependence of the superconducting transition temperature in ferromagnet-superconductor-ferromagnet trilayers},}\ }\href {\doibase 10.1103/PhysRevLett.105.207002} {\bibfield  {journal} {\bibinfo  {journal} {Phys. Rev. Lett.}\ }\textbf {\bibinfo {volume} {105}},\ \bibinfo {pages} {207002} (\bibinfo {year} {2010})}\BibitemShut {NoStop}%
\bibitem [{\citenamefont {Leksin}\ \emph {et~al.}(2012)\citenamefont {Leksin}, \citenamefont {Garif'yanov}, \citenamefont {Garifullin}, \citenamefont {Fominov}, \citenamefont {Schumann}, \citenamefont {Krupskaya}, \citenamefont {Kataev}, \citenamefont {Schmidt},\ and\ \citenamefont {B\"uchner}}]{Leksin2012}%
  \BibitemOpen
  \bibfield  {author} {\bibinfo {author} {\bibfnamefont {P.~V.}\ \bibnamefont {Leksin}}, \bibinfo {author} {\bibfnamefont {N.~N.}\ \bibnamefont {Garif'yanov}}, \bibinfo {author} {\bibfnamefont {I.~A.}\ \bibnamefont {Garifullin}}, \bibinfo {author} {\bibfnamefont {Ya.~V.}\ \bibnamefont {Fominov}}, \bibinfo {author} {\bibfnamefont {J.}~\bibnamefont {Schumann}}, \bibinfo {author} {\bibfnamefont {Y.}~\bibnamefont {Krupskaya}}, \bibinfo {author} {\bibfnamefont {V.}~\bibnamefont {Kataev}}, \bibinfo {author} {\bibfnamefont {O.~G.}\ \bibnamefont {Schmidt}}, \ and\ \bibinfo {author} {\bibfnamefont {B.}~\bibnamefont {B\"uchner}},\ }\bibfield  {title} {\enquote {\bibinfo {title} {Evidence for triplet superconductivity in a superconductor-ferromagnet spin valve},}\ }\href {\doibase 10.1103/PhysRevLett.109.057005} {\bibfield  {journal} {\bibinfo  {journal} {Phys. Rev. Lett.}\ }\textbf {\bibinfo {volume} {109}},\ \bibinfo {pages} {057005} (\bibinfo {year} {2012})}\BibitemShut {NoStop}%
\bibitem [{\citenamefont {Banerjee}\ \emph {et~al.}(2014)\citenamefont {Banerjee}, \citenamefont {Smiet}, \citenamefont {Smits}, \citenamefont {Ozaeta}, \citenamefont {Bergeret}, \citenamefont {Blamire},\ and\ \citenamefont {Robinson}}]{Banerjee2014}%
  \BibitemOpen
  \bibfield  {author} {\bibinfo {author} {\bibfnamefont {N.}~\bibnamefont {Banerjee}}, \bibinfo {author} {\bibfnamefont {C.~B.}\ \bibnamefont {Smiet}}, \bibinfo {author} {\bibfnamefont {R.~G.~J.}\ \bibnamefont {Smits}}, \bibinfo {author} {\bibfnamefont {A.}~\bibnamefont {Ozaeta}}, \bibinfo {author} {\bibfnamefont {F.~S.}\ \bibnamefont {Bergeret}}, \bibinfo {author} {\bibfnamefont {M.~G.}\ \bibnamefont {Blamire}}, \ and\ \bibinfo {author} {\bibfnamefont {J.~W.~A.}\ \bibnamefont {Robinson}},\ }\bibfield  {title} {\enquote {\bibinfo {title} {Evidence for spin selectivity of triplet pairs in superconducting spin valves},}\ }\href {\doibase 10.1038/ncomms4048} {\bibfield  {journal} {\bibinfo  {journal} {Nat. Commun.}\ }\textbf {\bibinfo {volume} {5}},\ \bibinfo {pages} {3048} (\bibinfo {year} {2014})}\BibitemShut {NoStop}%
\bibitem [{\citenamefont {Jara}\ \emph {et~al.}(2014)\citenamefont {Jara}, \citenamefont {Safranski}, \citenamefont {Krivorotov}, \citenamefont {Wu}, \citenamefont {Malmi-Kakkada}, \citenamefont {Valls},\ and\ \citenamefont {Halterman}}]{Jara2014}%
  \BibitemOpen
  \bibfield  {author} {\bibinfo {author} {\bibfnamefont {Alejandro~A.}\ \bibnamefont {Jara}}, \bibinfo {author} {\bibfnamefont {Christopher}\ \bibnamefont {Safranski}}, \bibinfo {author} {\bibfnamefont {Ilya~N.}\ \bibnamefont {Krivorotov}}, \bibinfo {author} {\bibfnamefont {Chien-Te}\ \bibnamefont {Wu}}, \bibinfo {author} {\bibfnamefont {Abdul~N.}\ \bibnamefont {Malmi-Kakkada}}, \bibinfo {author} {\bibfnamefont {Oriol~T.}\ \bibnamefont {Valls}}, \ and\ \bibinfo {author} {\bibfnamefont {Klaus}\ \bibnamefont {Halterman}},\ }\bibfield  {title} {\enquote {\bibinfo {title} {Angular dependence of superconductivity in superconductor/spin-valve heterostructures},}\ }\href {\doibase 10.1103/PhysRevB.89.184502} {\bibfield  {journal} {\bibinfo  {journal} {Phys. Rev. B}\ }\textbf {\bibinfo {volume} {89}},\ \bibinfo {pages} {184502} (\bibinfo {year} {2014})}\BibitemShut {NoStop}%
\bibitem [{\citenamefont {Singh}\ \emph {et~al.}(2015)\citenamefont {Singh}, \citenamefont {Voltan}, \citenamefont {Lahabi},\ and\ \citenamefont {Aarts}}]{Singh2015}%
  \BibitemOpen
  \bibfield  {author} {\bibinfo {author} {\bibfnamefont {A.}~\bibnamefont {Singh}}, \bibinfo {author} {\bibfnamefont {S.}~\bibnamefont {Voltan}}, \bibinfo {author} {\bibfnamefont {K.}~\bibnamefont {Lahabi}}, \ and\ \bibinfo {author} {\bibfnamefont {J.}~\bibnamefont {Aarts}},\ }\bibfield  {title} {\enquote {\bibinfo {title} {Colossal proximity effect in a superconducting triplet spin valve based on the half-metallic ferromagnet ${\mathrm{cro}}_{2}$},}\ }\href {\doibase 10.1103/PhysRevX.5.021019} {\bibfield  {journal} {\bibinfo  {journal} {Phys. Rev. X}\ }\textbf {\bibinfo {volume} {5}},\ \bibinfo {pages} {021019} (\bibinfo {year} {2015})}\BibitemShut {NoStop}%
\bibitem [{\citenamefont {Kamashev}\ \emph {et~al.}(2019)\citenamefont {Kamashev}, \citenamefont {Garif'yanov}, \citenamefont {Validov}, \citenamefont {Schumann}, \citenamefont {Kataev}, \citenamefont {B\"uchner}, \citenamefont {Fominov},\ and\ \citenamefont {Garifullin}}]{Kamashev2019}%
  \BibitemOpen
  \bibfield  {author} {\bibinfo {author} {\bibfnamefont {A.~A.}\ \bibnamefont {Kamashev}}, \bibinfo {author} {\bibfnamefont {N.~N.}\ \bibnamefont {Garif'yanov}}, \bibinfo {author} {\bibfnamefont {A.~A.}\ \bibnamefont {Validov}}, \bibinfo {author} {\bibfnamefont {J.}~\bibnamefont {Schumann}}, \bibinfo {author} {\bibfnamefont {V.}~\bibnamefont {Kataev}}, \bibinfo {author} {\bibfnamefont {B.}~\bibnamefont {B\"uchner}}, \bibinfo {author} {\bibfnamefont {Ya.~V.}\ \bibnamefont {Fominov}}, \ and\ \bibinfo {author} {\bibfnamefont {I.~A.}\ \bibnamefont {Garifullin}},\ }\bibfield  {title} {\enquote {\bibinfo {title} {Superconducting spin-valve effect in heterostructures with ferromagnetic heusler alloy layers},}\ }\href {\doibase 10.1103/PhysRevB.100.134511} {\bibfield  {journal} {\bibinfo  {journal} {Phys. Rev. B}\ }\textbf {\bibinfo {volume} {100}},\ \bibinfo {pages} {134511} (\bibinfo {year} {2019})}\BibitemShut {NoStop}%
\bibitem [{\citenamefont {Westerholt}\ \emph {et~al.}(2005)\citenamefont {Westerholt}, \citenamefont {Sprungmann}, \citenamefont {Zabel}, \citenamefont {Brucas}, \citenamefont {Hj\"orvarsson}, \citenamefont {Tikhonov},\ and\ \citenamefont {Garifullin}}]{Westerholt2005}%
  \BibitemOpen
  \bibfield  {author} {\bibinfo {author} {\bibfnamefont {K.}~\bibnamefont {Westerholt}}, \bibinfo {author} {\bibfnamefont {D.}~\bibnamefont {Sprungmann}}, \bibinfo {author} {\bibfnamefont {H.}~\bibnamefont {Zabel}}, \bibinfo {author} {\bibfnamefont {R.}~\bibnamefont {Brucas}}, \bibinfo {author} {\bibfnamefont {B.}~\bibnamefont {Hj\"orvarsson}}, \bibinfo {author} {\bibfnamefont {D.~A.}\ \bibnamefont {Tikhonov}}, \ and\ \bibinfo {author} {\bibfnamefont {I.~A.}\ \bibnamefont {Garifullin}},\ }\bibfield  {title} {\enquote {\bibinfo {title} {Superconducting spin valve effect of a v layer coupled to an antiferromagnetic $[\mathrm{Fe}/\mathrm{V}]$ superlattice},}\ }\href {\doibase 10.1103/PhysRevLett.95.097003} {\bibfield  {journal} {\bibinfo  {journal} {Phys. Rev. Lett.}\ }\textbf {\bibinfo {volume} {95}},\ \bibinfo {pages} {097003} (\bibinfo {year} {2005})}\BibitemShut {NoStop}%
\bibitem [{\citenamefont {Deutscher}\ and\ \citenamefont {Meunier}(1969)}]{Deutscher1969}%
  \BibitemOpen
  \bibfield  {author} {\bibinfo {author} {\bibfnamefont {G.}~\bibnamefont {Deutscher}}\ and\ \bibinfo {author} {\bibfnamefont {F.}~\bibnamefont {Meunier}},\ }\bibfield  {title} {\enquote {\bibinfo {title} {Coupling between ferromagnetic layers through a superconductor},}\ }\href {\doibase 10.1103/PhysRevLett.22.395} {\bibfield  {journal} {\bibinfo  {journal} {Phys. Rev. Lett.}\ }\textbf {\bibinfo {volume} {22}},\ \bibinfo {pages} {395--396} (\bibinfo {year} {1969})}\BibitemShut {NoStop}%
\bibitem [{\citenamefont {Gu}\ \emph {et~al.}(2002)\citenamefont {Gu}, \citenamefont {You}, \citenamefont {Jiang}, \citenamefont {Pearson}, \citenamefont {Bazaliy},\ and\ \citenamefont {Bader}}]{Gu2002}%
  \BibitemOpen
  \bibfield  {author} {\bibinfo {author} {\bibfnamefont {J.~Y.}\ \bibnamefont {Gu}}, \bibinfo {author} {\bibfnamefont {C.-Y.}\ \bibnamefont {You}}, \bibinfo {author} {\bibfnamefont {J.~S.}\ \bibnamefont {Jiang}}, \bibinfo {author} {\bibfnamefont {J.}~\bibnamefont {Pearson}}, \bibinfo {author} {\bibfnamefont {Ya.~B.}\ \bibnamefont {Bazaliy}}, \ and\ \bibinfo {author} {\bibfnamefont {S.~D.}\ \bibnamefont {Bader}},\ }\bibfield  {title} {\enquote {\bibinfo {title} {Magnetization-orientation dependence of the superconducting transition temperature in the ferromagnet-superconductor-ferromagnet system: $\mathrm{C}\mathrm{u}\mathrm{N}\mathrm{i}/\mathrm{N}\mathrm{b}/\mathrm{C}\mathrm{u}\mathrm{N}\mathrm{i}$},}\ }\href {\doibase 10.1103/PhysRevLett.89.267001} {\bibfield  {journal} {\bibinfo  {journal} {Phys. Rev. Lett.}\ }\textbf {\bibinfo {volume} {89}},\ \bibinfo {pages} {267001} (\bibinfo {year} {2002})}\BibitemShut {NoStop}%
\bibitem [{\citenamefont {Gu}\ \emph {et~al.}(2015)\citenamefont {Gu}, \citenamefont {Hal\'asz}, \citenamefont {Robinson},\ and\ \citenamefont {Blamire}}]{Gu2015}%
  \BibitemOpen
  \bibfield  {author} {\bibinfo {author} {\bibfnamefont {Yuanzhou}\ \bibnamefont {Gu}}, \bibinfo {author} {\bibfnamefont {G\'abor~B.}\ \bibnamefont {Hal\'asz}}, \bibinfo {author} {\bibfnamefont {J.~W.~A.}\ \bibnamefont {Robinson}}, \ and\ \bibinfo {author} {\bibfnamefont {M.~G.}\ \bibnamefont {Blamire}},\ }\bibfield  {title} {\enquote {\bibinfo {title} {Large superconducting spin valve effect and ultrasmall exchange splitting in epitaxial rare-earth-niobium trilayers},}\ }\href {\doibase 10.1103/PhysRevLett.115.067201} {\bibfield  {journal} {\bibinfo  {journal} {Phys. Rev. Lett.}\ }\textbf {\bibinfo {volume} {115}},\ \bibinfo {pages} {067201} (\bibinfo {year} {2015})}\BibitemShut {NoStop}%
\bibitem [{\citenamefont {Kamra}\ \emph {et~al.}(2023)\citenamefont {Kamra}, \citenamefont {Chourasia}, \citenamefont {Bobkov}, \citenamefont {Gordeeva}, \citenamefont {Bobkova},\ and\ \citenamefont {Kamra}}]{Johnsen_Kamra_2023}%
  \BibitemOpen
  \bibfield  {author} {\bibinfo {author} {\bibfnamefont {Lina~Johnsen}\ \bibnamefont {Kamra}}, \bibinfo {author} {\bibfnamefont {Simran}\ \bibnamefont {Chourasia}}, \bibinfo {author} {\bibfnamefont {G.~A.}\ \bibnamefont {Bobkov}}, \bibinfo {author} {\bibfnamefont {V.~M.}\ \bibnamefont {Gordeeva}}, \bibinfo {author} {\bibfnamefont {I.~V.}\ \bibnamefont {Bobkova}}, \ and\ \bibinfo {author} {\bibfnamefont {Akashdeep}\ \bibnamefont {Kamra}},\ }\bibfield  {title} {\enquote {\bibinfo {title} {Complete ${T}_{c}$ suppression and n\'eel triplets mediated exchange in antiferromagnet-superconductor-antiferromagnet trilayers},}\ }\href {\doibase 10.1103/PhysRevB.108.144506} {\bibfield  {journal} {\bibinfo  {journal} {Phys. Rev. B}\ }\textbf {\bibinfo {volume} {108}},\ \bibinfo {pages} {144506} (\bibinfo {year} {2023})}\BibitemShut {NoStop}%
\bibitem [{\citenamefont {Balatsky}\ \emph {et~al.}(2006)\citenamefont {Balatsky}, \citenamefont {Vekhter},\ and\ \citenamefont {Zhu}}]{Balatsky2006}%
  \BibitemOpen
  \bibfield  {author} {\bibinfo {author} {\bibfnamefont {A.~V.}\ \bibnamefont {Balatsky}}, \bibinfo {author} {\bibfnamefont {I.}~\bibnamefont {Vekhter}}, \ and\ \bibinfo {author} {\bibfnamefont {Jian-Xin}\ \bibnamefont {Zhu}},\ }\bibfield  {title} {\enquote {\bibinfo {title} {Impurity-induced states in conventional and unconventional superconductors},}\ }\href {\doibase 10.1103/RevModPhys.78.373} {\bibfield  {journal} {\bibinfo  {journal} {Rev. Mod. Phys.}\ }\textbf {\bibinfo {volume} {78}},\ \bibinfo {pages} {373--433} (\bibinfo {year} {2006})}\BibitemShut {NoStop}%
\bibitem [{\citenamefont {Luh}(1965)}]{Luh1965}%
  \BibitemOpen
  \bibfield  {author} {\bibinfo {author} {\bibfnamefont {Yu}~\bibnamefont {Luh}},\ }\bibfield  {title} {\enquote {\bibinfo {title} {Bound state in superconductors with paramagnetic impurities},}\ }\href {https://api.semanticscholar.org/CorpusID:123883367} {\bibfield  {journal} {\bibinfo  {journal} {Acta Physica Sinica}\ }\textbf {\bibinfo {volume} {21}},\ \bibinfo {pages} {75--91} (\bibinfo {year} {1965})}\BibitemShut {NoStop}%
\bibitem [{\citenamefont {Shiba}(1968)}]{Shiba1968}%
  \BibitemOpen
  \bibfield  {author} {\bibinfo {author} {\bibfnamefont {Hiroyuki}\ \bibnamefont {Shiba}},\ }\bibfield  {title} {\enquote {\bibinfo {title} {Classical spins in superconductors},}\ }\href {\doibase 10.1143/PTP.40.435} {\bibfield  {journal} {\bibinfo  {journal} {Progress of Theoretical Physics}\ }\textbf {\bibinfo {volume} {40}},\ \bibinfo {pages} {435--451} (\bibinfo {year} {1968})}\BibitemShut {NoStop}%
\bibitem [{\citenamefont {Rusinov}(1969)}]{Rusinov1969}%
  \BibitemOpen
  \bibfield  {author} {\bibinfo {author} {\bibfnamefont {A.~I.}\ \bibnamefont {Rusinov}},\ }\bibfield  {title} {\enquote {\bibinfo {title} {On the theory of gapless superconductivity in alloys containing paramagnetic impurities},}\ }\href@noop {} {\bibfield  {journal} {\bibinfo  {journal} {SOV. PHYS. JETP}\ }\textbf {\bibinfo {volume} {29}},\ \bibinfo {pages} {1101} (\bibinfo {year} {1969})}\BibitemShut {NoStop}%
\bibitem [{\citenamefont {Nadj-Perge}\ \emph {et~al.}(2014)\citenamefont {Nadj-Perge}, \citenamefont {Drozdov}, \citenamefont {Li}, \citenamefont {Chen}, \citenamefont {Jeon}, \citenamefont {Seo}, \citenamefont {MacDonald}, \citenamefont {Bernevig},\ and\ \citenamefont {Yazdani}}]{Nadj-Perge2014}%
  \BibitemOpen
  \bibfield  {author} {\bibinfo {author} {\bibfnamefont {Stevan}\ \bibnamefont {Nadj-Perge}}, \bibinfo {author} {\bibfnamefont {Ilya~K.}\ \bibnamefont {Drozdov}}, \bibinfo {author} {\bibfnamefont {Jian}\ \bibnamefont {Li}}, \bibinfo {author} {\bibfnamefont {Hua}\ \bibnamefont {Chen}}, \bibinfo {author} {\bibfnamefont {Sangjun}\ \bibnamefont {Jeon}}, \bibinfo {author} {\bibfnamefont {Jungpil}\ \bibnamefont {Seo}}, \bibinfo {author} {\bibfnamefont {Allan~H.}\ \bibnamefont {MacDonald}}, \bibinfo {author} {\bibfnamefont {B.~Andrei}\ \bibnamefont {Bernevig}}, \ and\ \bibinfo {author} {\bibfnamefont {Ali}\ \bibnamefont {Yazdani}},\ }\bibfield  {title} {\enquote {\bibinfo {title} {Observation of majorana fermions in ferromagnetic atomic chains on a superconductor},}\ }\href {\doibase 10.1126/science.1259327} {\bibfield  {journal} {\bibinfo  {journal} {Science}\ }\textbf {\bibinfo {volume} {346}},\ \bibinfo {pages} {602--607} (\bibinfo {year} {2014})}\BibitemShut {NoStop}%
\bibitem [{\citenamefont {Pawlak}\ \emph {et~al.}(2016)\citenamefont {Pawlak}, \citenamefont {Kisiel}, \citenamefont {Klinovaja}, \citenamefont {Meier}, \citenamefont {Kawai}, \citenamefont {Glatzel}, \citenamefont {Loss},\ and\ \citenamefont {Meyer}}]{Pawlak2016}%
  \BibitemOpen
  \bibfield  {author} {\bibinfo {author} {\bibfnamefont {R{\'e}my}\ \bibnamefont {Pawlak}}, \bibinfo {author} {\bibfnamefont {Marcin}\ \bibnamefont {Kisiel}}, \bibinfo {author} {\bibfnamefont {Jelena}\ \bibnamefont {Klinovaja}}, \bibinfo {author} {\bibfnamefont {Tobias}\ \bibnamefont {Meier}}, \bibinfo {author} {\bibfnamefont {Shigeki}\ \bibnamefont {Kawai}}, \bibinfo {author} {\bibfnamefont {Thilo}\ \bibnamefont {Glatzel}}, \bibinfo {author} {\bibfnamefont {Daniel}\ \bibnamefont {Loss}}, \ and\ \bibinfo {author} {\bibfnamefont {Ernst}\ \bibnamefont {Meyer}},\ }\bibfield  {title} {\enquote {\bibinfo {title} {Probing atomic structure and majorana wavefunctions in mono-atomic fe chains on superconducting pb surface},}\ }\href {\doibase 10.1038/npjqi.2016.35} {\bibfield  {journal} {\bibinfo  {journal} {npj Quantum Information}\ }\textbf {\bibinfo {volume} {2}},\ \bibinfo {pages} {16035} (\bibinfo {year} {2016})}\BibitemShut {NoStop}%
\bibitem [{\citenamefont {Schneider}\ \emph {et~al.}(2021)\citenamefont {Schneider}, \citenamefont {Beck}, \citenamefont {Posske}, \citenamefont {Crawford}, \citenamefont {Mascot}, \citenamefont {Rachel}, \citenamefont {Wiesendanger},\ and\ \citenamefont {Wiebe}}]{Schneider2021}%
  \BibitemOpen
  \bibfield  {author} {\bibinfo {author} {\bibfnamefont {Lucas}\ \bibnamefont {Schneider}}, \bibinfo {author} {\bibfnamefont {Philip}\ \bibnamefont {Beck}}, \bibinfo {author} {\bibfnamefont {Thore}\ \bibnamefont {Posske}}, \bibinfo {author} {\bibfnamefont {Daniel}\ \bibnamefont {Crawford}}, \bibinfo {author} {\bibfnamefont {Eric}\ \bibnamefont {Mascot}}, \bibinfo {author} {\bibfnamefont {Stephan}\ \bibnamefont {Rachel}}, \bibinfo {author} {\bibfnamefont {Roland}\ \bibnamefont {Wiesendanger}}, \ and\ \bibinfo {author} {\bibfnamefont {Jens}\ \bibnamefont {Wiebe}},\ }\bibfield  {title} {\enquote {\bibinfo {title} {Topological shiba bands in artificial spin chains on superconductors},}\ }\href {\doibase 10.1038/s41567-021-01234-y} {\bibfield  {journal} {\bibinfo  {journal} {Nature Physics}\ }\textbf {\bibinfo {volume} {17}},\ \bibinfo {pages} {943--948} (\bibinfo {year} {2021})}\BibitemShut {NoStop}%
\bibitem [{\citenamefont {Gor'kov}\ and\ \citenamefont {Rashba}(2001)}]{Gorkov2001}%
  \BibitemOpen
  \bibfield  {author} {\bibinfo {author} {\bibfnamefont {Lev~P.}\ \bibnamefont {Gor'kov}}\ and\ \bibinfo {author} {\bibfnamefont {Emmanuel~I.}\ \bibnamefont {Rashba}},\ }\bibfield  {title} {\enquote {\bibinfo {title} {Superconducting 2d system with lifted spin degeneracy: Mixed singlet-triplet state},}\ }\href {\doibase 10.1103/PhysRevLett.87.037004} {\bibfield  {journal} {\bibinfo  {journal} {Phys. Rev. Lett.}\ }\textbf {\bibinfo {volume} {87}},\ \bibinfo {pages} {037004} (\bibinfo {year} {2001})}\BibitemShut {NoStop}%
\bibitem [{\citenamefont {Annunziata}\ \emph {et~al.}(2012)\citenamefont {Annunziata}, \citenamefont {Manske},\ and\ \citenamefont {Linder}}]{Annunziata2012}%
  \BibitemOpen
  \bibfield  {author} {\bibinfo {author} {\bibfnamefont {Gaetano}\ \bibnamefont {Annunziata}}, \bibinfo {author} {\bibfnamefont {Dirk}\ \bibnamefont {Manske}}, \ and\ \bibinfo {author} {\bibfnamefont {Jacob}\ \bibnamefont {Linder}},\ }\bibfield  {title} {\enquote {\bibinfo {title} {Proximity effect with noncentrosymmetric superconductors},}\ }\href {\doibase 10.1103/PhysRevB.86.174514} {\bibfield  {journal} {\bibinfo  {journal} {Phys. Rev. B}\ }\textbf {\bibinfo {volume} {86}},\ \bibinfo {pages} {174514} (\bibinfo {year} {2012})}\BibitemShut {NoStop}%
\bibitem [{\citenamefont {Bergeret}\ and\ \citenamefont {Tokatly}(2013)}]{Bergeret2013}%
  \BibitemOpen
  \bibfield  {author} {\bibinfo {author} {\bibfnamefont {F.~S.}\ \bibnamefont {Bergeret}}\ and\ \bibinfo {author} {\bibfnamefont {I.~V.}\ \bibnamefont {Tokatly}},\ }\bibfield  {title} {\enquote {\bibinfo {title} {Singlet-triplet conversion and the long-range proximity effect in superconductor-ferromagnet structures with generic spin dependent fields},}\ }\href {\doibase 10.1103/PhysRevLett.110.117003} {\bibfield  {journal} {\bibinfo  {journal} {Phys. Rev. Lett.}\ }\textbf {\bibinfo {volume} {110}},\ \bibinfo {pages} {117003} (\bibinfo {year} {2013})}\BibitemShut {NoStop}%
\bibitem [{\citenamefont {Bergeret}\ and\ \citenamefont {Tokatly}(2014)}]{Bergeret2014}%
  \BibitemOpen
  \bibfield  {author} {\bibinfo {author} {\bibfnamefont {F.~S.}\ \bibnamefont {Bergeret}}\ and\ \bibinfo {author} {\bibfnamefont {I.~V.}\ \bibnamefont {Tokatly}},\ }\bibfield  {title} {\enquote {\bibinfo {title} {Spin-orbit coupling as a source of long-range triplet proximity effect in superconductor-ferromagnet hybrid structures},}\ }\href {\doibase 10.1103/PhysRevB.89.134517} {\bibfield  {journal} {\bibinfo  {journal} {Phys. Rev. B}\ }\textbf {\bibinfo {volume} {89}},\ \bibinfo {pages} {134517} (\bibinfo {year} {2014})}\BibitemShut {NoStop}%
\bibitem [{\citenamefont {Edelstein}(2003{\natexlab{a}})}]{Edelstein2003}%
  \BibitemOpen
  \bibfield  {author} {\bibinfo {author} {\bibfnamefont {Victor~M.}\ \bibnamefont {Edelstein}},\ }\bibfield  {title} {\enquote {\bibinfo {title} {Triplet superconductivity and magnetoelectric effect near the s-wave-superconductor-- normal-metal interface caused by local breaking of mirror symmetry},}\ }\href {\doibase 10.1103/PhysRevB.67.020505} {\bibfield  {journal} {\bibinfo  {journal} {Phys. Rev. B}\ }\textbf {\bibinfo {volume} {67}},\ \bibinfo {pages} {020505} (\bibinfo {year} {2003}{\natexlab{a}})}\BibitemShut {NoStop}%
\bibitem [{\citenamefont {Edelstein}(2003{\natexlab{b}})}]{Edelstein2003_JETPLett}%
  \BibitemOpen
  \bibfield  {author} {\bibinfo {author} {\bibfnamefont {V.~M.}\ \bibnamefont {Edelstein}},\ }\bibfield  {title} {\enquote {\bibinfo {title} {Influence of an interface double electric layer on the superconducting proximity effect in ferromagnetic metals},}\ }\href {\doibase 10.1134/1.1571878} {\bibfield  {journal} {\bibinfo  {journal} {JETP Letters}\ }\textbf {\bibinfo {volume} {77}},\ \bibinfo {pages} {182--186} (\bibinfo {year} {2003}{\natexlab{b}})}\BibitemShut {NoStop}%
\bibitem [{\citenamefont {Jacobsen}\ \emph {et~al.}(2015)\citenamefont {Jacobsen}, \citenamefont {Ouassou},\ and\ \citenamefont {Linder}}]{Jacobsen2015}%
  \BibitemOpen
  \bibfield  {author} {\bibinfo {author} {\bibfnamefont {Sol~H.}\ \bibnamefont {Jacobsen}}, \bibinfo {author} {\bibfnamefont {Jabir~Ali}\ \bibnamefont {Ouassou}}, \ and\ \bibinfo {author} {\bibfnamefont {Jacob}\ \bibnamefont {Linder}},\ }\bibfield  {title} {\enquote {\bibinfo {title} {Critical temperature and tunneling spectroscopy of superconductor-ferromagnet hybrids with intrinsic rashba-dresselhaus spin-orbit coupling},}\ }\href {\doibase 10.1103/PhysRevB.92.024510} {\bibfield  {journal} {\bibinfo  {journal} {Phys. Rev. B}\ }\textbf {\bibinfo {volume} {92}},\ \bibinfo {pages} {024510} (\bibinfo {year} {2015})}\BibitemShut {NoStop}%
\bibitem [{\citenamefont {Ouassou}\ \emph {et~al.}(2016)\citenamefont {Ouassou}, \citenamefont {Di~Bernardo}, \citenamefont {Robinson},\ and\ \citenamefont {Linder}}]{Ouassou2016}%
  \BibitemOpen
  \bibfield  {author} {\bibinfo {author} {\bibfnamefont {Jabir~Ali}\ \bibnamefont {Ouassou}}, \bibinfo {author} {\bibfnamefont {Angelo}\ \bibnamefont {Di~Bernardo}}, \bibinfo {author} {\bibfnamefont {Jason W.~A.}\ \bibnamefont {Robinson}}, \ and\ \bibinfo {author} {\bibfnamefont {Jacob}\ \bibnamefont {Linder}},\ }\bibfield  {title} {\enquote {\bibinfo {title} {Electric control of superconducting transition through a spin-orbit coupled interface},}\ }\href {\doibase 10.1038/srep29312} {\bibfield  {journal} {\bibinfo  {journal} {Scientific Reports}\ }\textbf {\bibinfo {volume} {6}},\ \bibinfo {pages} {29312} (\bibinfo {year} {2016})}\BibitemShut {NoStop}%
\bibitem [{\citenamefont {Simensen}\ and\ \citenamefont {Linder}(2018)}]{Simensen2018}%
  \BibitemOpen
  \bibfield  {author} {\bibinfo {author} {\bibfnamefont {Haakon~T.}\ \bibnamefont {Simensen}}\ and\ \bibinfo {author} {\bibfnamefont {Jacob}\ \bibnamefont {Linder}},\ }\bibfield  {title} {\enquote {\bibinfo {title} {Tunable superconducting critical temperature in ballistic hybrid structures with strong spin-orbit coupling},}\ }\href {\doibase 10.1103/PhysRevB.97.054518} {\bibfield  {journal} {\bibinfo  {journal} {Phys. Rev. B}\ }\textbf {\bibinfo {volume} {97}},\ \bibinfo {pages} {054518} (\bibinfo {year} {2018})}\BibitemShut {NoStop}%
\bibitem [{\citenamefont {Banerjee}\ \emph {et~al.}(2018)\citenamefont {Banerjee}, \citenamefont {Ouassou}, \citenamefont {Zhu}, \citenamefont {Stelmashenko}, \citenamefont {Linder},\ and\ \citenamefont {Blamire}}]{Banerjee2018}%
  \BibitemOpen
  \bibfield  {author} {\bibinfo {author} {\bibfnamefont {N.}~\bibnamefont {Banerjee}}, \bibinfo {author} {\bibfnamefont {J.~A.}\ \bibnamefont {Ouassou}}, \bibinfo {author} {\bibfnamefont {Y.}~\bibnamefont {Zhu}}, \bibinfo {author} {\bibfnamefont {N.~A.}\ \bibnamefont {Stelmashenko}}, \bibinfo {author} {\bibfnamefont {J.}~\bibnamefont {Linder}}, \ and\ \bibinfo {author} {\bibfnamefont {M.~G.}\ \bibnamefont {Blamire}},\ }\bibfield  {title} {\enquote {\bibinfo {title} {Controlling the superconducting transition by spin-orbit coupling},}\ }\href {\doibase 10.1103/PhysRevB.97.184521} {\bibfield  {journal} {\bibinfo  {journal} {Phys. Rev. B}\ }\textbf {\bibinfo {volume} {97}},\ \bibinfo {pages} {184521} (\bibinfo {year} {2018})}\BibitemShut {NoStop}%
\bibitem [{\citenamefont {Johnsen}\ \emph {et~al.}(2019)\citenamefont {Johnsen}, \citenamefont {Banerjee},\ and\ \citenamefont {Linder}}]{Jonsen2019}%
  \BibitemOpen
  \bibfield  {author} {\bibinfo {author} {\bibfnamefont {Lina~G.}\ \bibnamefont {Johnsen}}, \bibinfo {author} {\bibfnamefont {Niladri}\ \bibnamefont {Banerjee}}, \ and\ \bibinfo {author} {\bibfnamefont {Jacob}\ \bibnamefont {Linder}},\ }\bibfield  {title} {\enquote {\bibinfo {title} {Magnetization reorientation due to the superconducting transition in heavy-metal heterostructures},}\ }\href {\doibase 10.1103/PhysRevB.99.134516} {\bibfield  {journal} {\bibinfo  {journal} {Phys. Rev. B}\ }\textbf {\bibinfo {volume} {99}},\ \bibinfo {pages} {134516} (\bibinfo {year} {2019})}\BibitemShut {NoStop}%
\bibitem [{\citenamefont {Gonz\'alez-Ruano}\ \emph {et~al.}(2020)\citenamefont {Gonz\'alez-Ruano}, \citenamefont {Johnsen}, \citenamefont {Caso}, \citenamefont {Tiusan}, \citenamefont {Hehn}, \citenamefont {Banerjee}, \citenamefont {Linder},\ and\ \citenamefont {Aliev}}]{Gonzalez-Ruano2020}%
  \BibitemOpen
  \bibfield  {author} {\bibinfo {author} {\bibfnamefont {C\'esar}\ \bibnamefont {Gonz\'alez-Ruano}}, \bibinfo {author} {\bibfnamefont {Lina~G.}\ \bibnamefont {Johnsen}}, \bibinfo {author} {\bibfnamefont {Diego}\ \bibnamefont {Caso}}, \bibinfo {author} {\bibfnamefont {Coriolan}\ \bibnamefont {Tiusan}}, \bibinfo {author} {\bibfnamefont {Michel}\ \bibnamefont {Hehn}}, \bibinfo {author} {\bibfnamefont {Niladri}\ \bibnamefont {Banerjee}}, \bibinfo {author} {\bibfnamefont {Jacob}\ \bibnamefont {Linder}}, \ and\ \bibinfo {author} {\bibfnamefont {Farkhad~G.}\ \bibnamefont {Aliev}},\ }\bibfield  {title} {\enquote {\bibinfo {title} {Superconductivity-induced change in magnetic anisotropy in epitaxial ferromagnet-superconductor hybrids with spin-orbit interaction},}\ }\href {\doibase 10.1103/PhysRevB.102.020405} {\bibfield  {journal} {\bibinfo  {journal} {Phys. Rev. B}\ }\textbf {\bibinfo {volume} {102}},\ \bibinfo {pages} {020405} (\bibinfo {year} {2020})}\BibitemShut {NoStop}%
\bibitem [{\citenamefont {Gonz{\'a}lez-Ruano}\ \emph {et~al.}(2021)\citenamefont {Gonz{\'a}lez-Ruano}, \citenamefont {Caso}, \citenamefont {Johnsen}, \citenamefont {Tiusan}, \citenamefont {Hehn}, \citenamefont {Banerjee}, \citenamefont {Linder},\ and\ \citenamefont {Aliev}}]{Gonzalez-Ruano2021}%
  \BibitemOpen
  \bibfield  {author} {\bibinfo {author} {\bibfnamefont {C{\'e}sar}\ \bibnamefont {Gonz{\'a}lez-Ruano}}, \bibinfo {author} {\bibfnamefont {Diego}\ \bibnamefont {Caso}}, \bibinfo {author} {\bibfnamefont {Lina~G.}\ \bibnamefont {Johnsen}}, \bibinfo {author} {\bibfnamefont {Coriolan}\ \bibnamefont {Tiusan}}, \bibinfo {author} {\bibfnamefont {Michel}\ \bibnamefont {Hehn}}, \bibinfo {author} {\bibfnamefont {Niladri}\ \bibnamefont {Banerjee}}, \bibinfo {author} {\bibfnamefont {Jacob}\ \bibnamefont {Linder}}, \ and\ \bibinfo {author} {\bibfnamefont {Farkhad~G.}\ \bibnamefont {Aliev}},\ }\bibfield  {title} {\enquote {\bibinfo {title} {Superconductivity assisted change of the perpendicular magnetic anisotropy in v/mgo/fe junctions},}\ }\href {\doibase 10.1038/s41598-021-98079-5} {\bibfield  {journal} {\bibinfo  {journal} {Scientific Reports}\ }\textbf {\bibinfo {volume} {11}},\ \bibinfo {pages} {19041} (\bibinfo {year} {2021})}\BibitemShut {NoStop}%
\bibitem [{\citenamefont {Bobkov}\ \emph {et~al.}(2023{\natexlab{c}})\citenamefont {Bobkov}, \citenamefont {Bobkova},\ and\ \citenamefont {Golubov}}]{Bobkov2023_anisotropy}%
  \BibitemOpen
  \bibfield  {author} {\bibinfo {author} {\bibfnamefont {G.~A.}\ \bibnamefont {Bobkov}}, \bibinfo {author} {\bibfnamefont {I.~V.}\ \bibnamefont {Bobkova}}, \ and\ \bibinfo {author} {\bibfnamefont {A.~A.}\ \bibnamefont {Golubov}},\ }\bibfield  {title} {\enquote {\bibinfo {title} {Magnetic anisotropy of the superconducting transition in superconductor/antiferromagnet heterostructures with spin-orbit coupling},}\ }\href {\doibase 10.1103/PhysRevB.108.L060507} {\bibfield  {journal} {\bibinfo  {journal} {Phys. Rev. B}\ }\textbf {\bibinfo {volume} {108}},\ \bibinfo {pages} {L060507} (\bibinfo {year} {2023}{\natexlab{c}})}\BibitemShut {NoStop}%
\end{thebibliography}%

\end{document}